\documentclass[10pt,final,onecolumn]{IEEEtran}

\PassOptionsToPackage{}{tocbibind}
\PassOptionsToPackage{style=ieee}{biblatex}
\usepackage[tables=false,colorlinks,bibtex=true]{preamble}
\graphicspath{{/}}

\makeatletter
	\ifbool{@bibtex}{}{
		\addbibresource{IEEEabrv,UWA,PaperList,ownpub}
	}
\makeatother

\title{Identifiability Scaling Laws in Bilinear Inverse Problems}
\author{Sunav~Choudhary,~\IEEEmembership{Student~Member,~IEEE,} and~Urbashi~Mitra,~\IEEEmembership{Fellow,~IEEE}%
		\thanks{This work has been funded in part by the following grants and organizations:~ONR~N00014-09-1-0700, NSF~CNS-0832186 and NSF~CCF-1117896. Parts of this paper were presented at the IEEE International Conference on Acoustic, Speech, and Signal Processing (ICASSP), Vancouver, Canada, May 26-31, 2013~\cite{choudhary2012onidentifiability} and at the 47th Asilomar Conference on Signals, Systems and Computers, Pacific Grove, California, Nov. 3-6, 2013~\cite{choudhary2012identifiabilitybounds}.}%
		\thanks{S.~Choudhary and U.~Mitra are with the Ming Hsieh Department of Electrical Engineering, Viterbi School of Engineering, University of Southern California, Los Angeles CA 90089, USA (email: \href{mailto:sunavcho@usc.edu}{\protect\nolinkurl{sunavcho@usc.edu}}, \href{mailto:ubli@usc.edu}{\protect\nolinkurl{ubli@usc.edu}})}}

\begin{document}
	\maketitle

	\begin{abstract}
		A number of ill-posed inverse problems in signal processing, like blind deconvolution, matrix factorization, dictionary learning and blind source separation share the common characteristic of being bilinear inverse problems (BIPs), \ie~the observation model is a function of two variables and conditioned on one variable being known, the observation is a linear function of the other variable.
		A key issue that arises for such inverse problems is that of identifiability, \ie~whether the observation is sufficient to unambiguously determine the pair of inputs that generated the observation.
		Identifiability is a key concern for applications like blind equalization in wireless communications and data mining in machine learning.
		Herein, a unifying and flexible approach to identifiability analysis for general \emph{conic prior} constrained BIPs is presented, exploiting a connection to low-rank matrix recovery via `lifting'.
		We develop deterministic identifiability conditions on the input signals and examine their satisfiability in practice for three classes of signal distributions, \viz~dependent but uncorrelated, independent Gaussian, and independent Bernoulli.
		In each case, scaling laws are developed that trade-off probability of robust identifiability with the complexity of the rank two null space.
		An added appeal of our approach is that the rank two null space can be partly or fully characterized for many bilinear problems of interest (\eg~blind deconvolution).
		We present numerical experiments involving variations on the blind deconvolution problem that exploit a characterization of the rank two null space and demonstrate that the scaling laws offer good estimates of identifiability.
	\end{abstract}

	\IEEEpeerreviewmaketitle

	\section{Introduction}
		\label{sec:intro}
		\IEEEPARstart{W}{e} examine the problem of identifiability in bilinear inverse problems (BIPs), \ie~input signal pair recovery for systems where the output is a bilinear function of two unknown inputs.
		Important practical examples of BIPs include blind deconvolution~\cite{hopgood2003}, blind source separation~\cite{grady2005survey} and dictionary learning~\cite{xing2012dictionary} in signal processing, matrix factorization in machine learning~\cite{donoho2003whendoes}, blind equalization in wireless communications~\cite{johnson1998blind}, \etc{}
		Of particular interest are signal recovery problems from \textit{under-determined} systems of measurement where additional structure is needed in order to ensure recovery, and the observation model is \emph{non-linear} in the parametrization of the problem.
		
		Consider a discrete-time blind linear deconvolution problem.
		Let $\vec{x} \in \mathcal{D}_{\vec{x}}$ and $\vec{y} \in \mathcal{D}_{\vec{y}}$ be respectively $m$ and $n$ dimensional vectors from domains $\mathcal{D}_{\vec{x}} \subseteq \setR^{m}$ and $\mathcal{D}_{\vec{y}} \subseteq \setR^{n}$, and suppose that the noise free linear convolution of $\vec{x}$ and $\vec{y}$ is observed as $\vec{z}$.
		Then the blind linear deconvolution problem can be represented as the following feasibility problem.
		\find{\bb{\vec{x}, \vec{y}}}
		{\vec{x} \star \vec{y} = \vec{z}, \sep \vec{x} \in \mathcal{D}_{\vec{x}}, \vec{y} \in \mathcal{D}_{\vec{y}}.}
		{\label{prob:find_xy_deconv}}
		We draw the reader's attention to the observation/measurement model $\vec{z} = \vec{x} \star \vec{y}$.
		Notice that if either $\vec{x}$ or $\vec{y}$ was a fixed and known quantity, then we would have an observation model that is linear in the other variable.
		However, when both $\vec{x}$ and $\vec{y}$ are unknown variables, then the linear convolution measurement model $\vec{z} = \vec{x} \star \vec{y}$ is no longer linear in the variable pair $\bb{\vec{x}, \vec{y}}$.
		Such a structural characteristic is referred to as a \textit{bilinear} measurement structure (formally defined in \sectionname~\ref{sec:model}).
		The blind linear deconvolution problem~\eqref{prob:find_xy_deconv} is the resulting inverse problem.
		Such inverse problems arising from a bilinear measurement structure shall be referred to as \textit{bilinear inverse problems} (formally defined in \sectionname~\ref{sec:model}).

		A key issue in many under-determined inverse problems is that of \textit{identifiability}: ``Does a unique solution exist that satisfies the given observations?''
		Identifiability (and signal reconstruction) for \textit{linear} inverse problems with sparsity and low-rank structures has received considerable attention in the context of compressed sensing and low-rank matrix recovery, respectively, and are now quite well understood~\cite{chandrasekaran2012convex}.
		In a nutshell, both compressed sensing and low-rank matrix recovery theories guarantee that the unknown sparse/low-rank signal can be \textit{unambiguously} reconstructed from relatively few properly designed linear measurements using algorithms with runtime growing polynomially in the signal dimension.
		For non-linear inverse problems (including BIPs), however, characterization of identifiability (and signal reconstruction) still remains largely open.
		To illustrate that analyzing identifiability is nontrivial, we present a simple example.
		Consider the blind linear deconvolution problem represented by \problemname~\eqref{prob:find_xy_deconv}.
		Suppose that we have the observation $\vec{z} = \tpose{\bb{1,0,1,0,1,0,1,0,1,0,1,0,1,0,1,0,0}} \in \setR^{17}$ with $\mathcal{D}_{\vec{x}} = \setR^{7}$ and $\mathcal{D}_{\vec{y}} = \setR^{11}$.
		It is not difficult to verify that both
		\begin{subequations}
			\begin{alignat}{2}
				\SwapAboveDisplaySkip
				\vec{x}	& = \tpose{\bb{1,0,0,0,1,0,0}},	& \; \vec{y}	& = \tpose{\bb{1,0,1,0,0,0,0,0,1,0,1}} \\
				\shortintertext{and,}
				\vec{x}	& = \tpose{\bb{1,0,1,0,1,0,1}},	& \; \vec{y}	& = \tpose{\bb{1,0,0,0,0,0,0,0,1,0,0}}
			\end{alignat}
		\end{subequations}
		are valid solutions to \problemname~\eqref{prob:find_xy_deconv}.
		Furthermore, it is not immediately obvious as to what structural constraints would disambiguate between the above two solutions.
		We have showed identifiability and constructed fast recovery algorithms in a previous work~\cite{choudhary2012sparse} when $\vec{x}$ (possibly sparse) is in the non-negative orthant (modulo global sign flip), whereas we show negative results for the more general sparse (with respect to the canonical basis) blind deconvolution problem in~\cite{choudhary2014sbdidentifiability,choudhary2014identifiabilitylimitsSBD}.

		\subsection{Contributions}
			\label{sec:contributions}
			\begin{enumerate}
				\item	We cast conic prior constrained BIPs as low-rank matrix recovery problems, establish the validity of the `lifting' procedure (\sectionname~\ref{sec:lifting}) and develop deterministic sufficient conditions for identifiability (\sectionname~\ref{sec:deterministic identifiability}) while bridging the gap to necessary conditions in a special case.
						Our characterization agrees with the intuition that identifiability subject to priors should depend on the joint geometry of the signal space and the bilinear map.
						Our results are geared towards bilinear maps that admit a nontrivial rank two null space, as is the case with many important BIPs like blind deconvolution.
				\item	We develop trade-offs between probability of identifiability of a random instance and the complexity of the rank two null space of the lifted bilinear map under three classes of signal ensembles, \viz~dependent but uncorrelated, independent Gaussian, and independent Bernoulli (\sectionname~\ref{sec:random identifiability}).
						Specifically, we demonstrate that instance identifiability can be characterized by the complexity of restricted rank two null space, measured by the covering number of the set $\set{\bb{\mathcal{C}\argd{\mat{X}}, \mathcal{R}\argd{\mat{X}}}}{\mat{X} \in \mathcal{N}\argd{\mathscr{S}, 2} \bigcap \mathcal{M} \setminus \cc{\mat{0}}}$, where $\mathcal{C}\bb{\mat{X}}$ and $\mathcal{R}\argd{\mat{X}}$ denote, respectively, the column and row spaces of the matrix $\mat{X}$ and $\mathcal{N}\argd{\mathscr{S}, 2} \bigcap \mathcal{M}$ denotes the rank two null space of the lifted bilinear map $\mathscr{S}\bb{\cdot}$ restricted by the prior on the signal set to $\mathcal{M}$.
						To the best of our knowledge, this gives new structural results solely based on the bilinear measurement model and is thus applicable to general BIPs.
				\item	We demonstrate that the rank two null space of the lifted bilinear map can be partly characterized in at least one important case (blind deconvolution), and conjecture that the same should be possible for other bilinear maps of interest (dictionary learning, blind source separation, \etc).
						Based on this characterization, we present numerical simulations for selected variations on the blind deconvolution problem to demonstrate the tightness of our scaling laws (\sectionname~\ref{sec:numerical results}).
			\end{enumerate}

		\subsection{Related Work}
			\label{sec:prior art}
			Our treatment of BIPs draws on several different ideas.
			We employ `lifting' from optimization~\cite{balas2005projection} which enables the creation of good relaxations for intractable optimization problems.
			This can come at the expense of an increase in the ambient dimension of the optimization variables.
			Lifting was used in~\cite{candes2011phaselift} for analyzing the phase retrieval problem and in~\cite{ahmed2012blind} for the analysis of blind circular deconvolution.
			We employ lifting in the same spirit as~\cite{candes2011phaselift,ahmed2012blind} but our \textit{goals are different}.
			Firstly, we deal with general BIPs which include the linear convolution model of~\cite{asif2009random}, the circular convolution model of~\cite{hegde2011sampling,ahmed2012blind} and the compressed bilinear observation model of~\cite{walk2012compressed} as special cases.
			Secondly, we focus solely on identifiability (as opposed to recoverability by convex optimization~\cite{candes2011phaselift,ahmed2012blind}) enabling far milder assumptions on the distribution of the input signals.
			
			After lifting, we have a rank one matrix recovery problem, subject to inherited conic constraints.
			While encouraging results have been shown for low-rank matrix recovery using the nuclear norm heuristic~\cite{gross2011recovering}, quite stringent incoherence assumptions are needed between the sampling operator and the true matrix.
			Furthermore, the results do not generalize to an analysis of identifiability when the sampling operator admits rank two matrices in its null space.
			We are able to relax the incoherence assumptions in special cases for analyzing identifiability and also consider sampling operators with a non-trivial rank two null space.
			Since the works \cite{recht2011nullspace,candes2011tight,lee2013nearoptimalcompressed} can be interpreted as solving BIPs with the lifted map drawn from a Gaussian random ensemble, thus leading to a trivial rank two null space with high probability, the results therein are not directly comparable to our results.

			In~\cite{ahmed2012blind}, a recoverability analysis for the blind circular deconvolution problem is undertaken, but the knowledge of the sparsity pattern of one input signal is needed.
			Taking our \problemname~\eqref{prob:find_xy_deconv} as an example,~\cite{ahmed2012blind} assumes $\mathcal{D}_{\vec{x}} = \mathcal{C}\argd{\mat{B}}$ and $\mathcal{D}_{\vec{y}} = \mathcal{C}\argd{\mat{C}}$ for some \textit{tall} deterministic matrix $\mat{B}$ and a \textit{tall} Gaussian random matrix $\mat{C}$, where for any matrix $\mat{X}$, $\mathcal{C}\argd{\mat{X}}$ denotes the column space of $\mat{X}$.
			In contrast, we shall make the less stringent assumption on $\vec{x}$ and $\vec{y}$ and show that \textit{identifiability} holds with high probability in the presence of rank two matrices in the null space of the lifted linear operator (sampling operator).
			
			A closely related (but different) problem is that of retrieving the phase of a signal from the magnitude of its Fourier coefficients (the Fourier phase retrieval problem).
			This is equivalent to recovering the signal given its auto correlation function~\cite{jaganathan2013phase}.
			In terms of our example blind deconvolution problem~\eqref{prob:find_xy_deconv}, phase retrieval is equivalent to having the additional constraints $\mathcal{D}_{\vec{x}} = \mathcal{D}_{\vec{y}}$ and $\vec{y}$ being the time reversed version of $\vec{x}$.
			While the Fourier phase retrieval problem may seem superficially similar to the blind deconvolution problem, there are major differences between the two, so much as to ensure identifiability and efficient recoverability for the sparsity regularized (in the canonical basis) version of the former~\cite{jaganathan2013sparsephaseretrieval}, while one can explicitly show \textit{unidentifiability} for the sparsity regularized (in the canonical basis) version of the latter~\cite{choudhary2014sbdidentifiability,choudhary2014identifiabilitylimitsSBD} (even with oracle knowledge of the supports of both signals).
			The difference arises because the Fourier phase retrieval problem is a (non-convex) quadratic inverse problem rather than a BIP, and it satisfies additional properties (constant trace of the lifted variable) which make it better conditioned for efficient recovery algorithms~\cite{beck2009matrixQP}.

			For the dictionary learning problem, an identifiability analysis is developed in~\cite{kammoun2010robustness} leveraging results from~\cite{gribonval2010dictionary} on matrix factorization for sparse dictionary learning using the $\ell_1$ norm and $\ell_p$ quasi-norm for $0 < p < 1$.
			More recently, exact recoverability of over-complete dictionaries from training samples (only polynomially large in the dimensions of the dictionary) has been proved in~\cite{agarwal2013exactrecoveryof} assuming sparse (but unknown) coefficient matrix.
			While every BIP can be recast as a dictionary learning problem in principle, such a transformation would result in additional structural constraints on the dictionary that may or may not be trivial to incorporate in the existing analyses.
			This is especially true for bilinear maps over vector pairs.
			In contrast, we develop our methods to specifically target bilinear maps over vector pairs (\eg~convolution map) and thus obtain definitive results where the dictionary learning based formulations would most likely fail.

			Some identifiability results for blind deconvolution are summarized in~\cite{meraim1997blind}, but the treatment therein is inflexible to the inclusion of side information about the input signals.
			Identifiability for non-negative matrix factorization was examined in~\cite{donoho2003whendoes} exploiting geometric properties of the non-negative orthant.
			Although our results can be easily visualized in terms of geometry, they can also be stated purely in terms of linear algebra (\theoremname~\ref{thm:suff_ident}).
			Identifiability results for low-rank matrix completion~\cite{candes2009exact,candes2010thepower} are provided in~\cite{kiraly2012combinatorial} via algebraic and combinatorial conditions using graph theoretic tools, but there is no straightforward way to extend these results to more general lifted linear operators like the convolution map.
			Overall, to the best of our knowledge, a unified flexible treatment of identifiability in BIPs has not been developed till date.
			In this paper, we present such a framework incorporating conic constraints on the input signals (which includes sparse signals in particular).

		\subsection{Organization, Reading Guide and Notation}
			\label{sec:organization}
			The remainder of the paper is organized as follows.
			The first half of \sectionname~\ref{sec:model} formally introduces BIPs and a working definition of identifiability.
			\sectionname~\ref{sec:lifting} describes the lifting technique to reformulate BIPs as rank one matrix recovery problems, and characterizes the validity of the technique.
			\sectionname~\ref{sec:results} states our main results on both deterministic and random instance identifiability.
			\sectionname~\ref{sec:discussion} elaborates on the intuitions, ideas, assumptions and subtle implications associated with the results of \sectionname~\ref{sec:results}.
			\sectionname~\ref{sec:numerical results} is devoted to results of numerical verification and \sectionname~\ref{sec:conclusion} concludes the paper.
			Detailed proofs of all the results in the paper appear in the Appendices~\ref{sec:equivalence theorem proof}-\ref{sec:rank-2 null space proposition proof}.

			In order to maintain linearity of exposition to the greatest extent possible, we chose to create a separate section (\sectionname~\ref{sec:discussion}) for elaborating on intuitions, ideas, assumptions and implications associated with the important results of the paper.
			Thus, with the exception of \sectionname~\ref{sec:discussion}, rest of the paper can be read in a linear fashion.
			However, we recommend the reader to switch between \sectionsname~\ref{sec:results} and~\ref{sec:discussion} as necessary, to better interpret the results presented in \sectionname~\ref{sec:results}.

			We state the notational conventions used throughout rest of the paper.
			All vectors are assumed to be column vectors unless stated otherwise.
			We shall use lowercase boldface alphabets to denote column vectors~(\eg~$\vec{z}$) and uppercase boldface alphabets to denote matrices~(\eg~$\mat{A}$).
			The all zero (respectively all one) vector/matrix shall be denoted by $\vec{0}$ (respectively $\vec{1}$) and the identity matrix by $\eye$.
			The canonical base matrices for the space of $m \times n$ real matrices will be denoted by $\mat{E}_{i,j}$ for $1 \leq i \leq m$, $1 \leq j \leq n$ and is defined (element-wise) as
			\begin{equation}
				\bb{\mat{E}_{i,j}}_{k,l} =	\begin{cases}
												1,	& i = k, j = l, \\
												0,	& \text{otherwise}.
											\end{cases}
			\end{equation}
			For vectors and/or matrices, $\tpose{\bb{\cdot}}$, $\trace{\cdot}$ and $\rank{\cdot}$ respectively denote the transpose, trace and rank of their argument, whenever applicable.
			Special sets are denoted by uppercase blackboard bold font~(\eg~$\setR$ for real numbers).
			Other sets are denoted by uppercase calligraphic font~(\eg~$\mathcal{S}$).
%			For any set $\mathcal{S}$, $\card{\mathcal{S}}$ shall denote its cardinality.
			Linear operators on matrices are denoted by uppercase script font~(\eg~$\mathscr{S}$).
			The set of all matrices of rank at most $k$ in the null space of a linear operator $\mathscr{S}$ will be denoted by~$\mathcal{N}\argd{\mathscr{S}, k}$, defined as
			\begin{equation}
				\mathcal{N}\argd{\mathscr{S}, k} \triangleq \set{\mat{X} \in \setR^{m \times n}}{\rank{\mat{X}} \leq k,\, \mathscr{S}\argd{\mat{X}} = \vec{0}},
				\label{eqn:rank-k null space}
			\end{equation}
			and referred to as the `rank $k$ null space'.
			For any matrix $\mat{X}$, we denote the row and column spaces by $\mathcal{R}\argd{\mat{X}}$ and $\mathcal{C}\argd{\mat{X}}$ respectively.
			The projection matrix onto the column space (respectively row space) of $\mat{X}$ shall be denoted by $\mat{P}_{\mathcal{C}\bb{\mat{X}}}$ (respectively $\mat{P}_{\mathcal{R}\bb{\mat{X}}}$).
			For any rank one matrix $\mat{M}$, an expression of the form $\mat{M} = \sigma \vec{u} \tpose{\vec{v}}$ would denote the singular value decomposition of $\mat{M}$ with vectors $\vec{u}$ and $\vec{v}$ each admitting unit \ltwonorm.
			The standard Euclidean inner product on a vector space will be denoted by $\ip{\cdot}{\cdot}$ and the underlying vector space will be clear from the usage context.
			All logarithms are with respect to (\wrt) base $e$ unless specified otherwise.
			We shall use the $\BigOh{h}$, $\litOh{h}$ and $\Theta\argd{h}$ notation to denote order of growth of any function $f \fcolon \setR \to \setR$ of $h \in \setR$ \wrt~its argument.
			We have,
			\begin{subequations}
				\begin{align}
					f\argd{h} = \BigOh{h} & \iff \lim_{h \to \infty} \frac{f\argd{h}}{h} < \infty,	\\
					f\argd{h} = \litOh{h} & \iff \lim_{h \to \infty} \frac{f\argd{h}}{h} = 0,	\\
					f\argd{h} = \Theta\argd{h} & \iff \lim_{h \to \infty} \frac{f\argd{h}}{h} \in \bb{0,\infty}.
				\end{align}
			\end{subequations}

	\section{System Model}
		\label{sec:model}
		This section introduces the bilinear observation model and the associated bilinear inverse problem in \subsectionname~\ref{sec:bilinear maps} and our working definition of identifiability in \subsectionname~\ref{sec:identifiability defn}.
		\subsectionname~\ref{sec:lifting} describes the equivalent linear inverse problem obtained by lifting and conditions under which the equivalence holds.
		This equivalence is used to establish all of our identifiability results in \sectionname~\ref{sec:results}.

		\subsection{Bilinear Maps and Bilinear Inverse Problems (BIPs)}
			\label{sec:bilinear maps}
			
			\begin{definition}[Bilinear Map]
				\label{defn:bilinear map}
				A mapping $\vec{S} \fcolon \setR^{m} \times \setR^{n} \to \setR^{q}$ is called a bilinear map if $\vec{S}\argd{\cdot, \vec{y}} \fcolon \setR^{m} \to \setR^{q}$ is a linear map $\forall \vec{y} \in \setR^{n}$ and $\vec{S}\argd{\vec{x}, \cdot} \fcolon \setR^{n} \to \setR^{q}$ is a linear map $\forall \vec{x} \in \setR^{m}$.
			\end{definition}
			We shall consider the generic bilinear system/measurement model introduced in~\cite{choudhary2012onidentifiability},
			\begin{equation}
				\vec{z} = \vec{S}\argd{\vec{x}, \vec{y}},
				\label{eqn:model}
			\end{equation}
			where $\vec{z}$ is the vector of observations, $\vec{S} \fcolon \setR^{m} \times \setR^{n} \to \setR^{q}$ is a given bilinear map, and $\bb{\vec{x}, \vec{y}}$ denotes the pair of unknown signals with a given domain restriction $\bb{\vec{x}, \vec{y}} \in \mathcal{K}$.
			We are interested in solving for vectors $\vec{x}$ and $\vec{y}$ from the noiseless observation $\vec{z}$ as given by \eqref{eqn:model}.
			The BIP corresponding to the observation model \eqref{eqn:model} is represented by the following feasibility problem.
			\find{\bb{\vec{x}, \vec{y}}}
			{\vec{S}\argd{\vec{x}, \vec{y}} = \vec{z}, \sep \bb{\vec{x}, \vec{y}} \in \mathcal{K}.}
			{\label{prob:find_xy}}
			The non-negative matrix factorization problem~\cite{donoho2003whendoes} serves as an illustrative example of such a problem.
			Let $\mat{X} \in \setR^{m \times k}$ and $\mat{Y} \in \setR^{k \times n}$ be two element-wise non-negative, unknown matrices and suppose that we observe the matrix product $\mat{Z} = \mat{X} \mat{Y}$ which clearly has a bilinear structure.
			The non-negative matrix factorization problem is represented by the feasibility problem
			\find{\bb{\mat{X}, \mat{Y}}}
			{\mat{Z} = \mat{X} \mat{Y}, \sep \mat{X} \geq \mat{0}, \mat{Y} \geq \mat{0}.}
			{\label{prob:NMF_XY}}
			where the expressions $\mat{X} \geq \mat{0}$ and $\mat{Y} \geq \mat{0}$ constrain the matrices $\mat{X}$ and $\mat{Y}$ to be elementwise non-negative.
			The elementwise non-negativity constraints $\mat{X} \geq \mat{0}, \mat{Y} \geq \mat{0}$ form a domain restriction in \problemname~\eqref{prob:NMF_XY}, in the same way as the constraint $\bb{\vec{x}, \vec{y}} \in \mathcal{K}$ serves to restrict the feasible set in \problemname~\eqref{prob:find_xy}.

		\subsection{Identifiability Definition}
			\label{sec:identifiability defn}
			Notice that \textit{every} BIP has an inherent scaling ambiguity due to the identity
			\begin{equation}
				\vec{S}\argd{\vec{x}, \vec{y}} = \vec{S}\argd{\alpha \vec{x}, \frac{1}{\alpha} \vec{y}}, \quad \forall \alpha \neq 0,
				\label{eqn:scaling ambiguity}
			\end{equation}
			where $\vec{S}\argd{\cdot, \cdot}$ represents the bilinear map.
			Thus, a meaningful definition of identifiability, in the context of BIPs, must disregard this type of scaling ambiguity.
			This leads us to the following definition of identifiability.

			\begin{definition}[Identifiability]
				\label{defn:identifiability}
				A vector pair $\bb{\vec{x}, \vec{y}} \in \mathcal{K} \subseteq \setR^{m} \times \setR^{n}$ is identifiable \wrt~the bilinear map $\vec{S} \fcolon \setR^{m} \times \setR^{n} \to \setR^{q}$ if $\forall \bb{\vec{x}', \vec{y}'} \in \mathcal{K} \subseteq \setR^{m} \times \setR^{n}$ satisfying $\vec{S}\argd{\vec{x}, \vec{y}} = \vec{S}\argd{\vec{x}', \vec{y}'}$, $\exists \alpha \neq 0$ such that $\bb{\vec{x}', \vec{y}'} = \bb{\alpha \vec{x}, \frac{1}{\alpha} \vec{y}}$.
			\end{definition}

			\begin{remark}
				\label{rem:equivalence classes}
				It is straightforward to see that our definition of identifiability in turn defines an equivalence class of solutions.
				Thus, we seek to identify the equivalence class induced by the observation $\vec{z}$ in \eqref{eqn:model}.
				Later, in \sectionname~\ref{sec:lifting}, we shall `lift' \problemname~\eqref{prob:find_xy} to \problemname~\eqref{prob:rank} where, every equivalence class in the domain $\bb{\vec{x}, \vec{y}} \in \mathcal{K}$ of the former problem maps to a single point in the domain $\mat{W} \in \mathcal{K}'$ of the latter problem.
			\end{remark}

			\begin{remark}
				\label{rem:ambiguities}
				The scaling ambiguity represented by \eqref{eqn:scaling ambiguity} is common to all BIPs and our definition of identifiability (\definitionname~\ref{defn:identifiability}) only allows for this kind of ambiguity.
				There may be other types of ambiguities depending on the specific BIP.
				For example, the forward system model associated with \problemname~\eqref{prob:NMF_XY} is given by the matrix product operation $\mat{S}\argd{\mat{X}, \mat{Y}} = \mat{X} \mat{Y}$ which shows the following matrix multiplication ambiguity.
				\begin{equation}
					\mat{S}\argd{\mat{X}, \mat{Y}} = \mat{S}\argd{\mat{X} \mat{T}, \inv{\mat{T}} \mat{Y}}
				\end{equation}
				where $\inv{\mat{T}}$ is the right inverse of $\mat{T}$.
				It is possible to define weaker notions of identifiability to allow for this kind of ambiguity.
				In this paper, we shall not address this question any further and limit ourselves to the stricter notion of identifiability as given by \definitionname~\ref{defn:identifiability}.
			\end{remark}

		\subsection{Lifting}
			\label{sec:lifting}
			While \problemname~\eqref{prob:find_xy} is an accurate representation of the class of BIPs, the formulation does not easily lend itself to an identifiability analysis.
			We next rewrite \problemname~\eqref{prob:find_xy} to facilitate analysis, subject to some technical conditions~(see \theoremname~\ref{thm:equivalence} and \corollaryname~\ref{cor:equivalence}).
			The equivalent problem is a matrix rank minimization problem subject to linear equality constraints
			\minimize{\mat{W}}
			{\rank{\mat{W}}}
			{\mathscr{S}\argd{\mat{W}} = \vec{z}, \sep \mat{W} \in \mathcal{K}',}
			{\label{prob:rank}}
			where $\mathcal{K}' \subseteq \setR^{m \times n}$ is \textit{any} set satisfying
			\begin{equation}
				\mathcal{K}' \bigcap \set{\mat{W} \in \setR^{m \times n}}{\rank{\mat{W}} \leq 1}
				= \set{\vec{x} \tpose{\vec{y}}}{\bb{\vec{x}, \vec{y}} \in \mathcal{K}},
				\label{eqn:set change}
			\end{equation}
			and $\mathscr{S} \fcolon \setR^{m \times n} \to \setR^{q}$ is a linear operator that can be \textit{deterministically} constructed from the bilinear map $\vec{S}\argd{\cdot, \cdot}$ with the optimization variable $\mat{W}$ in \problemname~\eqref{prob:rank} being related to the optimization variable pair $\bb{\vec{x}, \vec{y}}$ in \problemname~\eqref{prob:find_xy} by the relation $\mat{W} = \vec{x} \tpose{\vec{y}}$.
			The transformation of \problemname~\eqref{prob:find_xy} to \problemname~\eqref{prob:rank} is an example of `lifting' and we shall refer to $\mathscr{S}\argd{\cdot}$ as the `lifted linear operator' \wrt~the bilinear map $\vec{S}\argd{\cdot, \cdot}$.
			Other examples on lifting can be found in~\cite{candes2011phaselift,ahmed2012blind}.
			Before stating the equivalence results between \problemsname~\eqref{prob:find_xy} and \eqref{prob:rank} we describe the construction of $\mathscr{S}\argd{\cdot}$ from $\vec{S}\argd{\cdot, \cdot}$.

			Let $\phi_j \fcolon \setR^{q} \to \setR$ be the $j^{\thp}$ coordinate projection operator of $q$ dimensional vectors to scalars, \ie~if $\vec{z} = \bb{z_{1},z_{2},\dotsc,z_{q}}$ then $\phi_{j}\argd{\vec{z}} = z_{j}$.
			Clearly, $\phi_j$ is a linear operator and hence the composition $\phi_j \circ \vec{S} \fcolon \setR^{m} \times \setR^{n} \to \setR$ is a bilinear map.
			As $\vec{S}$ is a finite dimensional operator, it is a bounded operator, hence by the Riesz Representation Theorem~\cite{rudin1987realandcomplex}, $\exists \mat{S}_j \in \setR^{m \times n}$ such that $\mat{S}_j$ is the unique linear operator satisfying
			\begin{equation}
				\phi_j \circ \vec{S}\argd{\vec{x},\vec{y}} = \ip{\vec{x}}{\mat{S}_j \vec{y}}, \quad \forall \vec{x} \in \setR^{m}, \vec{y} \in \setR^{n},
				\label{eqn:Sj}
			\end{equation}
			where $\ip{\cdot}{\cdot}$ denotes an inner product operation in $\setR^{m}$.
			Using \eqref{eqn:Sj}, we can convert the bilinear equality constraint in \problemname~\eqref{prob:find_xy} into a set of $q$ linear equality constraints as follows:
			\begin{equation}
				z_j = \phi_j \circ \vec{S}\argd{\vec{x}, \vec{y}} = \tpose{\vec{x}} \mat{S}_j\vec{y} = \ip{\vec{x}\tpose{\vec{y}}}{\mat{S}_j}
				\label{eqn:coordinate proj}
			\end{equation}
			for each $1 \leq j \leq q$, where the last inner product in \eqref{eqn:coordinate proj} is the trace inner product in the space $\setR^{m \times n}$ and $z_j$ denotes the $j^{\thp}$ coordinate of the observation vector $\vec{z}$.
			Setting $\mat{W} = \vec{x}\tpose{\vec{y}}$ in \eqref{eqn:coordinate proj}, the $q$ linear equality constraints in \eqref{eqn:coordinate proj} can be compactly represented, using operator notation, by the vector equality constraint~$\mathscr{S}\argd{\mat{W}} = \vec{z}$, where $\mathscr{S} \fcolon \setR^{m \times n} \to \setR^{q}$ is a linear operator acting on $\mat{W} \in \setR^{m \times n}$.
			This derivation uniquely specifies $\mathscr{S}\argd{\cdot}$ using the matrices $\mat{S}_{j}$, $1 \leq j \leq q$, and we have the identity
			\begin{equation}
				\mathscr{S}\argd{\vec{x} \tpose{\vec{y}}} = \vec{S}\argd{\vec{x}, \vec{y}}, \quad \forall \bb{\vec{x}, \vec{y}} \in \setR^{m} \times \setR^{n}.
				\label{eqn:lifted op}
			\end{equation}
			
			For the sake of completeness, we state the definitions of \textit{equivalence} and \textit{feasibility} in the context of optimization problems (\definitionsname~\ref{defn:equivalence} and~\ref{defn:feasibility}).
			Thereafter, the connection between \problemsname~\eqref{prob:find_xy} and \eqref{prob:rank} is described via the statements of \theoremname~\ref{thm:equivalence} and \corollaryname~\ref{cor:equivalence}.
			
			\begin{definition}[Equivalence of optimization problems]
				\label{defn:equivalence}
				Two optimization problems P and Q are said to be equivalent if every solution to P gives a solution to Q and every solution to Q gives a solution to P.
			\end{definition}
			
			\begin{definition}[Feasibility]
				\label{defn:feasibility}
				An optimization problem is said to be feasible, if the domain of the optimization variable is non-empty.
			\end{definition}

			\begin{theorem}
				\label{thm:equivalence}
				Let \problemname~\eqref{prob:find_xy} be feasible and let $\mathcal{K}_{\opt}$ and $\mathcal{K}'_{\opt}$ denote the set of solutions to \problemsname~\eqref{prob:find_xy} and~\eqref{prob:rank}, respectively.
				Then the following are true.
				\begin{enumerate}
					\item	\problemname~\eqref{prob:rank} is feasible with solution(s) of rank at most one.
					\item	$\mathcal{K}'_{\opt} \subseteq \set{\vec{x} \tpose{\vec{y}}}{\bb{\vec{x}, \vec{y}} \in \mathcal{K}_{\opt}}$.
					\item	$\mathcal{K}'_{\opt} = \set{\vec{x} \tpose{\vec{y}}}{\bb{\vec{x}, \vec{y}} \in \mathcal{K}_{\opt}}$ if and only if $\cc{\mat{0}} \subsetneq \set{\vec{x} \tpose{\vec{y}}}{\bb{\vec{x}, \vec{y}} \in \mathcal{K}_{\opt}}$ does not hold.
				\end{enumerate}
			\end{theorem}

			\begin{IEEEproof}
				\appendixname~\ref{sec:equivalence theorem proof}.
			\end{IEEEproof}

			Notice that $\mathcal{K}_{\opt}$ and $\mathcal{K}'_{\opt}$ in \theoremname~\ref{thm:equivalence} depend on the observation vector $\vec{z}$, so that the statements of \theoremname~\ref{thm:equivalence} have a hidden dependence on $\vec{z}$.
			Since the observation vector $\vec{z}$ is a function of the input signal pair $\bb{\vec{x}, \vec{y}}$ it is desirable to have statements analogous to \theoremname~\ref{thm:equivalence} that do not depend on the observation vector $\vec{z}$.
			This is the purpose of \corollaryname~\ref{cor:equivalence} below which makes use of $\mathcal{N}\argd{\mathscr{S}, 1}$, the rank one null space of the lifted operator $\mathscr{S}\argd{\cdot}$ (see \eqref{eqn:rank-k null space}).

			\begin{corollary}
				\label{cor:equivalence}
				Let \problemname~\eqref{prob:find_xy} be feasible and let $\mathcal{K}_{\opt}\argd{\vec{z}}$ and $\mathcal{K}'_{\opt}\argd{\vec{z}}$ respectively denote the set of optimal solutions to \problemsname~\eqref{prob:find_xy} and \eqref{prob:rank} for a given observation vector $\vec{z}$.
				\problemsname~\eqref{prob:find_xy} and \eqref{prob:rank} are equivalent, \ie~$\mathcal{K}'_{\opt}\argd{\vec{z}} = \set{\vec{x} \tpose{\vec{y}}}{\bb{\vec{x}, \vec{y}} \in \mathcal{K}_{\opt}\argd{\vec{z}}}$, for every $\vec{z} \in \set{\vec{S}\argd{\vec{x}, \vec{y}}}{\bb{\vec{x}, \vec{y}} \in \mathcal{K}}$ if and only if $\cc{\mat{0}} \subsetneq \mathcal{K}' \bigcap \mathcal{N}\argd{\mathscr{S}, 1}$ does not hold.
			\end{corollary}
			
			\begin{IEEEproof}
				\appendixname~\ref{sec:equivalence corollary proof}.
			\end{IEEEproof}

			\begin{remark}
				\label{rem:need for validating lifting}
				The statements of \theoremname~\ref{thm:equivalence} and \corollaryname~\ref{cor:equivalence} are needed to establish the validity of lifting for general BIPs with $\mathcal{N}\argd{\mathscr{S}, 1} \neq \cc{\mat{0}}$.
				In case $\mathcal{N}\argd{\mathscr{S}, 1} = \cc{\mat{0}}$ (\eg~blind deconvolution), \corollaryname~\ref{cor:equivalence} immediately implies that lifting is valid.
			\end{remark}

			\begin{remark}
				\label{rem:relax}
				Notice that lifting \problemname~\eqref{prob:find_xy} to \problemname~\eqref{prob:rank} allows us some freedom in the choice of the set $\mathcal{K}'$.
				Also, we have the additional \textit{side information} that the optimal solution to \problemname~\eqref{prob:rank} is a rank one matrix.
				These factors could be potentially helpful to develop tight and tractable relaxations to \problemname~\eqref{prob:rank}, that work better than the simple nuclear norm heuristic \cite{recht2007guaranteed} (\eg~see \cite{agarwal2013exactrecoveryof}).
				We do not pursue this question here.
			\end{remark}

			\begin{figure}
				\centering
				\includegraphics[width=\figwidth]{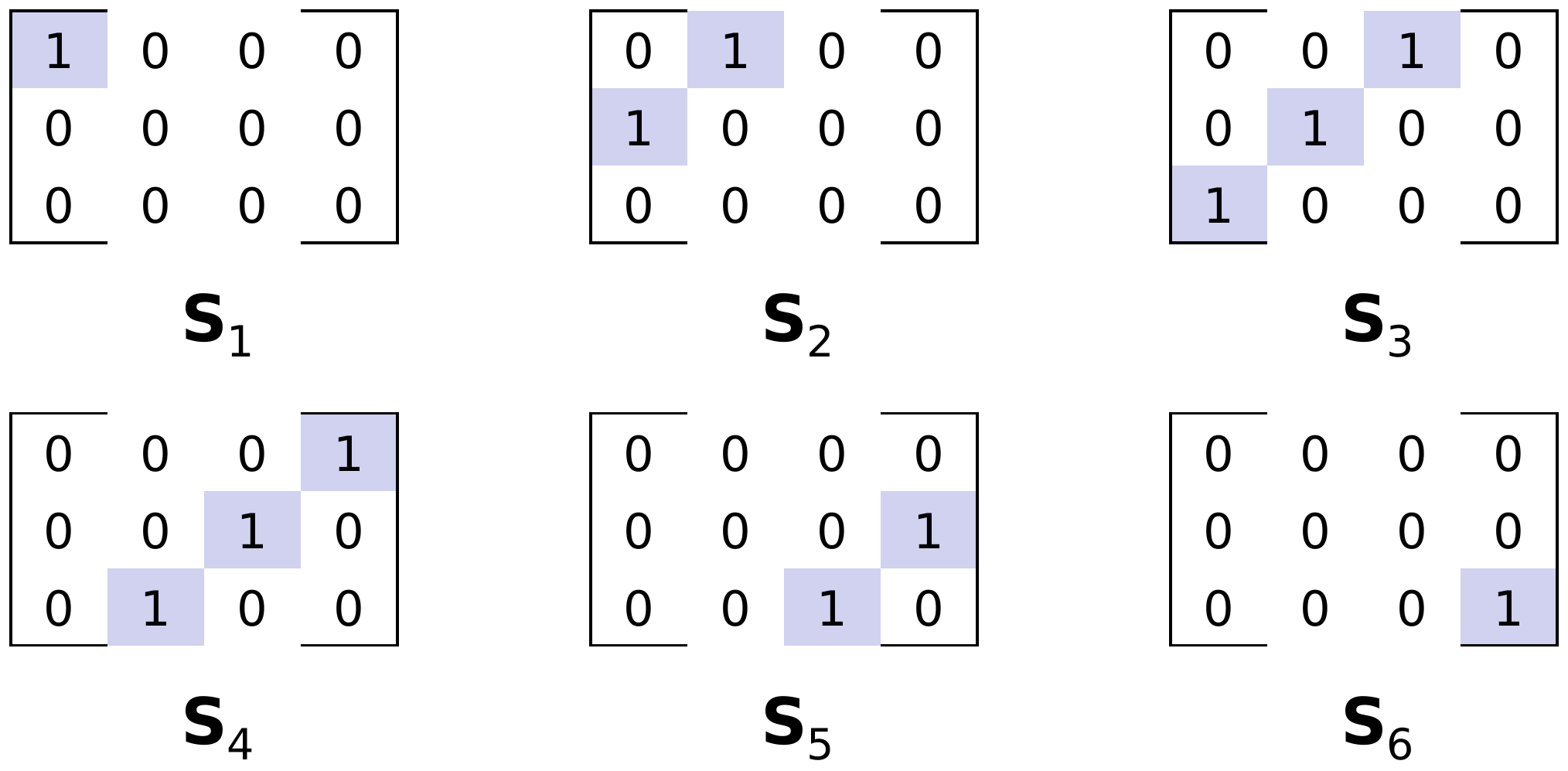}
				\caption{Lifted matrices $\mat{S}_{k} \in \setR^{m \times n}$ for linear convolution map with $m = 3$, $n = 4$, $q = m + n - 1 = 6$ and $1 \leq k \leq q$.}
				\label{fig:lifting example}
			\end{figure}

			This transformation from \problemname~\eqref{prob:find_xy} to \problemname~\eqref{prob:rank} gives us several advantages,
			\begin{enumerate}
				\item	\problemname~\eqref{prob:rank} has linear equality constraints as opposed to the bilinear equality constraints of \problemname~\eqref{prob:find_xy}.
						The former is much easier to handle from an optimization as well as algorithmic perspective than the latter.
				\item	Convex relaxation for the nonconvex rank constraint in \problemname~\eqref{prob:rank} is well known \cite{recht2007guaranteed}, which is an important requirement from an algorithmic perspective.
						In contrast, convex relaxation for a generic bilinear constraint is not known.
				\item	The bilinear map is completely determined by the set of matrices $\mat{S}_j$ and is separated from the variable $\mat{W}$ in \problemname~\eqref{prob:rank}.
						Thus, \problemname~\eqref{prob:rank} can be used to study generic BIPs.
						\figurename~\ref{fig:lifting example} illustrates a toy example involving the linear convolution map.
				\item	For every BIP there is an inherent scaling ambiguity (see \eqref{eqn:scaling ambiguity}) associated with the bilinear constraint.
						However, in \problemname~\eqref{prob:rank}, this scaling ambiguity has been taken care of implicitly when $\mat{W} = \vec{x}\tpose{\vec{y}}$ is the variable to be determined.
						Clearly, $\mat{W}$ is unaffected by the type of scaling ambiguity described in \eqref{eqn:scaling ambiguity}.
						Norm constraints on $\vec{x}$ or $\vec{y}$ can be used to recover $\vec{x}$ and $\vec{y}$ from $\mat{W}$ but these constraints do not affect \problemname~\eqref{prob:rank}.
				\item	If $\vec{x}$ and/or $\vec{y}$ are sparse in some known dictionary (possibly over-complete) then they can be absorbed into the mapping matrices $\mat{S}_j$ without altering the structure of \problemname~\eqref{prob:rank}.
						Indeed, if $\mat{A}$ and $\mat{B}$ are dictionaries such that $\vec{x} = \mat{A} \vec{\beta}$ and $\vec{y} = \mat{B} \vec{\gamma}$ then we have
						\begin{equation}
							\tpose{\vec{x}} \mat{S}_j\vec{y}
							= \tpose{\vec{\beta}} \bb{\tpose{\mat{A}} \mat{S}_j \mat{B}} \vec{\gamma}
							= \ip{\vec{\beta} \tpose{\vec{\gamma}}}{\tpose{\mat{A}} \mat{S}_j \mat{B}}
							\label{eqn:sparse basis transform}
						\end{equation}
						for each $1 \leq j \leq q$.
						It is clear that \problemname~\eqref{prob:rank} can be rewritten with $\mat{W} = \vec{\beta} \tpose{\vec{\gamma}}$ as the optimization variable (with a corresponding modification to $\mathcal{K}'$), and comparing \eqref{eqn:sparse basis transform} and \eqref{eqn:coordinate proj} we see that the matrix $\tpose{\mat{A}} \mat{S}_j \mat{B}$ can be designated to play the same role in the \emph{rewritten} \problemname~\eqref{prob:rank} as $\mat{S}_{j}$ played in the original \problemname~\eqref{prob:rank}.
						Thus, without loss of generality, we can consider \problemname~\eqref{prob:rank} to be our lifted problem that retains all available prior information from \problemname~\eqref{prob:find_xy} (assuming that the equivalence conditions in \corollaryname~\ref{cor:equivalence} are satisfied).
			\end{enumerate}

	\section{Identifiability Results}
		\label{sec:results}
		We state our main results in this section starting with deterministic characterizations of identifiability in \subsectionsname~\ref{sec:universal identifiability} and~\ref{sec:deterministic identifiability} that are simple to state but computationally hard to check for a given BIP.
		Subsequently, in \subsectionname~\ref{sec:random identifiability} we investigate whether identifiability holds for most inputs if the input is drawn from some distribution over the domain.

		Since we have some freedom of choice in the selection of the set $\mathcal{K}'$ according to \remarkname~\ref{rem:relax}, we will work with an arbitrary $\mathcal{K}'$ satisfying \eqref{eqn:set change}.
		The extreme cases of $\mathcal{K}' = \set{\vec{x} \tpose{\vec{y}}}{\bb{\vec{x}, \vec{y}} \in \mathcal{K}}$ and $\mathcal{K}' = \setR^{m \times n}$ will sometimes be used for examples and to build intuition.
		Also, for some of the results, we have converse statements only for one of the extreme cases.
		We shall use the set $\mathcal{M}$ to denote the difference $\mathcal{K}' - \mathcal{K}'$, defined as
		\begin{equation}
			\mathcal{M} = \mathcal{K}' - \mathcal{K}' \triangleq \set{\mat{X}_{1} - \mat{X}_{2}}{\mat{X}_{1}, \mat{X}_{2} \in \mathcal{K}'}.
		\end{equation}

		\subsection{Universal Identifiability}
			\label{sec:universal identifiability}
			As a straightforward consequence of lifting, we have the following necessary and sufficient condition for \problemname~\eqref{prob:rank} to succeed for all values of the observation $\vec{z} = \vec{S}\argd{\vec{x}, \vec{y}}$.

			\begin{proposition}
				\label{prop:ident_easy}
				Let $\mathcal{K}' = \set{\vec{x} \tpose{\vec{y}}}{\bb{\vec{x}, \vec{y}} \in \mathcal{K}}$. The solution to \problemname~\eqref{prob:rank} will be correct for every observation $\vec{z} = \vec{S}\argd{\vec{x}, \vec{y}}$ if and only if $\mathcal{N}\argd{\mathscr{S}, 2} \bigcap \mathcal{M} = \cc{\mat{0}}$.
			\end{proposition}

			\begin{IEEEproof}
				\appendixname~\ref{sec:ident_easy proposition proof}.
			\end{IEEEproof}

			\begin{remark}
				\label{rem:converse in proposition}
				Notice that the ``only if'' part of \propositionname~\ref{prop:ident_easy} requires uniqueness of an observation $\vec{z}$ that is valid for \problemname~\eqref{prob:find_xy} as well and not just for \problemname~\eqref{prob:rank}.
				The latter could have observations that arise because of the freedom in the choice of $\mathcal{K}'$, but those may not be valid for the former.
				As a result, the conclusion of the ``only if'' part of \propositionname~\ref{prop:ident_easy} is somewhat weaker in that it does not imply $\mathcal{N}\argd{\mathscr{S}, 2} = \cc{\mat{0}}$.
			\end{remark}
			
			When $\mathcal{K} = \setR^{m} \times \setR^{n}$, $\mathcal{M}$ represents the set of all rank two matrices in $\setR^{m \times n}$ so that \propositionname~\ref{prop:ident_easy} reduces to the more familiar result: $\mathcal{N}\argd{\mathscr{S}, 2} = \cc{\mat{0}}$ is necessary and sufficient for the action of the linear operator $\mathscr{S}$ to be invertible on the set of all rank one matrices, where the inversion of the action of $\mathscr{S}$ is achieved as the solution to \problemname~\eqref{prob:rank}.
			
			While the characterization of $\mathcal{N}\argd{\mathscr{S}, 2}$ for arbitrary linear operators $\mathscr{S}\argd{\cdot}$ is challenging, it has been shown that if $\mathscr{S}\argd{\cdot}$ is picked as a realization from some desirable distribution then $\mathcal{N}\argd{\mathscr{S}, 2} = \cc{\mat{0}}$ (implies $\mathcal{N}\argd{\mathscr{S}, 2} \bigcap \mathcal{M} = \cc{\mat{0}}$) is satisfied with high probability.
			As an example,~\cite{recht2011nullspace,candes2011tight} show that if $\mathscr{S} \fcolon \setR^{m \times n} \to \setR^{q}$ is picked from a Gaussian random ensemble, then $\mathcal{N}\argd{\mathscr{S}, 2} = \cc{\mat{0}}$ is satisfied with high probability for $q = \BigOh{\max\bb{m, n}}$.

		\subsection{Deterministic Instance Identifiability}
			\label{sec:deterministic identifiability}
			When $\mathscr{S}\argd{\cdot}$ is sampled from less desirable distributions, as for matrix completion~\cite{candes2009exact,candes2010thepower} or matrix recovery for a specific given basis~\cite{gross2011recovering}, one does \textit{not} have $\mathcal{N}\argd{\mathscr{S}, 2} = \cc{\mat{0}}$ with high probability.
			To guarantee identifiability (and unique reconstruction) for such realizations of $\mathscr{S}\argd{\cdot}$, significant domain restrictions via the set $\mathcal{K}$ (or $\mathcal{K}'$) are usually needed, so that $\mathcal{N}\argd{\mathscr{S}, 2} \bigcap \mathcal{M} = \cc{\mat{0}}$ and \propositionname~\ref{prop:ident_easy} comes into effect.
			Unfortunately, for many important BIPs (blind deconvolution, blind source separation, matrix factorization, \etc) the lifted linear operator $\mathscr{S}\argd{\cdot}$ \textit{does have} a non-trivial $\mathcal{N}\argd{\mathscr{S}, 2}$ set.
			This makes identifiability an important issue in practice.
			Fortunately, we still have $\mathcal{N}\argd{\mathscr{S}, 1} = \cc{\mat{0}}$ in many of these cases so that \corollaryname~\ref{cor:equivalence} implies that lifting is valid.
			For such maps, we have the following deterministic sufficient condition (\theoremname~\ref{thm:suff_ident}) for a rank one matrix $\mat{M} \in \mathcal{K}' \subseteq \setR^{m \times n}$ to be identifiable as a solution of \problemname~\eqref{prob:rank}.
			\theoremname~\ref{thm:suff_ident} is heavily used for the results in the sequel.

			\begin{theorem}
				\label{thm:suff_ident}
				Let $\mathcal{N}\argd{\mathscr{S}, 1} \bigcap \mathcal{M} = \cc{\mat{0}}$ and $\mat{M} = \sigma \vec{u} \tpose{\vec{v}}$ be a rank one matrix in $\mathcal{K}' \subseteq \setR^{m \times n}$.
				Suppose that for every $\mat{X} \in \mathcal{N}\argd{\mathscr{S}, 2} \bigcap \mathcal{M} \setminus \cc{\mat{0}}$ either $\vec{u} \notin \mathcal{C}\argd{\mat{X}}$ or $\vec{v} \notin \mathcal{R}\argd{\mat{X}}$ is true, then
				given the observation $\vec{z} = \mathscr{S}\argd{\mat{M}}$, $\mat{M}$ can be successfully recovered by solving \problemname~\eqref{prob:rank}.
			\end{theorem}

			\begin{IEEEproof}
				\appendixname~\ref{sec:suff_ident theorem proof}.
			\end{IEEEproof}

			\theoremname~\ref{thm:suff_ident} is only a sufficient condition for identifiability.
			We bridge the gap to the necessary conditions under a special case in \corollaryname~\ref{cor:equal singular values} below.
			We use the notation $\mat{M} - \mathcal{K}'$ to denote the set $\set{\mat{M} - \mat{Y}}{\mat{Y} \in \mathcal{K}'}$.
			
			\begin{corollary}
				\label{cor:equal singular values}
				Let $\mathcal{N}\argd{\mathscr{S}, 1} \bigcap \mathcal{M} = \cc{\mat{0}}$ and $\mat{M} = \sigma \vec{u} \tpose{\vec{v}}$ be a rank one matrix in $\mathcal{K}' \subseteq \setR^{m \times n}$.
				Suppose that every matrix $\mat{X} \in \mathcal{N}\argd{\mathscr{S}, 2} \bigcap \bb{\mat{M} - \mathcal{K}'} \setminus \cc{\mat{0}}$ admits a singular value decomposition with $\sigma_1\argd{\mat{X}} = \sigma_2\argd{\mat{X}}$.
				Let us denote such a decomposition as $\mat{X} = \sigma_{\ast} \vec{u}_1 \tpose{\vec{v}_1} + \sigma_{\ast} \vec{u}_2 \tpose{\vec{v}_2}$, and let $\vec{u} = \alpha_1 \vec{u}_1 + \alpha_2 \vec{u}_2$ and $\vec{v} = \alpha_3 \vec{v}_1 + \alpha_4 \vec{v}_2$ for some $\alpha_1, \alpha_2, \alpha_3, \alpha_4 \in \setR$ with $\alpha_1^2 + \alpha_2^2 = \alpha_3^2 + \alpha_4^2 = 1$.
				Given the observation $\vec{z} = \mathscr{S}\argd{\mat{M}}$, \problemname~\eqref{prob:rank} successfully recovers $\mat{M}$ if and only if for every $\mat{X} \in \mathcal{N}\argd{\mathscr{S}, 2} \bigcap \bb{\mat{M} - \mathcal{K}'} \setminus \cc{\mat{0}}$, $\alpha_1 \alpha_3 + \alpha_2 \alpha_4 \leq 0$.
			\end{corollary}

			\begin{IEEEproof}
				\appendixname~\ref{sec:equal singular values corollary proof}.
			\end{IEEEproof}

			Intuitively, \corollaryname~\ref{cor:equal singular values} exploits the fact that all nonzero singular values of a matrix are of the same sign.
			Indeed, $\bb{\alpha_{1}, \alpha_{2}}$ (respectively $\bb{\alpha_{3}, \alpha_{4}}$) is an element of the two dimensional space of representation coefficients of $\vec{u}$ \wrt~$\mathcal{C}\argd{\mat{X}}$ (respectively $\vec{v}$ \wrt~$\mathcal{R}\argd{\mat{X}}$) with a fixed representation basis.
			\corollaryname~\ref{cor:equal singular values} says that identifiability of $\mat{M}$ holds if and only if the vectors $\bb{\alpha_{1}, \alpha_{2}}$ and $\bb{\alpha_{3}, \alpha_{4}}$ do not form an acute angle between them.
			The assumption of $\sigma_1\argd{\mat{X}} = \sigma_2\argd{\mat{X}}$ has been made in \corollaryname~\ref{cor:equal singular values} for ease of intuition.
			Although we do not state it here, an analogous result holds for $\sigma_1\argd{\mat{X}} \neq \sigma_2\argd{\mat{X}}$ with the condition on the inner product $\ip{\bb{\alpha_{1}, \alpha_{2}}}{\bb{\alpha_{3}, \alpha_{4}}} \leq 0$ replaced by the same condition on a weighted inner product, where the weights depend on the ratio of $\sigma_1\argd{\mat{X}}$ to $\sigma_2\argd{\mat{X}}$.

			For arbitrary lifted linear operators $\mathscr{S}\argd{\cdot}$, checking \theoremname~\ref{thm:suff_ident} for a given rank one matrix $\mat{M}$ is usually hard, unless a simple characterization of $\mathcal{N}\argd{\mathscr{S}, 2}$ or $\mathcal{N}\argd{\mathscr{S}, 2} \bigcap \mathcal{M}$ has been provided.
			It is reasonable to ask ``How many rank one matrices $\mat{M}$ are identifiable?'', given any particular lifted linear operator $\mathscr{S}\argd{\cdot}$ and assuming that the rank one matrices $\mat{M}$ are drawn at random from some distribution.
			It is highly desirable if most rank one matrices $\mat{M}$ are identifiable.
			Before we can show such a result we need to define a random model for the rank one matrix $\mat{M}$.

		\subsection{A Random Rank One Model}
			\label{sec:random model}
			We consider $\mat{M} = \vec{x} \tpose{\vec{y}}$ as a random rank one matrix drawn from an ensemble with the following properties:
			\begin{enumerate}[({A}1)]
				\item	\label{itm:cov id}
						$\vec{x}$ (and $\vec{y}$) is a zero mean random vector with an identity covariance matrix.
				\item	\label{itm:mutual indep}
						$\vec{x}$ and $\vec{y}$ are mutually independent.
			\end{enumerate}
			As a practical motivation for this random model, we consider a blind channel estimation problem where the transmitted signal $\vec{x}$ passes through an unknown linear time invariant channel impulse response $\vec{y}$.
			In the absence of measurement noise, the observed signal at the receiver would be the linear convolution $\vec{z} = \vec{x} \star \vec{y}$, which is a bilinear map.
			A practical modeling choice puts the channel realization $\vec{y}$ statistically independent of the transmitted signal $\vec{x}$.
			Furthermore, if channel phase is rapidly varying, then the sign of each entry for $\vec{y}$ is equally likely to be positive or negative with resultant mean as zero.
			The transmitted signal $\vec{x}$ can be assumed to be zero mean with independent and identically distributed entries (and thus identical variance per entry) under Binary-Phase-Shift-Keying and other balanced Phase-Shift-Keying modulation schemes.
			The assumption of equal variance per tap is somewhat idealistic for channel $\vec{y}$, but strictly speaking, this requirement is \textit{not} absolutely necessary for our identifiability results.

			\subsubsection{Dependent Entries}
				\label{sec:dependent entries}
				First, we consider the case when the elements of $\vec{x}$ (respectively $\vec{y}$) are \textit{not independent}.
				We shall be interested in the following two possible properties of $\vec{x}$ and $\vec{y}$:
				\begin{enumerate}[({A}1)]
					\setcounter{enumi}{2}
					\item	\label{itm:factor marginal}
							The distribution of $\vec{x}$ (respectively, $\vec{y}$) factors into a product of marginal distributions of $\twonorm{\vec{x}}$ and $\vec{x}/\twonorm{\vec{x}}$ (respectively, $\twonorm{\vec{y}}$ and $\vec{y}/\twonorm{\vec{y}}$).
					\item	\label{itm:abs lower bound}
							$\exists r > 0$ such that $\twonorm{\vec{x}} \geq r$ (respectively $\twonorm{\vec{y}} \geq r$) a.s.
				\end{enumerate}
	
				We state the following technical lemmas that will be needed in the proofs of \theoremname~\ref{thm:whp_suff_ident} and \corollaryname~\ref{cor:whp_suff_ident}.
				\lemmaname~\ref{lem:Markov estimate special} is mainly useful when the assumption~\aref{itm:factor marginal} cannot be satisfied but one needs bounds that closely resemble that of \lemmaname~\ref{lem:Markov estimate}.
				These lemmas allow us to upper bound the probability that $\vec{x}$ (respectively $\vec{y}$) is close to one of the key subspaces in \theoremname~\ref{thm:suff_ident}, \ie~$\mathcal{C}\bb{\mat{X}}$ (respectively $\mathcal{R}\bb{\mat{X}}$) where $\mat{X}$ is in the appropriately constrained subset of $\mathcal{N}\bb{\mathscr{S}, 2}$.
				\begin{lemma}
					\label{lem:Markov estimate}
					Given any $m \times n$ real matrix $\mat{X} \in \mathcal{N}\argd{\mathscr{S}, 2} \bigcap \mathcal{M} \setminus \cc{\mat{0}}$ and a constant $\delta \in \bb{0,1}$, a rank one random matrix $\mat{M} = \vec{x} \tpose{\vec{y}} = \sigma \vec{u} \tpose{\vec{v}}$ satisfying assumptions~\aref{itm:cov id}-\aref{itm:factor marginal} also satisfies,
					\begin{subequations}
						\begin{align}
							\Pr\bb{\twonorm{\mat{P}_{\mathcal{C}\bb{\mat{X}}} \vec{u}}^{2} \geq 1 - \delta}
							& \leq \frac{2}{m \bb{1 - \delta}}	\label{eqn:Markov 1} \\
							\shortintertext{and,}
							\Pr\bb{\twonorm{\mat{P}_{\mathcal{R}\bb{\mat{X}}} \vec{v}}^{2} \geq 1 - \delta}
							& \leq \frac{2}{n \bb{1 - \delta}}.	\label{eqn:Markov 2}
						\end{align}
					\end{subequations}
				\end{lemma}
				
				\begin{IEEEproof}
					\appendixname~\ref{sec:Markov estimate lemma proof}.
				\end{IEEEproof}

				\begin{lemma}
					\label{lem:Markov estimate special}
					Given any $m \times n$ real matrix $\mat{X} \in \mathcal{N}\argd{\mathscr{S}, 2} \bigcap \mathcal{M} \setminus \cc{\mat{0}}$ and a constant $\delta \in \bb{0,1}$, a rank one random matrix $\mat{M} = \vec{x} \tpose{\vec{y}} = \sigma \vec{u} \tpose{\vec{v}}$ satisfying assumptions~\aref{itm:cov id}-\aref{itm:mutual indep}, with $\vec{x}$ (respectively $\vec{y}$) satisfying~\aref{itm:abs lower bound} for a constant $r = r_{\vec{x}}$ (respectively $r=r_{\vec{y}}$), also satisfies,
					\begin{subequations}
						\begin{align}
							\Pr\bb{\twonorm{\mat{P}_{\mathcal{C}\bb{\mat{X}}} \vec{u}}^{2} \geq 1 - \delta}
							& \leq \frac{2}{r_{\vec{x}}^{2} \bb{1 - \delta}}	\label{eqn:Markov 1 special} \\
							\shortintertext{and,}
							\Pr\bb{\twonorm{\mat{P}_{\mathcal{R}\bb{\mat{X}}} \vec{v}}^{2} \geq 1 - \delta}
							& \leq \frac{2}{r_{\vec{y}}^{2} \bb{1 - \delta}}.	\label{eqn:Markov 2 special}
						\end{align}
					\end{subequations}
				\end{lemma}
				
				\begin{IEEEproof}
					\appendixname~\ref{sec:Markov estimate special lemma proof}.
				\end{IEEEproof}
				
				\begin{remark}
					\label{rem:when is corollary 3 useful}
					\lemmaname~\ref{lem:Markov estimate special} will give non-trivial bounds if $r_{\vec{x}}$ (respectively $r_{\vec{y}}$) go to $\infty$ fast enough as $m$ (respectively $n$) goes to $\infty$, and this growth rate could be slower than $\Theta\argd{\sqrt{m}}$ (respectively $\Theta\argd{\sqrt{n}}$).
				\end{remark}
				
				An example where \lemmaname~\ref{lem:Markov estimate special} is applicable but \lemmaname~\ref{lem:Markov estimate} is not, can be constructed as follows.
				As before, let $\vec{y}$ represent a channel impulse response independent of $\vec{x}$, so that~\aref{itm:mutual indep} is satisfied.
				Let $\vec{x}$ represent a coded data stream under Pulse-Amplitude-Modulation such that $\twonorm{\vec{x}} \in \cc{\sqrt{m/3}, \sqrt{2m/3}}$ with equal probability, $\expect{\vec{x}} = \vec{0}$ and $x_{m}$ is coded as a function of $\twonorm{\vec{x}}$ yielding the following conditional correlation matrices:
				\begin{subequations}
					\begin{align}
						\expect{\vec{x} \tpose{\vec{x}} \, \middle| \, \twonorm{\vec{x}} = \sqrt{\frac{\vphantom{1} m}{3}}}
						& = \frac{1}{3} \eye + \frac{1}{6} \bb{\mat{E}_{1,m} + \mat{E}_{m,1}}	\\
						\shortintertext{and,}
						\expect{\vec{x} \tpose{\vec{x}} \, \middle| \, \twonorm{\vec{x}} = \sqrt{\frac{2m}{3}}}
						& = \frac{2}{3} \eye - \frac{1}{6} \bb{\mat{E}_{1,m} + \mat{E}_{m,1}}
					\end{align}
					\label{eqn:mag azimuth depend}%
				\end{subequations}
				where $\mat{E}_{i,j} \in \setR^{m \times m}$ is the matrix with elements given by
				\begin{equation}
					\bb{\mat{E}_{i,j}}_{k,l} =	\begin{cases}
													1,	& i = k, j = l, \\
													0,	& \text{otherwise},
												\end{cases}
				\end{equation}
				for every $1 \leq i,j \leq m$.
				The expressions in \eqref{eqn:mag azimuth depend} clearly imply that $\twonorm{\vec{x}}$ and $\vec{x}/\twonorm{\vec{x}}$ are dependent so that~\aref{itm:factor marginal} does not hold.
				Nonetheless, by construction, we have
				\begin{equation}
					\Pr\argd{\twonorm{\vec{x}} = \sqrt{\frac{m}{3}}}
					= \Pr\argd{\twonorm{\vec{x}} = \sqrt{\frac{2m}{3}}}
					= \frac{1}{2},
				\end{equation}
				so that \eqref{eqn:mag azimuth depend} implies $\expect{\vec{x} \tpose{\vec{x}}} = \eye$, thus satisfying~\aref{itm:cov id}.
				Also, $\twonorm{\vec{x}} \geq \sqrt{m/3}$ a.s. so that~\aref{itm:abs lower bound} is satisfied.
				Thus, \lemmaname~\ref{lem:Markov estimate special} is applicable with $r_{\vec{x}} = \sqrt{m/3}$.

			\subsubsection{Independent Entries}
				\label{sec:independent entries}
				While \lemmaname~\ref{lem:Markov estimate} provides useful bounds, it does not suffice for many problems where $\mathcal{N}\argd{\mathscr{S}, 2} \bigcap \mathcal{M}$ is large.
				We can get much stronger bounds than \lemmaname~\ref{lem:Markov estimate} if the elements of vector $\vec{x}$ (respectively $\vec{y}$) come from independent distributions, by utilizing the \textit{concentration of measure} phenomenon~\cite{ledoux2011probability}.
				We shall consider the standard Gaussian and the symmetric Bernoulli distributions, and sharpen the bounds of \lemmaname~\ref{lem:Markov estimate} in the two technical lemmas to follow.
				Note that a zero mean independent and identically distributed assumption on the elements of $\vec{x}$ and $\vec{y}$ already implies the assumptions~\aref{itm:cov id}-\aref{itm:factor marginal}.
				The bounds of \lemmasname~\ref{lem:Bernoulli Chernoff estimate} and~\ref{lem:Gaussian Chernoff estimate} have an interpretation similar to the \textit{restricted isometry property}~\cite{candes2006nearoptimal} and are used in the proofs for \theoremsname~\ref{thm:whp_infinite} and~\ref{thm:whp_infinite_Gaussian}, respectively.
				We retain the assumption $\mathcal{N}\argd{\mathscr{S}, 1} \bigcap \mathcal{M} = \cc{\mat{0}}$ from \theoremname~\ref{thm:suff_ident} and follow the convention that a random variable $Z$ has a symmetric Bernoulli distribution if $\Pr\bb{Z = +1} = \Pr\bb{Z = -1} = 1/2$.

				\begin{lemma}
					\label{lem:Gaussian Chernoff estimate}
					Let $\mathcal{N}\argd{\mathscr{S}, 1} \bigcap \mathcal{M} = \cc{\mat{0}}$.
					Given any $m \times n$ real matrix $\mat{X} \in \mathcal{N}\argd{\mathscr{S}, 2} \bigcap \mathcal{M} \setminus \cc{\mat{0}}$ and a constant $\delta \in \bb{0,1}$, a random vector $\vec{x} \in \setR^{m}$ with each element drawn independently from a standard normal distribution satisfies
					\makeatletter
						\if@twocolumn
							\begin{multline}
								\Pr\bb{\twonorm{\mat{P}_{\mathcal{C}\bb{\mat{X}}} \vec{x}}^{2} \geq \bb{1 - \delta} \twonorm{\vec{x}}^{2}} \\
								\leq \exp\BB{-m \log \frac{1}{\sqrt{\delta}} + 2 \log m - \frac{2}{m} + 2 - \log \frac{2 \delta}{1 - \delta}}.
								\label{eqn:Gaussian Chernoff}
							\end{multline}
						\else
							\begin{equation}
								\Pr\bb{\twonorm{\mat{P}_{\mathcal{C}\bb{\mat{X}}} \vec{x}}^{2} \geq \bb{1 - \delta} \twonorm{\vec{x}}^{2}}
								\leq \exp\BB{-m \log \frac{1}{\sqrt{\delta}} + 2 \log m - \frac{2}{m} + 2 - \log \frac{2 \delta}{1 - \delta}}.
								\label{eqn:Gaussian Chernoff}
							\end{equation}
						\fi
					\makeatother
				\end{lemma}

				\begin{IEEEproof}
					\appendixname~\ref{sec:Gaussian Chernoff estimate proof}.
				\end{IEEEproof}

				\begin{lemma}
					\label{lem:Bernoulli Chernoff estimate}
					Let $\mathcal{N}\argd{\mathscr{S}, 1} \bigcap \mathcal{M} = \cc{\mat{0}}$.
					Given any $m \times n$ real matrix $\mat{X} \in \mathcal{N}\argd{\mathscr{S}, 2} \bigcap \mathcal{M} \setminus \cc{\mat{0}}$ and a constant $\delta \in \bb{0,1}$, a random vector $\vec{x} \in \setR^{m}$ with each element drawn independently from a symmetric Bernoulli distribution satisfies
					\makeatletter
						\if@twocolumn
							\begin{multline}
								\Pr\bb{\twonorm{\mat{P}_{\mathcal{C}\bb{\mat{X}}} \vec{x}}^{2} \geq \bb{1 - \delta} \twonorm{\vec{x}}^{2}} \\
								\leq \exp\BB{-\frac{m\bb{1 - \delta}}{4} + \log 4}.
								\label{eqn:Bernoulli Chernoff}
							\end{multline}
						\else
							\begin{equation}
								\Pr\bb{\twonorm{\mat{P}_{\mathcal{C}\bb{\mat{X}}} \vec{x}}^{2} \geq \bb{1 - \delta} \twonorm{\vec{x}}^{2}}
								\leq \exp\BB{-\frac{m\bb{1 - \delta}}{4} + \log 4}.
								\label{eqn:Bernoulli Chernoff}
							\end{equation}
						\fi
					\makeatother
				\end{lemma}

				\begin{IEEEproof}
					\appendixname~\ref{sec:Bernoulli Chernoff estimate proof}.
				\end{IEEEproof}
				\sectionname~\ref{rem:constant factor loss} provides additional remarks on \lemmaname~\ref{lem:Bernoulli Chernoff estimate}.

		\subsection{Random Instance Identifiability}
			\label{sec:random identifiability}
			We first consider the special case where the size of the set $\mathcal{N}\argd{\mathscr{S}, 2} \bigcap \mathcal{M}$ is small \wrt~$mn$, in \sectionname~\ref{sec:finite cardinality}.
			We use the same intuition in \sectionname~\ref{sec:infinite cardinality} to appropriately partition the set $\mathcal{N}\argd{\mathscr{S}, 2} \bigcap \mathcal{M}$ when its size is large (possibly infinite) with respect to $m+n$.
			
			\subsubsection{Small Complexity of \texorpdfstring{$\mathcal{N}\argd{\mathscr{S}, 2} \bigcap \mathcal{M}$}{Restricted Rank Two Null Space}}
				\label{sec:finite cardinality}
				It is intuitive to expect that the number of rank one matrices $\mat{M}$ that are identifiable as optimal solutions to \problemname~\eqref{prob:rank} should depend inversely on the \textit{size/complexity} of $\mathcal{N}\argd{\mathscr{S}, 2} \bigcap \mathcal{M}$.
				Below, we shall make this notion precise.
				We shall do so by lower bounding the probability of satisfaction of the sufficient conditions in \theoremname~\ref{thm:suff_ident}.

				\begin{theorem}
					\label{thm:whp_suff_ident}
					Let $\mathcal{N}\argd{\mathscr{S}, 1} \bigcap \mathcal{M} = \cc{\mat{0}}$ and $\mat{M} = \sigma \vec{u} \tpose{\vec{v}} \in \mathcal{K}' \subseteq \setR^{m \times n}$ be a rank one random matrix satisfying assumptions~\aref{itm:cov id}-\aref{itm:factor marginal}.
					Suppose that the set $\set{\bb{\mathcal{C}\argd{\mat{X}}, \mathcal{R}\argd{\mat{X}}}}{\mat{X} \in \mathcal{N}\argd{\mathscr{S}, 2} \bigcap \mathcal{M} \setminus \cc{\mat{0}}}$ is finite with cardinality $f_{\mathscr{S}, \mathcal{M}}\argd{m,n}$.
					For any constant $\delta \in \bb{0, 1}$, the sufficient conditions of \theoremname~\ref{thm:suff_ident} are satisfied with probability greater than $\bb{1 - \dfrac{4 f_{\mathscr{S}, \mathcal{M}}\argd{m,n}}{m n \bb{1 - \delta}}}$.
				\end{theorem}
	
				\begin{IEEEproof}
					We describe the basic idea behind the proof and defer the full proof to \appendixname~\ref{sec:whp_suff_ident theorem proof}.
					The proof consists of the following important steps.
					\begin{enumerate}[(a)]
						\item	We fix the matrix $\mat{X} \in \mathcal{N}\argd{\mathscr{S}, 2} \bigcap \mathcal{M} \setminus \cc{\mat{0}}$ and then relax the ``hard'' event of subspace membership $\cc{\vec{u} \in \mathcal{C}\argd{\mat{X}}}$ to the ``soft'' event of being close to the subspace in \ltwonorm{} $\cc{\twonorm{\vec{u} - \mat{P}_{\mathcal{C}\bb{\mat{X}}} \vec{u}}^{2} \leq \delta}$.
						This ``soft'' event describes a body of nonzero volume in $\setR^{m}$.
						A similar argument holds for the vector $\vec{v}$ as well.
						\item	Next, the volumes (probabilities) of both these bodies (events) is computed individually and utilizing independence between realizations of $\vec{u}$ and $\vec{v}$, the probability of the intersection of these events is easily computed.
						The bounds of \lemmaname~\ref{lem:Markov estimate} are used in this step.
						\item	Lastly, we employ a union bound over the set of valid matrices $\mat{X} \in \mathcal{N}\argd{\mathscr{S}, 2} \bigcap \mathcal{M} \setminus \cc{\mat{0}}$ to make our results universal in nature.
					\end{enumerate}
				\end{IEEEproof}
				\sectionsname~\ref{rem:measuring row column space pairs},~\ref{rem:role of conic prior} and~\ref{rem:role of delta} provide additional remarks on \theoremname~\ref{thm:whp_suff_ident}.

				In \theoremname~\ref{thm:whp_suff_ident}, we can drive the probability of identifiability $\bb{1 - \frac{4 f_{\mathscr{S}, \mathcal{M}}\argd{m,n}}{m n \bb{1 - \delta}}}$ arbitrarily close to one by increasing $m$ and/or $n$ provided that $f_{\mathscr{S}, \mathcal{M}}\argd{m,n}$ grows as $\litOh{m n}$.
				For many important BIPs (blind deconvolution, blind source separation, matrix factorization, \etc) this growth rate requirement on $f_{\mathscr{S}, \mathcal{M}}\argd{m,n}$ is too pessimistic.
				Tighter versions of \theoremname~\ref{thm:whp_suff_ident}, with more optimistic growth rate requirements on $f_{\mathscr{S}, \mathcal{M}}\argd{m,n}$, are possible if the assumptions of \lemmaname~\ref{lem:Gaussian Chernoff estimate} or~\ref{lem:Bernoulli Chernoff estimate} are satisfied.
				This is the content of \theoremsname~\ref{thm:whp_infinite} and~\ref{thm:whp_infinite_Gaussian} described in \sectionname~\ref{sec:infinite cardinality}.

				We provide a corollary to \theoremname~\ref{thm:whp_suff_ident} when assumption~\aref{itm:factor marginal} does not hold so that \lemmaname~\ref{lem:Markov estimate} is inapplicable.
				The result uses \lemmaname~\ref{lem:Markov estimate special} in place of \lemmaname~\ref{lem:Markov estimate} for the proof.
				The bound is asymptotically useful if $f_{\mathscr{S}, \mathcal{M}}\argd{m,n}$ grows as $\litOh{r_{\vec{x}}^{2}\argd{m} r_{\vec{y}}^{2}\argd{n}}$.
				
				\begin{corollary}
					\label{cor:whp_suff_ident}
					Let $\mathcal{N}\argd{\mathscr{S}, 1} \bigcap \mathcal{M} = \cc{\mat{0}}$ and $\mat{M} = \sigma \vec{u} \tpose{\vec{v}} \in \mathcal{K}' \subseteq \setR^{m \times n}$ be a rank one random matrix satisfying assumptions~\aref{itm:cov id}-\aref{itm:mutual indep} with $\vec{x}$ (respectively $\vec{y}$) satisfying~\aref{itm:abs lower bound} for a constant $r = r_{\vec{x}} \argd{m}$ (respectively $r=r_{\vec{y}} \argd{n}$).
					Suppose that the set $\set{\bb{\mathcal{C}\argd{\mat{X}}, \mathcal{R}\argd{\mat{X}}}}{\mat{X} \in \mathcal{N}\argd{\mathscr{S}, 2} \bigcap \mathcal{M} \setminus \cc{\mat{0}}}$ is finite with cardinality $f_{\mathscr{S}, \mathcal{M}}\argd{m,n}$.
					For any constant $\delta \in \bb{0, 1}$, the sufficient conditions of \theoremname~\ref{thm:suff_ident} are satisfied with probability greater than $\bb{1 - \dfrac{4 f_{\mathscr{S}, \mathcal{M}}\argd{m,n}}{r_{\vec{x}}^{2}\argd{m} r_{\vec{y}}^{2}\argd{n} \bb{1 - \delta}}}$.
				\end{corollary}

				\begin{IEEEproof}
					\appendixname~\ref{sec:whp_suff_ident corollary proof}.
				\end{IEEEproof}

			\subsubsection{Large/Infinite Complexity of \texorpdfstring{$\mathcal{N}\argd{\mathscr{S}, 2} \bigcap \mathcal{M}$}{Restricted Rank Two Null Space}}
				\label{sec:infinite cardinality}
				When the complexity of $\mathcal{N}\argd{\mathscr{S}, 2} \bigcap \mathcal{M}$ is infinite or exponentially large in $m + n$, the bounds of \sectionname~\ref{sec:finite cardinality} become trivially true for large enough $m$ or $n$.
				We investigate an alternative bounding technique for this situation using covering numbers.
				Intuitively speaking, covering numbers measure the size of discretized versions of uncountable sets.
				The advantage of using such an approach is that the results are not contingent upon the exact geometry of $\mathcal{N}\argd{\mathscr{S}, 2} \bigcap \mathcal{M}$.
				Thus, like \theoremname~\ref{thm:whp_suff_ident}, the technique and subsequent results are applicable to every bilinear map.
				We shall see that to arrive at any sensible results, we will need to use the tighter estimates given by \lemmasname~\ref{lem:Gaussian Chernoff estimate} and~\ref{lem:Bernoulli Chernoff estimate}
				that are only possible when our signals $\vec{x}$ and $\vec{y}$ are component-wise independent.

				\begin{definition}[Covering Number and Metric Entropy~\cite{vershynin2009lecturesingeometric}]
					For any two sets $\mathcal{B}, \mathcal{D} \subseteq \setR^{n}$, the minimum number of translates of $\mathcal{B}$ needed to cover $\mathcal{D}$ is called the \uline{covering number} of $\mathcal{D}$ \wrt~$\mathcal{B}$ and is denoted by $N\argd{\mathcal{D}, \mathcal{B}}$.
					The quantity $\log N\argd{\mathcal{D}, \mathcal{B}}$ is known as the \uline{metric entropy} of $\mathcal{D}$ \wrt~$\mathcal{B}$.
				\end{definition}

				It is known that if $\mathcal{D} \subseteq \setR^{n}$ is a bounded convex body that is symmetric about the origin, and we let $\mathcal{B} = \epsilon \mathcal{D} \triangleq \set{\epsilon \vec{x}}{\vec{x} \in \mathcal{D}}$ for some $0 < \epsilon < 1$, then the covering number $N\argd{\mathcal{D}, \epsilon\mathcal{D}}$ obeys~\cite{vershynin2009lecturesingeometric}
				\begin{equation}
					\bb{\frac{1}{\epsilon}}^{n} \leq N\argd{\mathcal{D}, \epsilon\mathcal{D}} \leq \bb{2 + \frac{1}{\epsilon}}^{n}.
					\label{eqn:covering number}
				\end{equation}
				We can \textit{equivalently} say that the metric entropy $\log N\argd{\mathcal{D}, \epsilon\mathcal{D}}$ equals $n \log \Theta\argd{1/\epsilon}$.
				We shall use this notation for specifying metric entropies of key sets in the theorems to follow.

				We state a technical lemma needed to prove \theoremsname~\ref{thm:whp_infinite} and~\ref{thm:whp_infinite_Gaussian}.
				The lemma bounds the difference between norms of topologically close projection operators as a function of the covering resolution, thus providing a characterization of the sets used to cover over the space of interest.

				\begin{lemma}
					\label{lem:metric entropy}
					Let $\mathcal{G}\argd{m} = \set{\mat{Y} \in \setR^{m \times 2}}{\tpose{\mat{Y}} \mat{Y} = \eye}$, $\mathcal{D}\argd{m} = \set{\BB{\vec{y}_{1}, \vec{y}_{2}} \in \setR^{m \times 2}}{\displaystyle \max_{j = 1,2}\twonorm{\vec{y}_{j}} \leq 1}$ and $0 < \epsilon < 1$.
					There exists a covering of $\mathcal{G}\argd{m}$ with metric entropy $\leq 2m \log \Theta\argd{1/\epsilon}$ \wrt~$\epsilon\mathcal{D}\argd{m}$ such that for any $\mat{Y}, \mat{Z} \in \mathcal{G}\argd{m}$ satisfying $\mat{Y} - \mat{Z} \in \epsilon \mathcal{D}\argd{m}$ we have
					\begin{equation}
						\abs{\twonorm{\mat{P}_{\mathcal{C}\bb{\mat{Y}}} \vec{x}} - \twonorm{\mat{P}_{\mathcal{C}\bb{\mat{Z}}} \vec{x}}} \leq \sqrt{2} \epsilon \twonorm{\vec{x}}
						\label{eqn:projection bound}
					\end{equation}
					for all $\vec{x} \in \setR^{m}$.
				\end{lemma}

				\begin{IEEEproof}
					\appendixname~\ref{sec:epsilon net proof}.
				\end{IEEEproof}
				\sectionname~\ref{rem:meaning of covering lemma} provides additional remarks on \lemmaname~\ref{lem:metric entropy}.

				We are now ready to extend \theoremname~\ref{thm:whp_suff_ident} to the case where the complexity of $\mathcal{N}\argd{\mathscr{S}, 2} \bigcap \mathcal{M}$ is large (possibly infinite).
				We shall do so for Bernoulli and Gaussian priors (as illustrative distributions) in \theoremsname~\ref{thm:whp_infinite} and~\ref{thm:whp_infinite_Gaussian} respectively.
				The proofs for both these theorems follow on the same lines as that of \theoremname~\ref{thm:whp_suff_ident}, except that the probability bounds of \lemmaname~\ref{lem:Markov estimate} are replaced by those of \lemmasname~\ref{lem:Bernoulli Chernoff estimate} and~\ref{lem:Gaussian Chernoff estimate} for Bernoulli and Gaussian priors, respectively.

				\makeatletter
					\if@twocolumn
						\begin{theorem}
							\label{thm:whp_infinite}
							Let $\mathcal{N}\argd{\mathscr{S}, 1} \bigcap \mathcal{M} = \cc{\mat{0}}$, the sets $\mathcal{G}\argd{m}, \mathcal{G}\argd{n}$ and $\mathcal{D}\argd{m}, \mathcal{D}\argd{n}$ be defined according to \lemmaname~\ref{lem:metric entropy}, and $\mat{M} = \vec{x} \tpose{\vec{y}} \in \setR^{m \times n}$ be a rank one random matrix with components of $\vec{x}$ (respectively $\vec{y}$) drawn independently from a symmetric Bernoulli distribution with $\mathcal{K}'$ chosen as
							\begin{equation}
								\mathcal{K}' = \set{\lambda \vec{x} \tpose{\vec{y}}}{\vec{x} \in \cc{-1,1}^{m}, \vec{y} \in \cc{-1,1}^{n}, \lambda \in \setR}.
							\end{equation}
							and $\mathcal{M} = \mathcal{K}' - \mathcal{K}'$.
							Let $p_{c} \log \Theta\argd{1/\epsilon}$ (respectively $p_{r} \log \Theta\argd{1/\epsilon}$) denote the metric entropy of the set $\mathcal{G}\argd{m} \bigcap \set{\mathcal{C} \bb{\mat{X}}}{\mat{X} \in \mathcal{N} \argd{\mathscr{S}, 2} \bigcap \mathcal{M} \setminus \cc{\mat{0}}}$ \wrt~$\epsilon \mathcal{D}\argd{m}$ (respectively $\mathcal{G}\argd{n} \bigcap \set{\mathcal{R}\bb{\mat{X}}}{\mat{X} \in \mathcal{N}\argd{\mathscr{S}, 2} \bigcap \mathcal{M} \setminus \cc{\mat{0}}}$ \wrt~$\epsilon \mathcal{D}\argd{n}$) for any $1 > \epsilon \geq \epsilon_{0} > 0$ and let $p = p_{c} + p_{r}$.
							For any constant $\delta' \in \bb{0, 1-2\epsilon^{2}}$, the sufficient conditions of \theoremname~\ref{thm:suff_ident} are satisfied with probability greater than ${\displaystyle \bb{1 - 16\exp\BB{p \log \Theta\argd{\frac{1}{\epsilon}} - \bb{m + n} \frac{1 - \delta}{4}}}}$ with $\delta = 1 - \bb{\sqrt{1 - \delta'} - \sqrt{2}\epsilon}^{2}$.
						\end{theorem}
					\else
						\begin{theorem}
							\label{thm:whp_infinite}
							Let $\mathcal{N}\argd{\mathscr{S}, 1} \bigcap \mathcal{M} = \cc{\mat{0}}$, the sets $\mathcal{G}\argd{m}, \mathcal{G}\argd{n}$ and $\mathcal{D}\argd{m}, \mathcal{D}\argd{n}$ be defined according to \lemmaname~\ref{lem:metric entropy}, and $\mat{M} = \vec{x} \tpose{\vec{y}} \in \setR^{m \times n}$ be a rank one random matrix with components of $\vec{x}$ (respectively $\vec{y}$) drawn independently from a symmetric Bernoulli distribution with $\mathcal{K}'$ chosen as
							\begin{equation}
								\mathcal{K}' = \set{\lambda \vec{x} \tpose{\vec{y}}}{\vec{x} \in \cc{-1,1}^{m}, \vec{y} \in \cc{-1,1}^{n}, \lambda \in \setR}.
							\end{equation}
							and $\mathcal{M} = \mathcal{K}' - \mathcal{K}'$.
							Let $p_{c} \log \Theta\argd{1/\epsilon}$ denote the metric entropy of the set $\mathcal{G}\argd{m} \bigcap \set{\mathcal{C} \bb{\mat{X}}}{\mat{X} \in \mathcal{N} \argd{\mathscr{S}, 2} \bigcap \mathcal{M} \setminus \cc{\mat{0}}}$ \wrt~$\epsilon \mathcal{D}\argd{m}$, $p_{r} \log \Theta\argd{1/\epsilon}$ denote the metric entropy of the set $\mathcal{G}\argd{n} \bigcap \set{\mathcal{R}\bb{\mat{X}}}{\mat{X} \in \mathcal{N}\argd{\mathscr{S}, 2} \bigcap \mathcal{M} \setminus \cc{\mat{0}}}$ \wrt~$\epsilon \mathcal{D}\argd{n}$, for any $1 > \epsilon \geq \epsilon_{0} > 0$ and let $p = p_{c} + p_{r}$.
							For any constant $\delta' \in \bb{0, 1-2\epsilon^{2}}$, the sufficient conditions of \theoremname~\ref{thm:suff_ident} are satisfied with probability greater than ${\displaystyle \bb{1 - 16\exp\BB{p \log \Theta\argd{\frac{1}{\epsilon}} - \bb{m + n} \frac{1 - \delta}{4}}}}$ with $\delta = 1 - \bb{\sqrt{1 - \delta'} - \sqrt{2}\epsilon}^{2}$.
						\end{theorem}
					\fi
				\makeatother

				\begin{IEEEproof}
					\appendixname~\ref{sec:whp_infinite theorem proof}.
				\end{IEEEproof}

				\makeatletter
					\if@twocolumn
						\begin{theorem}
							\label{thm:whp_infinite_Gaussian}
							Let $\mathcal{N}\argd{\mathscr{S}, 1} = \cc{\mat{0}}$, the sets $\mathcal{G}\argd{m}, \mathcal{G}\argd{n}$ and $\mathcal{D}\argd{m}, \mathcal{D}\argd{n}$ be defined according to \lemmaname~\ref{lem:metric entropy}, and $\mat{M} = \vec{x} \tpose{\vec{y}} \in \setR^{m \times n}$ be a rank one random matrix with components of $\vec{x}$ (respectively $\vec{y}$) drawn independently from a standard Gaussian distribution.
							Let $p_{c} \log \Theta\argd{1/\epsilon}$ (respectively $p_{r} \log \Theta\argd{1/\epsilon}$) denote the metric entropy of the set $\mathcal{G}\argd{m} \bigcap \set{\mathcal{C} \bb{\mat{X}}}{\mat{X} \in \mathcal{N}\argd{\mathscr{S}, 2} \setminus \cc{\mat{0}}}$ \wrt~$\epsilon \mathcal{D}\argd{m}$ (respectively $\mathcal{G}\argd{n} \bigcap \set{\mathcal{R}\bb{\mat{X}}}{\mat{X} \in \mathcal{N}\argd{\mathscr{S}, 2} \setminus \cc{\mat{0}}}$ \wrt~$\epsilon \mathcal{D}\argd{n}$) for any $0 < \epsilon < 1$ and let $p = p_{c} + p_{r}$.
							For any constant $\delta' \in \bb{0, 1-2\epsilon^{2}}$, the sufficient conditions of \theoremname~\ref{thm:suff_ident} are satisfied with probability greater than ${\displaystyle \bb{1 - C\argd{m,n,\delta} \exp\!\BB{p \log \Theta\argd{\frac{1}{\epsilon}} - \bb{m + n} \log \frac{1}{\sqrt{\delta}} }}}$ where
							\begin{equation}
								\begin{split}
									C\argd{m,n,\delta}
									& = \exp\!\BB{2\log mn + 4 - 2\log \frac{2\delta}{1-\delta}}	\\
									& = \bb{\frac{1}{\delta} - 1}^{2} \Theta\argd{m^{2} n^{2}}.
								\end{split}
							\end{equation}
							and $\delta = 1 - \bb{\sqrt{1 - \delta'} - \sqrt{2}\epsilon}^{2}$.
						\end{theorem}
					\else
						\begin{theorem}
							\label{thm:whp_infinite_Gaussian}
							Let $\mathcal{N}\argd{\mathscr{S}, 1} = \cc{\mat{0}}$, the sets $\mathcal{G}\argd{m}, \mathcal{G}\argd{n}$ and $\mathcal{D}\argd{m}, \mathcal{D}\argd{n}$ be defined according to \lemmaname~\ref{lem:metric entropy}, and $\mat{M} = \vec{x} \tpose{\vec{y}} \in \setR^{m \times n}$ be a rank one random matrix with components of $\vec{x}$ (respectively $\vec{y}$) drawn independently from a standard Gaussian distribution.
							Let $p_{c} \log \Theta\argd{1/\epsilon}$ denote the metric entropy of the set $\mathcal{G}\argd{m} \bigcap \set{\mathcal{C} \bb{\mat{X}}}{\mat{X} \in \mathcal{N}\argd{\mathscr{S}, 2} \setminus \cc{\mat{0}}}$ \wrt~$\epsilon \mathcal{D}\argd{m}$, $p_{r} \log \Theta\argd{1/\epsilon}$ denote the metric entropy of the set $\mathcal{G}\argd{n} \bigcap \set{\mathcal{R}\bb{\mat{X}}}{\mat{X} \in \mathcal{N}\argd{\mathscr{S}, 2} \setminus \cc{\mat{0}}}$ \wrt~$\epsilon \mathcal{D}\argd{n}$ for any $0 < \epsilon < 1$ and let $p = p_{c} + p_{r}$.
							For any constant $\delta' \in \bb{0, 1-2\epsilon^{2}}$, the sufficient conditions of \theoremname~\ref{thm:suff_ident} are satisfied with probability greater than ${\displaystyle \bb{1 - C\argd{m,n,\delta} \exp\!\BB{p \log \Theta\argd{\frac{1}{\epsilon}} - \bb{m + n} \log \frac{1}{\sqrt{\delta}} }}}$ where
							\begin{equation}
								C\argd{m,n,\delta}
								= \exp\!\BB{2\log mn + 4 - 2\log \frac{2\delta}{1-\delta}}
								= \bb{\frac{1}{\delta} - 1}^{2} \Theta\argd{m^{2} n^{2}}.
							\end{equation}
							and $\delta = 1 - \bb{\sqrt{1 - \delta'} - \sqrt{2}\epsilon}^{2}$.
						\end{theorem}
					\fi
				\makeatother
				
				\begin{IEEEproof}
					\appendixname~\ref{sec:whp_infinite_Gaussian theorem proof}.
				\end{IEEEproof}
				\sectionsname~\ref{rem:bernoulli example} and~\ref{rem:range of epsilon} provide additional remarks on \theoremsname~\ref{thm:whp_infinite} and~\ref{thm:whp_infinite_Gaussian}.

\begin{figure}
					\centering
					\includegraphics[width=\figwidth]{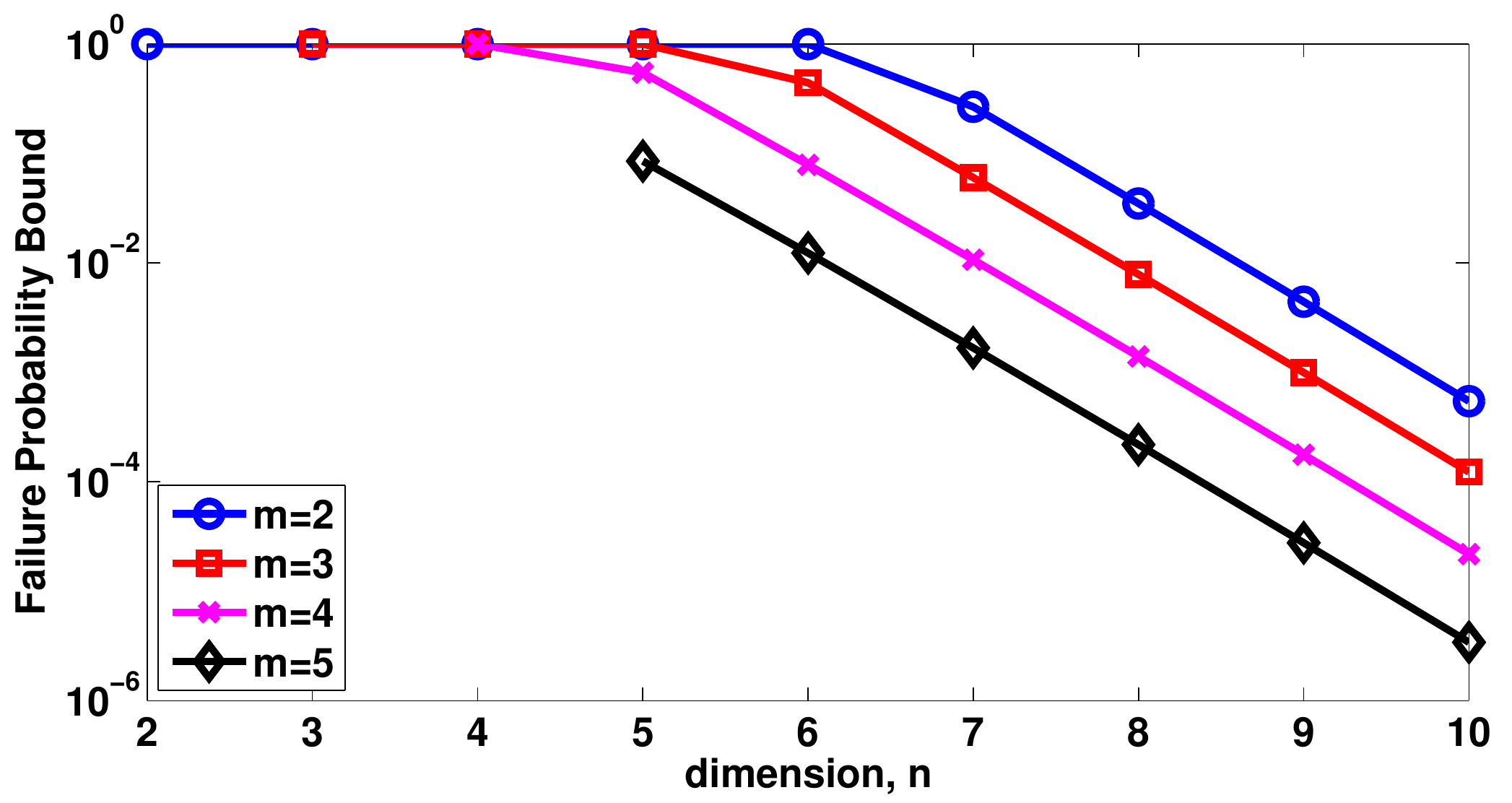}
					\caption{Exponentially decaying behavior of the theoretically predicted failure probability bound in \theoremname~\ref{thm:whp_infinite_Gaussian}, \wrt~$n$ for fixed values of $m$, for parameters $\epsilon = 0.1$, $\delta = 10^{-4}$ and $p = m+n-3$ for the lifted linear convolution map ($p$ defined as in \theoremname~\ref{thm:whp_infinite_Gaussian}).}
					\label{fig:theory bounds}
				\end{figure}

				A non-trivial illustration of the theoretical scaling law bound of \theoremname~\ref{thm:whp_infinite_Gaussian} is provided in \figurename~\ref{fig:theory bounds}, with $\mathscr{S}\bb{\cdot}$ as the lifted linear convolution map.
				Since the bound is parametrized by $\bb{\epsilon, \delta}$, we choose $\epsilon = 0.1$ and $\delta = 10^{-4}$ for the illustration.
				Quite surprisingly (and fortunately), the metric entropy $p$ in \theoremname~\ref{thm:whp_infinite_Gaussian} can be exactly characterized when $\mathscr{S}\bb{\cdot}$ represents the lifted linear convolution map.
				Specifically, we have $p = m+n-3$.
				We refer the reader to \propositionname~\ref{prop:rank-2 nullspace} in \sectionname~\ref{sec:null space of linear convolution} for details.
				
				\begin{remark}
					\label{rem:mixed prior distributions}
					We can obtain results analogous to \theoremsname~\ref{thm:whp_infinite} and~\ref{thm:whp_infinite_Gaussian} when $\vec{x}$ and $\vec{y}$ are drawn from non-identical distributions, \eg~$\vec{x}$ is component-wise \iid~symmetric Bernoulli and $\vec{y}$ is component-wise \iid~standard Gaussian.
					The argument is a straightforward modification of the proof.
				\end{remark}

	\section{Discussion}
		\label{sec:discussion}
		In this section, we elaborate on the intuitions, ideas, assumptions and subtle implications associated with the main results of this paper that were presented in \sectionname~\ref{sec:results}.
		
		\subsection{A Measure of Geometric Complexity}
			\label{rem:measuring row column space pairs}
			For the purpose of measuring the size/complexity of $\mathcal{N}\argd{\mathscr{S}, 2} \bigcap \mathcal{M}$ in \theoremname~\ref{thm:whp_suff_ident}, we used the cardinality $f_{\mathscr{S}, \mathcal{M}}\argd{m,n}$ of the set $\set{\bb{\mathcal{C}\argd{\mat{X}}, \mathcal{R}\argd{\mat{X}}}}{\mat{X} \in \mathcal{N}\argd{\mathscr{S}, 2} \bigcap \mathcal{M} \setminus \cc{\mat{0}}}$ as a surrogate.
			This set essentially lists the distinct pairs of row and column spaces in the rank two null space of the lifted linear operator $\mathscr{S}\argd{\cdot}$ that are not excluded by the domain restriction $\mat{M} \in \mathcal{K}'$.
			We note that the cardinality of the set $\mathcal{N}\argd{\mathscr{S}, 2} \bigcap \mathcal{M} \bigcap \set{\mat{X} \in \setR^{m \times n}}{\fronorm{\mat{X}} = 1}$ could be infinite while its complexity could be finite in the sense just described.
			The same measure of complexity is used for the extensions of \theoremname~\ref{thm:whp_suff_ident} in \theoremsname~\ref{thm:whp_infinite} and~\ref{thm:whp_infinite_Gaussian}.
			Throughout rest of the paper, any reference to the complexity of a set of matrices $\mathcal{M}' \subseteq \setR^{m \times n}$ is in the sense just described, \ie~through the cardinality of the set $\set{\bb{\mathcal{C}\argd{\mat{X}}, \mathcal{R}\argd{\mat{X}}}}{\mat{X} \in \mathcal{M}' \setminus \cc{\mat{0}}}$.

		\subsection{The Role of Conic Prior}
			\label{rem:role of conic prior}
			There are three distinct aspects to the prior knowledge in terms of the conic constraint $\mat{M} \in \mathcal{K}'$ on the unknown signal.
			\begin{enumerate}
				\item	\textit{Probability Bounds:} A key advantage of prior knowledge about the signal is apparent from the union bounding step in the proof of \theoremname~\ref{thm:whp_suff_ident}.
				Union bounding over the set $\mathcal{N}\argd{\mathscr{S}, 2} \bigcap \mathcal{M} \setminus \cc{\mat{0}}$ always gives better bounds than union bounding over the superset $\mathcal{N}\argd{\mathscr{S}, 2} \setminus \cc{\mat{0}}$, the quantitative difference being the number $f_{\mathscr{S}, \mathcal{M}}\argd{m,n}$ in the bound of \theoremname~\ref{thm:whp_suff_ident}.
				In general, the difference could be exponentially large in $m$ or $n$ (see \theoremsname~\ref{thm:whp_infinite} and~\ref{thm:whp_infinite_Gaussian}).
				We also note that $\mathcal{K}'$ does not need to be a cone in order to exploit this approach to improve the probability bounds.
				
				\item	\textit{Computational Trade-offs:} Recalling \remarkname~\ref{rem:relax}, the size of $\mathcal{K}'$ also trades off the ease of computation and the identifiability bounds of \theoremname~\ref{thm:whp_suff_ident}.
				If the size of $\mathcal{K}'$ needs to be increased to ease computation, an effort must be made to not suffer a substantial increase in the size/complexity of the set $\mathcal{N}\argd{\mathscr{S}, 2} \bigcap \mathcal{M} \setminus \cc{\mat{0}}$.
				For high dimensional problems ($m$ or $n$ is large), non-convex conic priors like the sparse cone in compressed sensing~\cite{donoho2006compressed} and the low-rank cone in matrix completion~\cite{candes2009exact} have been shown to admit good computationally tractable relaxations.

				\item	\textit{Geometric Complexity Measure:} The measure of geometric complexity described in \sectionname~\ref{rem:measuring row column space pairs} followed naturally from \theoremname~\ref{thm:suff_ident} in an effort to describe the identifiability of a BIP in terms of quantities like row and column spaces familiar from linear algebra.
				This measure of complexity is \textit{invariant \wrt~conic extensions} in the following way.
				Let $\mathcal{M}' \subseteq \setR^{m \times n}$ be any set of matrices and let $\mathcal{M}''$ denote its conic extension, defined as
				\begin{equation}
					\mathcal{M}'' \triangleq \set{\lambda \mat{X}}{\mat{X} \in \setR^{m \times n}, \lambda \in \setR^{+}}.
				\end{equation}
				Then, we have
				\makeatletter
					\if@twocolumn
						\begin{multline}
							\set{\bb{\mathcal{C}\argd{\mat{X}}, \mathcal{R}\argd{\mat{X}}}}{\mat{X} \in \mathcal{M}' \setminus \cc{\mat{0}}}	\\
							= \set{\bb{\mathcal{C}\argd{\mat{X}}, \mathcal{R}\argd{\mat{X}}}}{\mat{X} \in \mathcal{M}'' \setminus \cc{\mat{0}}}.
						\end{multline}%
					\else
						\begin{equation}
							\set{\bb{\mathcal{C}\argd{\mat{X}}, \mathcal{R}\argd{\mat{X}}}}{\mat{X} \in \mathcal{M}' \setminus \cc{\mat{0}}} = \set{\bb{\mathcal{C}\argd{\mat{X}}, \mathcal{R}\argd{\mat{X}}}}{\mat{X} \in \mathcal{M}'' \setminus \cc{\mat{0}}}.
						\end{equation}%
					\fi
				\makeatother
				Qualitatively speaking, the flavor of results in this paper could also be derived for non-conic priors but the measure of geometric complexity that is used is implicitly based on conic extensions.
				Thus, there is no significant loss of generality in restricting ourselves to conic priors.
			\end{enumerate}

		\subsection{\texorpdfstring{The Role of $\delta$}{A Tolerance Parameter}}
			\label{rem:role of delta}
			Although the parameter $\delta \in \bb{0,1}$ appears in \theoremname~\ref{thm:whp_suff_ident} as an artifact of our proof strategy, it has an important practical consequence.
			It represents a tolerance parameter for approximate versus exact prior information on the input signals.
			Specifically, \theoremname~\ref{thm:whp_suff_ident} is a statement about identifiability up to a $\delta$-neighborhood around the true signal $\bb{\vec{x}, \vec{y}}$.
			The same holds true for \theoremsname~\ref{thm:whp_infinite} and~\ref{thm:whp_infinite_Gaussian} describing the large/infinite complexity case.
		
		\subsection{Interpretation of \lemmaname~\ref{lem:metric entropy}}
			\label{rem:meaning of covering lemma}
			\lemmaname~\ref{lem:metric entropy} can be informally restated as follows.
			Keeping \eqref{eqn:projection bound} satisfied, $\mathcal{G}\argd{m}$ can always be covered by $\epsilon\mathcal{D}\argd{m}$ with metric entropy $\leq 2m \log \Theta\argd{1/\epsilon}$.
			In \theoremsname~\ref{thm:whp_infinite} and~\ref{thm:whp_infinite_Gaussian} below, we are interested in covering the subset $\mathcal{G}\argd{m} \bigcap \set{\mathcal{C} \bb{\mat{X}}}{\mat{X} \in \mathcal{N} \argd{\mathscr{S}, 2} \bigcap \mathcal{M} \setminus \cc{\mat{0}}} \subseteq \mathcal{G}\argd{m}$ by $\epsilon\mathcal{D}\argd{m}$ and suppose that the resulting metric entropy is $p_{c} \log \Theta\argd{1/\epsilon}$.
			In a sense, \lemmaname~\ref{lem:metric entropy} represents the worst case scenario that $p_{c}$ is upper bounded by $2m$ and no better upper bound is known.
			In the worst case, the aforementioned subset of $\mathcal{G}\argd{m}$ has nearly the same complexity as $\mathcal{G}\argd{m}$ and this happens when the set $\mathcal{N} \argd{\mathscr{S}, 2} \bigcap \mathcal{M}$ does not represent a large enough structural restriction on the set of rank two matrices in $\setR^{m \times n}$.
			For large $m$, to guarantee identifiability for most inputs, we would (realistically) want $p_{c}$ to be less than $m$ by at least a constant factor.
			This is implied by \theoremsname~\ref{thm:whp_infinite} and~\ref{thm:whp_infinite_Gaussian}.
			Informally, smaller or more structured $\mathcal{N} \argd{\mathscr{S}, 2} \bigcap \mathcal{M}$ implies a smaller value of $p_{c}$ which in turn implies identifiability for a greater fraction of the input ensemble.
		
		\subsection{The Gaussian and Bernoulli Special Cases}
			\label{rem:bernoulli example}
			A standard Gaussian prior on the elements of $\vec{x}$ and $\vec{y}$ gives an example of the set $\mathcal{N}\argd{\mathscr{S}, 2} \bigcap \mathcal{M}$ with infinite complexity, provided that $\mathcal{N}\argd{\mathscr{S}, 2}$ is complex enough.
			In this case, $\mathcal{K} = \set{\bb{\vec{x},\vec{y}}}{\vec{x} \in \setR^{m}, \vec{y} \in \setR^{n}}$ in \problemname~\eqref{prob:find_xy} implying that $\mathcal{K}' \supseteq \set{\mat{W} \in \setR^{m \times n}}{\rank{\mat{W}} \leq 1}$ from \eqref{eqn:set change}.
			Thus, $\mathcal{M} \supseteq \set{\mat{W} \in \setR^{m \times n}}{\rank{\mat{W}} \leq 2} \supseteq \mathcal{N}\argd{\mathscr{S}, 2}$ and hence $\mathcal{N}\argd{\mathscr{S}, 2} \bigcap \mathcal{M} = \mathcal{N}\argd{\mathscr{S}, 2}$.
			Since $\mathcal{M}$ is superfluous in this case, \theoremname~\ref{thm:whp_infinite_Gaussian} omits all references to it.
			If the row or column spaces of matrices in $\mathcal{N}\argd{\mathscr{S}, 2}$ are parametrized by one or more real parameters (see \sectionname~\ref{sec:null space of linear convolution} for an example involving the linear convolution operator), then $\mathcal{N}\argd{\mathscr{S}, 2}$ has infinite complexity.

			The scenario of a Bernoulli prior on elements of $\vec{x}$ and $\vec{y}$ gives an example of the set $\mathcal{N}\argd{\mathscr{S}, 2} \bigcap \mathcal{M}$ with finite (but exponentially large in $m + n$) complexity, provided that $\mathcal{N}\argd{\mathscr{S}, 2}$ is complex enough.
			The precise statement requires a little more care than the Gaussian case described above.
			The motivation behind considering Bernoulli priors is to restrict the unit vectors $\vec{x}/\twonorm{\vec{x}}$ and $\vec{y}/\twonorm{\vec{y}}$ to take values from a large but finite set while adhering to the requirement of a conic prior on $\bb{\vec{x}, \vec{y}}$ according to \problemname~\eqref{prob:find_xy}.
			Thus, in this case we have $\mathcal{K} = \set{\bb{\lambda_{1} \vec{x}, \lambda_{2} \vec{y}}}{\vec{x} \in \cc{-1,1}^{m}, \vec{y} \in \cc{-1,1}^{n}, \lambda_{1} \in \setR, \lambda_{2} \in \setR}$.
			Let us select $\mathcal{K}'$ according to \eqref{eqn:set change}, but without any relaxation, as
			\begin{equation}
			\mathcal{K}' = \set{\lambda \vec{x} \tpose{\vec{y}}}{\vec{x} \in \cc{-1,1}^{m}, \vec{y} \in \cc{-1,1}^{n}, \lambda \in \setR}.
			\end{equation}
			Clearly, matrices in $\mathcal{K}'$ can account for at most $2^{m-1}$ distinct column spaces and $2^{n-1}$ distinct row spaces, thus implying that matrices in $\mathcal{M} = \mathcal{K}' - \mathcal{K}'$ are generated by at most $\binom{2^{m-1}}{2} \leq 2^{2m-2}$ distinct column spaces and at most $\binom{2^{n-1}}{2} \leq 2^{2n-2}$ distinct row spaces.
			Thus, $\mathcal{N}\argd{\mathscr{S}, 2} \bigcap \mathcal{M} \subseteq \mathcal{M}$ is of finite complexity.
			It is clear that the complexity of $\mathcal{M}$ is $\exp\bb{\Theta\bb{m+n}}$ so that if $\mathcal{M} \setminus \mathcal{N}\bb{\mathscr{S}, 2}$ is small enough then the complexity of $\mathcal{N}\bb{\mathscr{S}, 2} \bigcap \mathcal{M}$ is exponentially large in $m+n$.
		
		\subsection{Distinctions between \theoremsname~\ref{thm:whp_infinite} and~\ref{thm:whp_infinite_Gaussian}}
			\subsubsection{Assumptions on \texorpdfstring{$\epsilon$}{Covering Radius}}
				\label{rem:range of epsilon}
				We prevent an arbitrarily small $\epsilon$ for \theoremname~\ref{thm:whp_infinite} by imposing a strictly positive lower bound $\epsilon_{0} > 0$.
				This is necessary for Bernoulli priors on $\vec{x}$ and $\vec{y}$ since $\mathcal{N}\bb{\mathscr{S}, 2} \bigcap \mathcal{M}$ has a finite complexity, implying that the covering numbers of $\mathcal{G}\argd{m} \bigcap \set{\mathcal{C} \bb{\mat{X}}}{\mat{X} \in \mathcal{N}\argd{\mathscr{S}, 2} \setminus \cc{\mat{0}}}$ \wrt~$\epsilon \mathcal{D}\argd{m}$ and $\mathcal{G}\argd{n} \bigcap \set{\mathcal{R}\bb{\mat{X}}}{\mat{X} \in \mathcal{N}\argd{\mathscr{S}, 2} \setminus \cc{\mat{0}}}$ \wrt~$\epsilon \mathcal{D}\argd{n}$ have an absolute upper bound independent of $\epsilon$.
				Thus, the logarithmic dependence (of the key metric entropies) on $1/\epsilon$ cannot hold unless $\epsilon$ is lower bounded away from zero.
				\theoremname~\ref{thm:whp_infinite_Gaussian}, in contrast, allows for arbitrarily small $\epsilon$ since $\mathcal{N}\bb{\mathscr{S}, 2} \bigcap \mathcal{M} = \mathcal{N}\bb{\mathscr{S}, 2}$ has infinite complexity, for Gaussian priors on $\vec{x}$ and $\vec{y}$.
				Despite this distinction between \theoremsname~\ref{thm:whp_infinite} and~\ref{thm:whp_infinite_Gaussian}, we choose to present our results in the stated form to emphasize similarity in the theorem statements and proofs.

			\subsubsection{A constant factor loss}
				\label{rem:constant factor loss}
				We loose a constant factor of approximately 2 in the exponent on the r.h.s. of \eqref{eqn:Bernoulli Chernoff} as compared to \eqref{eqn:Gaussian Chernoff} for a fixed $\delta \in \bb{0,1}$ (compared using first order approximation of $\log \delta$).
				While this seems to be an artifact of the proof strategy, it is unclear whether a better constant can be obtained for the symmetric Bernoulli distribution (or more generally, for subgaussian distributions~\cite{baraniuk2011introduction}).
				Indeed, for the proof of \lemmaname~\ref{lem:Gaussian Chernoff estimate} in \appendixname~\ref{sec:Gaussian Chernoff estimate proof}, we have used the rotational invariance property of the multivariate standard normal distribution.
				This property does not carry over to general subgaussian distributions.

	\section{Numerical Results on Blind Deconvolution}
		\label{sec:numerical results}
		We observe that if $\mathcal{N}\argd{\mathscr{S}, 2} \bigcap \mathcal{M} = \cc{\mat{0}}$ then $f_{\mathscr{S}, \mathcal{M}}\argd{m,n} = 0$ and \theoremname~\ref{thm:whp_suff_ident} correctly predicts that the input signals are identifiable with probability one (in agreement with \propositionname~\ref{prop:ident_easy}).
		Below, we consider example bilinear maps and input distributions with $\mathcal{N}\argd{\mathscr{S}, 2} \bigcap \mathcal{M} \neq \cc{\mat{0}}$ and numerically examine the scaling behavior suggested by \lemmasname~\ref{lem:Markov estimate}, \ref{lem:Bernoulli Chernoff estimate} and~\ref{lem:Gaussian Chernoff estimate} and \theoremsname~\ref{thm:whp_suff_ident}, \ref{thm:whp_infinite} and~\ref{thm:whp_infinite_Gaussian}.
		Since \lemmaname~\ref{lem:Markov estimate} and \theoremname~\ref{thm:whp_suff_ident} impose only broad constraints on the input distribution, for the purpose of numerical simulations, we construct a specific input distribution that satisfies assumptions~\aref{itm:cov id}-\aref{itm:factor marginal} in \sectionname~\ref{sec:dependant but uncorrelated}.
		Since this research was motivated by our interest to understand the cone constrained blind deconvolution problem~\eqref{prob:find_xy_deconv}, our selection of example bilinear maps are closely related to the linear convolution map.
		We provide a partial description of the rank two null space for the linear convolution map in \sectionname~\ref{sec:null space of linear convolution}.

		\subsection{Bi-orthogonally Supported Uniform Distributions}
			\label{sec:dependant but uncorrelated}
			A bi-orthogonal set of vectors is a collection of orthonormal vectors and their additive inverses.
			It is widely used for signal representation in image processing and as a modulation scheme in communication systems.
			We can construct a uniform distribution over a bi-orthogonal set and it would satisfy assumptions~\aref{itm:cov id}-\aref{itm:factor marginal} as shown below.

			Let $\cc{\vec{e}_{1}, \vec{e}_{2}, \dots, \vec{e}_{m}}$ be an orthonormal basis for $\setR^{m}$ and the random unit vector $\vec{u} \in \cc{\pm \vec{e}_{1}, \pm \vec{e}_{2}, \dots, \pm \vec{e}_{m}}$ be drawn according to the law
			\begin{equation}
				\Pr\bb{\vec{u} = +\vec{e}_{j}} = \Pr\bb{\vec{u} = -\vec{e}_{j}} = \frac{1}{2m}, \quad \forall 1 \leq j \leq m,
				\label{eqn:biorthogonal uniform distribution}
			\end{equation}
			where $\vec{u}$ has the same meaning as in \lemmaname~\ref{lem:Markov estimate} and \theoremname~\ref{thm:whp_suff_ident}.
			Let $\twonorm{\vec{x}}$ be drawn from a distribution (independent of $\vec{u}$) supported on the non-negative real axis with $\expect{\twonorm{\vec{x}}^{2}} = m$.
			Then, by construction, $\vec{x} = \twonorm{\vec{x}} \cdot \vec{u}$ satisfies assumption~\aref{itm:factor marginal} and it also satisfies assumption~\aref{itm:mutual indep} if $\twonorm{\vec{y}}$ and $\vec{v}$ are drawn analogously but independent of $\vec{u}$ and $\twonorm{\vec{x}}$.
			Using \eqref{eqn:biorthogonal uniform distribution}, we further observe that
			\begin{equation}
				\expect{\vec{u}} = \sum_{j=1}^{m} \BB{\Pr\bb{\vec{u} = +\vec{e}_{j}} - \Pr\bb{\vec{u} = -\vec{e}_{j}}} \cdot \vec{e}_{j} = \vec{0}
				\label{eqn:mean zero unit vector}
			\end{equation}
			and,
			\makeatletter
				\if@twocolumn
					\begin{equation}
						\begin{split}
							\expect{\vec{u} \tpose{\vec{u}}}
							& = \sum_{j=1}^{m} \BB{\Pr\bb{\vec{u} = +\vec{e}_{j}} + \Pr\bb{\vec{u} = -\vec{e}_{j}}} \cdot \vec{e}_{j} \tpose{\vec{e}_{j}}	\\
							& = \frac{1}{m} \sum_{j = 1}^{m} \vec{e}_{j} \tpose{\vec{e}_{j}} = \frac{1}{m} \eye
							\label{eqn:orthonormal sum}
						\end{split}
					\end{equation}
				\else
					\begin{equation}
						\expect{\vec{u} \tpose{\vec{u}}}
						= \sum_{j=1}^{m} \BB{\Pr\bb{\vec{u} = +\vec{e}_{j}} + \Pr\bb{\vec{u} = -\vec{e}_{j}}} \cdot \vec{e}_{j} \tpose{\vec{e}_{j}}
						= \frac{1}{m} \sum_{j = 1}^{m} \vec{e}_{j} \tpose{\vec{e}_{j}}
						= \frac{1}{m} \eye
						\label{eqn:orthonormal sum}
					\end{equation}
				\fi
			\makeatother
			where the last equality in \eqref{eqn:orthonormal sum} is true since $\cc{\vec{e}_{1}, \vec{e}_{2}, \dots, \vec{e}_{m}}$ is an orthonormal basis for $\setR^{m}$.
			By independence of $\twonorm{\vec{x}}$ from $\vec{u}$ we have
			\begin{equation}
				\expect{\vec{x}} = \expect{\twonorm{\vec{x}}} \cdot \expect{\vec{u}} = \vec{0}
			\end{equation}
			from \eqref{eqn:mean zero unit vector}, and
			\begin{equation}
				\expect{\vec{x} \tpose{\vec{x}}}
				= \expect{\twonorm{\vec{x}}^{2}} \cdot \expect{\vec{u} \tpose{\vec{u}}}
				= m \cdot \frac{1}{m} \eye = \eye
			\end{equation}
			from \eqref{eqn:orthonormal sum}.
			Hence, $\vec{x}$ is a zero mean random vector with an identity covariance matrix and thus satisfies assumption~\aref{itm:cov id}.

			Following the same line of reasoning as in \sectionname~\ref{rem:bernoulli example}, we can show that a bi-orthogonally supported uniform prior on $\vec{x}$ and $\vec{y}$ gives an example of the set $\mathcal{N}\argd{\mathscr{S}, 2} \bigcap \mathcal{M}$ with small complexity in the sense described in \sectionname~\ref{rem:measuring row column space pairs}.
			Indeed, we have $\mathcal{K} = \set{\bb{\lambda_{1} \vec{e}_{i}, \lambda_{2} \vec{f}_{j}}}{1 \leq i \leq m, 1 \leq j \leq n, \lambda_{1} \in \setR, \lambda_{2} \in \setR}$ where $\cc{\vec{e}_{1}, \vec{e}_{2}, \dots, \vec{e}_{m}}$ and $\cc{\vec{f}_{1}, \vec{f}_{2}, \dots, \vec{f}_{n}}$ respectively form an orthonormal basis for $\setR^{m}$ and $\setR^{n}$.
			Let us select $\mathcal{K}'$ according to \eqref{eqn:set change}, but without any relaxation, as $\mathcal{K}' = \set{\lambda \vec{e}_{i} \tpose{\vec{f}_{j}}}{1 \leq i \leq m, 1 \leq j \leq n, \lambda \in \setR}$.
			It is clear that matrices in $\mathcal{K}'$ can account for at most $m$ distinct column spaces and $n$ distinct row spaces, thus implying that matrices in $\mathcal{M} = \mathcal{K}' - \mathcal{K}'$ are generated by at most $\binom{m}{2} \leq m^{2}$ distinct column spaces and by at most $\binom{n}{2} \leq n^{2}$ distinct row spaces.
			Thus, $\mathcal{N}\argd{\mathscr{S}, 2} \bigcap \mathcal{M} \subseteq \mathcal{M}$ is of small complexity (only polynomially large in $m$ and $n$).
			In fact, \emph{exhaustive search for \problemname~\eqref{prob:rank} is tractable for any bi-orthogonally supported uniform prior, owing to the small complexity of $\mathcal{N}\argd{\mathscr{S}, 2} \bigcap \mathcal{M}$}.

		\subsection{Null Space of Linear Convolution}
			\label{sec:null space of linear convolution}
			The following proposition establishes a parametric representation of a subset of $\mathcal{N}\argd{\mathscr{S}, 2}$ where $\mathscr{S}\fcolon \setR^{m \times n} \to \setR^{m+n-1}$ denotes the lifted equivalent of the linear convolution map in \problemname~\eqref{prob:find_xy_deconv}.
			As described by \eqref{eqn:coordinate proj} in \sectionname~\ref{sec:lifting}, let $\mat{S}_k \in \setR^{m \times n}$, $1 \leq k \leq m+n-1$ denote a basis for $\mathscr{S}\argd{\cdot}$.
			For $1 \leq i \leq m$, $1 \leq j \leq n$ and $1 \leq k \leq m+n-1$, we have the description
			\begin{equation}
				\bb{\mat{S}_k}_{ij} =	\begin{cases}
												1,	& i + j = k + 1, \\
												0,	& \text{otherwise}.
											\end{cases}
			\end{equation}
			\figurename~\ref{fig:lifting example} illustrates a toy example of the linear convolution map with $m=3$ and $n=4$.

			\begin{proposition}
				\label{prop:rank-2 nullspace}
				If $\mat{X} \in \setR^{m \times n}$ admits a factorization of the form
				\begin{equation}
					\mat{X} =	\begin{bmatrix}
									\vec{u}	&	0	\\
									0	&	-\vec{u}
								\end{bmatrix}
								\begin{bmatrix}
									0	&	\tpose{\vec{v}}	\\
									\tpose{\vec{v}}	&	0
								\end{bmatrix}
					\label{eqn:rank-2 nullspace}
				\end{equation}
				for some $\vec{v} \in \setR^{n-1}$ and $\vec{u} \in \setR^{m-1}$, then $\mat{X} \in \mathcal{N}\argd{\mathscr{S}, 2}$.
			\end{proposition}

			\begin{IEEEproof}
				\appendixname~\ref{sec:rank-2 null space proposition proof}.
			\end{IEEEproof}

			Since the set of $m \times n$ dimensional rank two matrices has $2(m+n-2)$ DoF and $\mathscr{S}\argd{\cdot}$ maps $\setR^{m \times n}$ to $\setR^{m+n-1}$ with $\mathcal{N}\argd{\mathscr{S}, 1} = \cc{\mat{0}}$, $\mathcal{N}\bb{\mathscr{S}, 2}$ has at most $(2m+2n-4) - (m+n-1) = (m+n-3)$ DoF.
			We see that the representation on the r.h.s. of \eqref{eqn:rank-2 nullspace} also has $(m+n-3)$ DoF, so that our parametrization is tight up to DoF.
			The converse of \propositionname~\ref{prop:rank-2 nullspace} is false in general~\cite{choudhary2014identifiabilitylimitsSBD}.

		\subsection{Verification Methodology}
			\label{sec:verification technique}
			We test identifiability by (approximately) solving the following optimization problem,
			\minimize{\mat{X}}
			{\rank{\mat{X}}}
			{\fronorm{\mat{X} - \mat{M}} \leq \mu, \sep \mathscr{S}\argd{\mat{X}} = \vec{0},}
			{\label{prob:ReWtNucNormIdentifiability}}
			where $\mat{M} = \vec{x} \tpose{\vec{y}}$ is the true matrix and $\epsilon$ is a tuning parameter.
			The rationale behind solving \problemname~\eqref{prob:ReWtNucNormIdentifiability} is as follows.
			If the sufficient conditions of \theoremname~\ref{thm:suff_ident} are not satisfied, then $\exists \mat{X} \in \mathcal{N}\argd{\mathscr{S}, 2} \bigcap \mathcal{M}$ such that both $\vec{x} \in \mathcal{C}\argd{\mat{X}}$ and $\vec{y} \in \mathcal{R}\argd{\mat{X}}$ are true.
			We approximate the event
			\begin{subequations}
				\begin{align}
					\mathcal{E}_{1}	& = \cc{\exists \mat{X} \in \mathcal{N}\argd{\mathscr{S}, 2} \text{ with } \vec{u} \in \mathcal{C}\bb{\mat{X}}, \vec{v} \in \mathcal{R}\bb{\mat{X}}}	\\
					\shortintertext{by the event}
					\mathcal{E}_{2}	& = \cc{\exists \mat{X} \in \mathcal{N}\argd{\mathscr{S}, 2} \text{ such that } \fronorm{\mat{X} - \mat{M}} \leq \mu}.	\label{eqn:event E2}
				\end{align}
			\end{subequations}
			As \problemname~\eqref{prob:ReWtNucNormIdentifiability} is itself NP-hard to solve exactly, we can employ the re-weighted nuclear norm heuristic \cite{mohan2010reweighted} to solve \problemname~\eqref{prob:ReWtNucNormIdentifiability} approximately.
			If the resulting solution to \problemname~\eqref{prob:ReWtNucNormIdentifiability} has rank two then we declare that event $\mathcal{E}_{2}$ has happened.
			Clearly, we have $\mathcal{E}_{2} \subseteq \mathcal{E}_{1}$ so that sufficient conditions for identifiability by \theoremname~\ref{thm:suff_ident} fail if event $\mathcal{E}_{2}$ took place.

			The examples we consider in \sectionsname~\ref{sec:small finite null space} to \ref{sec:infinite null space} are, however, motivated from the representation in \eqref{eqn:rank-2 nullspace} and share the same parametrization structure for $\mathcal{N}\argd{\mathscr{S}, 2}$.
			This enables us to use approximate verification techniques that are faster than the re-weighted nuclear norm heuristic, especially if the search space is discrete and finite.
			The re-weighted nuclear norm heuristic is still useful if no parametrization structure is available for $\mathcal{N}\argd{\mathscr{S}, 2}$.

		\subsection{Small Complexity of \texorpdfstring{$\mathcal{N}\argd{\mathscr{S}, 2} \bigcap \mathcal{M}$}{Restricted Rank Two Null Space}}
			\label{sec:small finite null space}
\begin{figure}
				\centering
				\includegraphics[width=\figwidth]{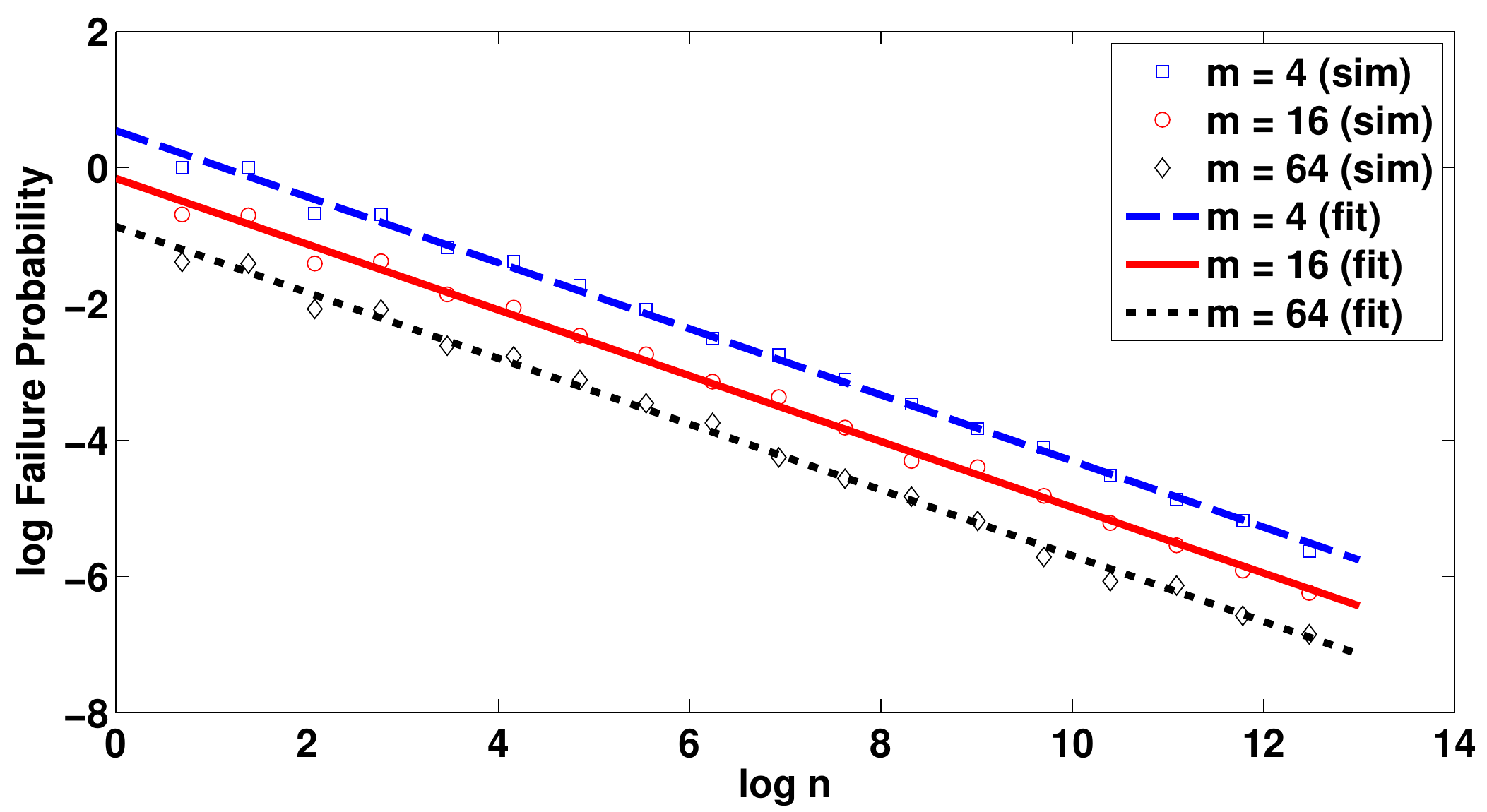}
				\caption{Linear Scaling behavior of $\log\argd{\text{Failure Probability}}$ with $\log n$ for fixed values of $m$. The absolute value of the fitted slope is 0.48.}
				\label{fig:Example3}
			\end{figure}

			Let $\vec{x} \in \cc{\vec{e}_{1}', \vec{e}_{2}', \dots, \vec{e}_{m}'}$ and $\vec{y} \in \cc{\vec{f}_{1}', \vec{f}_{2}', \dots, \vec{f}_{n}'}$ be drawn from bi-orthogonally supported uniform distributions, as described in \sectionname~\ref{sec:dependant but uncorrelated}, where $\cc{\vec{e}_{1}', \vec{e}_{2}', \dots, \vec{e}_{m}'}$ and $\cc{\vec{f}_{1}', \vec{f}_{2}', \dots, \vec{f}_{n}'}$ respectively represent the canonical bases for $\setR^{m}$ and $\setR^{n}$.
			We consider a lifted linear operator $\mathscr{S}\bb{\cdot}$ with the following description:
			$\mathcal{N}\argd{\mathscr{S}, 2} \bigcap \mathcal{M}$ consists of $\floor{\sqrt{m}} \cdot \floor{\sqrt{n}}$ parts and the $\bb{i,j}^{\thp}$ part $\mathcal{P}_{ij}$, $1 \leq i \leq \floor{\sqrt{m}}$, $1 \leq j \leq \floor{\sqrt{n}}$ is given by
			\begin{equation}
				\mathcal{P}_{ij}
				=	\set{\lambda	\begin{bmatrix}
										\vec{e}_{i}	&	0	\\
												0	&	-\vec{e}_{i}
									\end{bmatrix}
									\begin{bmatrix}
																0	&	\tpose{\vec{f}_{j}}	\\
										\tpose{\vec{f}_{j}}	&	0
									\end{bmatrix}}%
						{\lambda \in \setR}
				\label{eqn:small finite null space}
			\end{equation}
			where $\cc{\vec{e}_{1}, \vec{e}_{2}, \dots, \vec{e}_{m-1}}$ and $\cc{\vec{f}_{1}, \vec{f}_{2}, \dots, \vec{f}_{n-1}}$ respectively denote the canonical basis for $\setR^{m-1}$ and $\setR^{n-1}$, and $\floor{\cdot}$ is the floor function.
			Clearly, the elements of $\mathcal{P}_{ij}$ are closely related to the representation in \eqref{eqn:rank-2 nullspace}.
			For this lifted linear operator, the bound of \theoremname~\ref{thm:whp_suff_ident} is applicable with $f_{\mathscr{S}, \mathcal{M}}\argd{m,n} = \floor{\sqrt{m}} \cdot \floor{\sqrt{n}}$ implying that the probability of failure to satisfy the sufficient conditions of \theoremname~\ref{thm:suff_ident} decreases as $\BigOh{1/\sqrt{mn}}$.
			Since exhaustive search for event $\mathcal{E}_{2}$ is tractable (see \sectionname~\ref{sec:dependant but uncorrelated}), we employ the same to compute the failure probability.
			The results are plotted in \figurename~\ref{fig:Example3} on a log-log scale.
			Note that we have plotted the best linear fit for the simulated parameter values, since the probabilities can be locally discontinuous in $\log n$ due to the appearance of $\floor{\cdot}$ function in the expression of $f_{\mathscr{S}, \mathcal{M}}\argd{m,n}$.
			We see that the simulated order of growth of the failure probability is $\BigOh{n^{-0.48}}$ for every fixed value of $m$ (exponent determined by slope of plot in \figurename~\ref{fig:Example3}) almost exactly matches the theoretically predicted order of growth (equals $\BigOh{n^{-0.5}}$).
			
		\subsection{Large Complexity of \texorpdfstring{$\mathcal{N}\argd{\mathscr{S}, 2} \bigcap \mathcal{M}$}{Restricted Rank Two Null Space}}
			\label{sec:large finite null space}
\begin{figure}
				\centering
				\includegraphics[width=\figwidth]{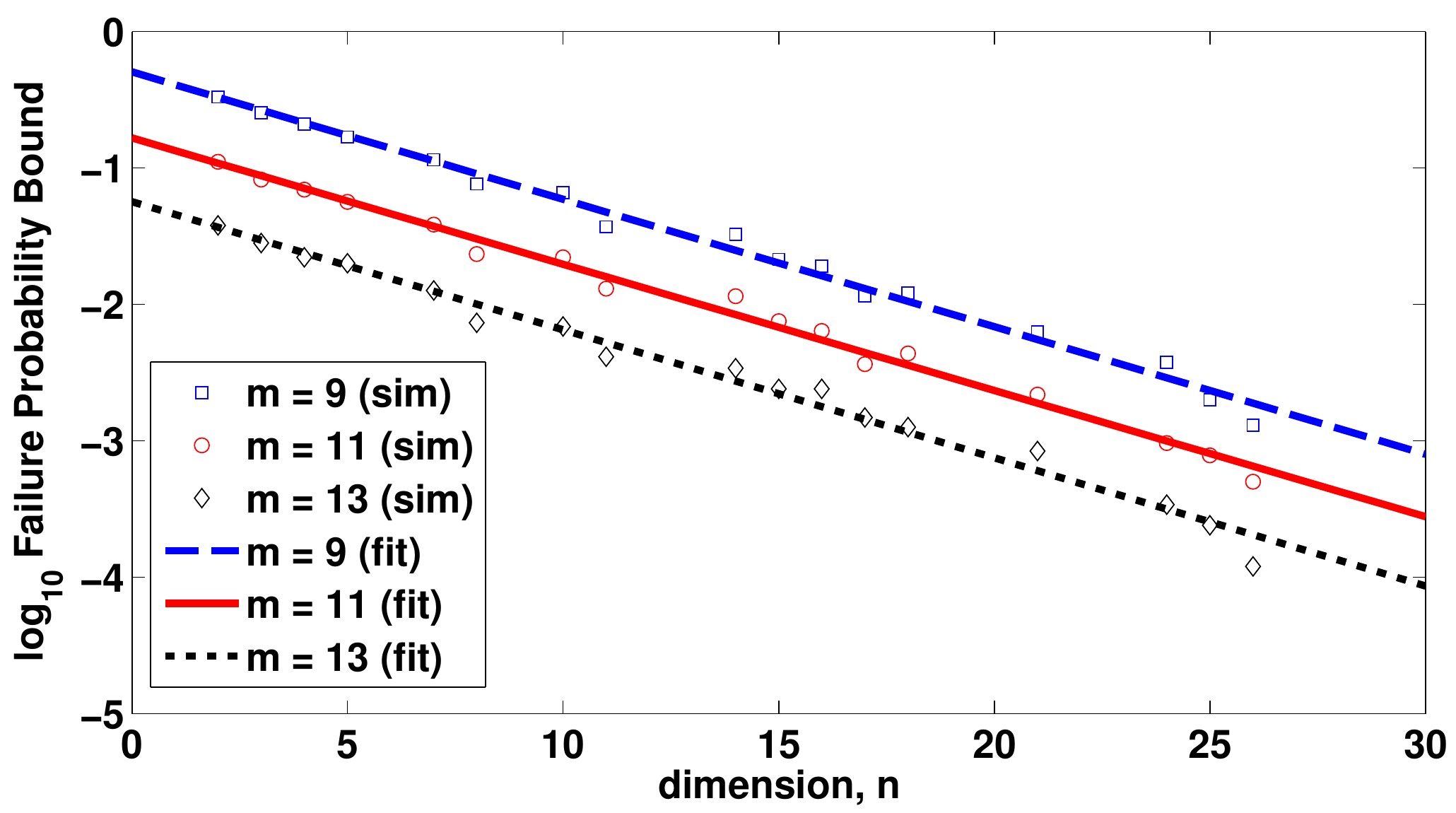}
				\caption{Linear Scaling behavior of $\log\argd{\text{Failure Probability}}$ with problem dimension $n$ for fixed values of $m$. The absolute value of the fitted slopes are between 0.093 and 0.094.}
				\label{fig:Example4}
			\end{figure}

			Let $\vec{x} \in \setR^{m}$ and $\vec{y} \in \setR^{n}$ be drawn component-wise independently from a symmetric Bernoulli distribution (see \sectionname~\ref{rem:bernoulli example}) and let $\tau \in \bb{0,1}$ be a constant.
			Following our guiding representation \eqref{eqn:rank-2 nullspace}, we consider a lifted linear operator $\mathscr{S}\argd{\cdot}$ with the following description: $\mathcal{N}\argd{\mathscr{S}, 2}$ consists of $2^{\floor{\tau m}} \times 2^{\floor{\tau n}}$ parts and the $\bb{i,j}^{\thp}$ part $\mathcal{P}_{ij}$, $0 \leq i \leq 2^{\floor{\tau m}} - 1$, $0 \leq j \leq 2^{\floor{\tau n}} - 1$, is given by
			\begin{equation}
				\mathcal{P}_{ij}
				= \set{\lambda	\begin{bmatrix}
									\vec{g}_{i}	&	0	\\
										\vec{1}	&	-\vec{g}_{i}	\\
											0	&	-\vec{1}
								\end{bmatrix}
								\begin{bmatrix}
									0	&	\tpose{\vec{h}_{j}}	&	\tpose{\vec{1}}	\\
									\tpose{\vec{h}_{j}}	&	\tpose{\vec{1}}	&	0
								\end{bmatrix}}%
						{\lambda \in \setR}
				\label{eqn:large finite null space}
			\end{equation}
			where $\vec{g}_{i} \in \cc{-1,1}^{\floor{\tau m}}$ (respectively $\vec{h}_{j} \in \cc{-1,1}^{\floor{\tau n}}$) denotes the binary representation of $i$ (respectively $j$) of length $\floor{\tau m}$ bits (respectively $\floor{\tau n}$ bits) expressed in the alphabet set $\cc{-1,1}$, and the all one column vectors in \eqref{eqn:large finite null space} are of appropriate dimensions so that the elements of $\mathcal{P}_{ij}$ are matrices in $\setR^{m \times n}$.
			The bound in \theoremname~\ref{thm:whp_infinite} is applicable to this example.
			We employ exhaustive search for event $\mathcal{E}_{2}$ for small values of $m$ and $n$ (it is computationally intractable for large $m$ or $n$).
			The results are plotted in \figurename~\ref{fig:Example4} on a semilog scale, where we have used $\tau = 0.2$ and $\delta' = 0.3$ and $\delta'$ is as in the statement of \theoremname~\ref{thm:whp_infinite}.
			As in the case of \figurename~\ref{fig:Example3}, we plot the best linear fit for the simulated parameter values to disregard local discontinuities introduced due to the use of the $\floor{\cdot}$ function.

			Since it is hard to analytically compute the metric entropies $p_{c}$ and $p_{r}$, we shall settle for a numerical verification of the scaling law with problem dimension and an approximate argument as to the validity of predictions made by \theoremname~\ref{thm:whp_infinite} for this example.
			By construction, we have the bounds $p_{c} \leq \floor{\tau m}$ and $p_{r} \leq \floor{\tau n}$ but the careful reader will note that because of the element-wise constant magnitude property of a symmetric Bernoulli random vector, it does not lie in the column span of any of the matrices in $\mathcal{N}\argd{\mathscr{S}, 2}$, as described by the generative description in \eqref{eqn:large finite null space}, but can be arbitrarily close to such a span as $m$ increases.
			We thus expect that $p_{c} = \epsilon_{c} m$ and $p_{r} = \epsilon_{r} n$ for some parameters $\epsilon_{c}$ and $\epsilon_{r}$ close to zero.
			By choice of parameters, $\epsilon \leq \sqrt{(1-\delta')/2} = 0.59$.
			With $\epsilon_{c} = \epsilon_{r} = 0$ and setting $\epsilon = 0.01$ the theoretical prediction on the absolute value of the slope  is 0.073 which is quite close to the simulated value of 0.093.
			We clearly recover the linear scaling behavior of the logarithm of failure probability with the problem dimension $n$.

		\subsection{Infinite Complexity of \texorpdfstring{$\mathcal{N}\argd{\mathscr{S}, 2} \bigcap \mathcal{M}$}{Restricted Rank Two Null Space}}
			\label{sec:infinite null space}
\begin{figure}
				\centering
				\includegraphics[width=\figwidth]{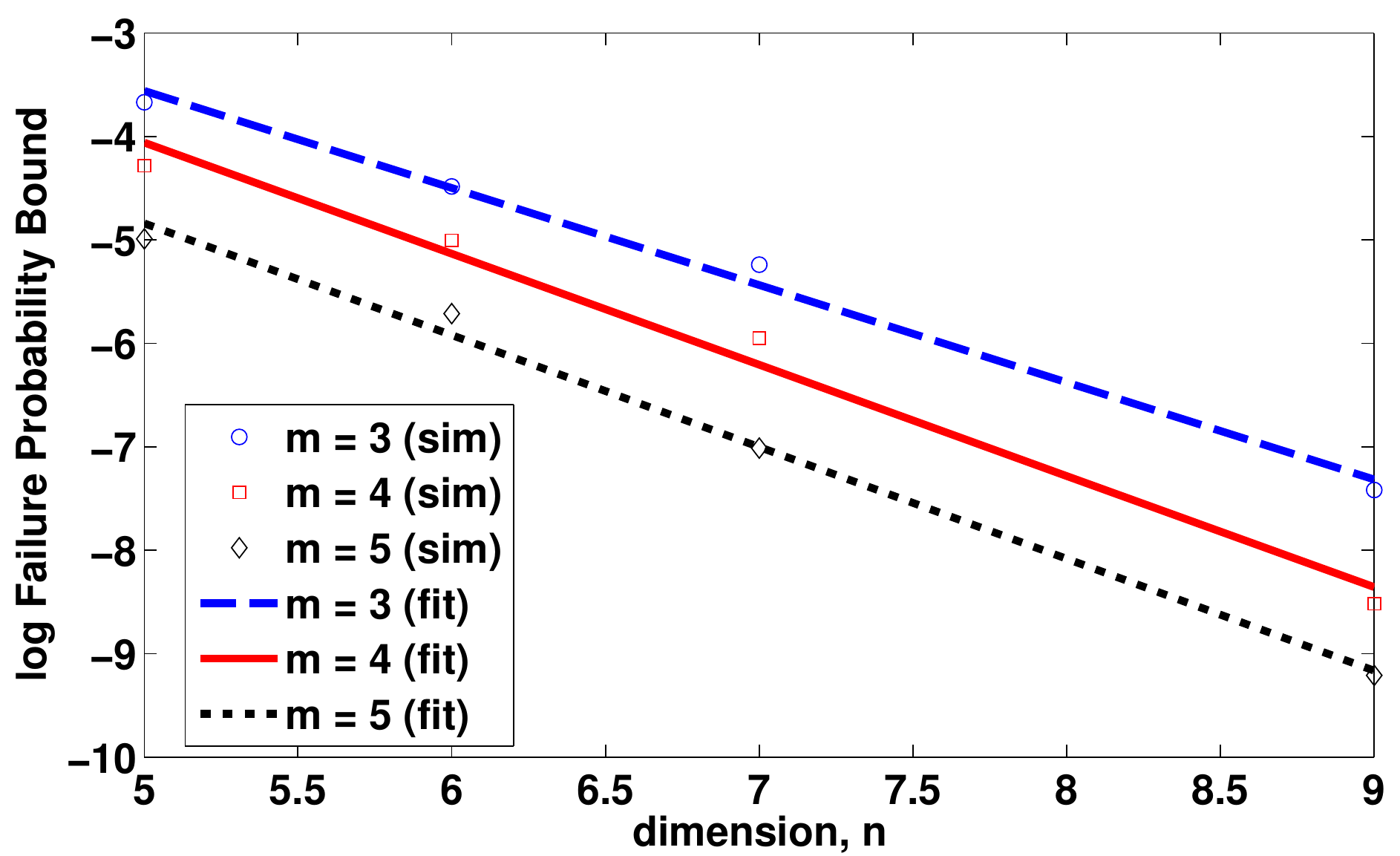}
				\caption{Exponentially decaying behavior of the simulated failure probability \wrt~$n$ for fixed values of $m$, for parameter $\mu = 0.8$ and the lifted linear convolution map. The absolute value of the fitted slopes are between 0.94 and 1.08.}
				\label{fig:Example2}
			\end{figure}

			Let $\vec{x} \in \setR^{m}$ and $\vec{y} \in \setR^{n}$ be drawn component-wise independently from a standard Normal distribution.
			We consider the linear convolution operator from \problemname~\eqref{prob:find_xy_deconv}, letting $\mathscr{S}\bb{\cdot}$ denote the lifted linear convolution map.
			A representation of $\mathscr{S}\bb{\cdot}$ and a description of the rank two null space $\mathcal{N}\bb{\mathscr{S}, 2}$ has been mentioned in the prequel (\sectionname~\ref{sec:null space of linear convolution}).
			The bound in \theoremname~\ref{thm:whp_infinite_Gaussian} is applicable to this example.
			However, unlike the examples in \sectionsname~\ref{sec:small finite null space} and~\ref{sec:large finite null space}, we cannot employ exhaustive search over $\mathcal{N}\argd{\mathscr{S}, 2}$ to test identifiability, since the search space is uncountably infinite by \propositionname~\ref{prop:rank-2 nullspace}.
			We resort to the method described in \sectionname~\ref{sec:verification technique} relying on the re-weighted nuclear norm heuristic.
			The results are plotted in \figurename~\ref{fig:Example2} on a semilog scale, where we have used $\mu = 0.8$ to detect the occurrence of the event $\mathcal{E}_{2}$ as described by \eqref{eqn:event E2}, and $\mat{M}$ in \problemname~\eqref{prob:ReWtNucNormIdentifiability} is normalized such that $\fronorm{\mat{M}} = 1$.
			A relatively high value of $\mu = 0.8$ is used to ensure that the rare event $\mathcal{E}_{2}$ admits a large enough probability of occurrence.
			Only data points that satisfy $n \geq m$ are plotted since the behavior of the convolution operator is symmetric \wrt~the order of its inputs.
			Since the re-weighted nuclear norm heuristic does not always converge monotonically in a small number of steps, we stopped execution after a finite number of steps, which might explain the small deviation from linearity, observed in \figurename~\ref{fig:Example2}, as compared to the respective best linear fits on the same plot.
			Nonetheless, we approximately recover the theoretically predicted qualitative linear scaling law of the logarithm of the failure probability with the problem dimension $n$, for fixed values of $m$.
			There does not seem to be an easy way of comparing the constants involved in the simulated result to their theoretical counterparts as predicted by \theoremname~\ref{thm:whp_infinite_Gaussian}.

	\section{Conclusions}
		\label{sec:conclusion}
		Bilinear transformations occur in a number of signal processing problems like linear and circular convolution, matrix product, linear mixing of multiple sources, \etc{}
		Identifiability and signal reconstruction for the corresponding inverse problems are important in practice and identifiability is a precursor to establishing any form of reconstruction guarantee.
		In the current work, we determined a series of sufficient conditions for identifiability in conic prior constrained Bilinear Inverse Problems (BIPs) and investigated the probability of achieving those conditions under three classes of random input signal ensembles, \viz~dependent but uncorrelated, independent Gaussian, and independent Bernoulli.
		The theory is \emph{unified} in the sense that it is applicable to all BIPs, and is specifically developed for bilinear maps over vector pairs with non-trivial rank two null space.
		Universal identifiability is absent for many interesting and important BIPs owing to the non-triviality of the rank two null space, but a deterministic characterization of the input instance identifiability is still possible (may be hard to check).
		Our probabilistic results were formulated as scaling laws that trade-off probability of identifiability with the complexity of the restricted rank two null space of the bilinear map in question, and results were derived for three different levels of complexity, \viz~small (polynomial in the signal dimension), large (exponential in the signal dimension) and infinite.
		In each case, identifiability can hold with high probability depending on the relative geometry of the null space of the bilinear map and the signal space.
		Overall, most random input instances are identifiable, with the probability of identifiability scaling inversely with the complexity of the rank two null space of the bilinear map.
		An especially appealing aspect of our approach is that the rank two null space can be partly or fully characterized for many bilinear problems of interest.
		We demonstrated this by partly characterizing the rank two null space of the linear convolution map, and presented numerical verification of the derived scaling laws on examples that were based on variations of the blind deconvolution problem, exploiting the representation of its rank two null space.
		Overall, the results in this paper indicate that lifting is a powerful technique for identifiability analysis of general cone constrained BIPs.

	\makeatletter
		\ifbool{@bibtex}{
			\bibliographystyle{IEEEtran}
			\bibliography{IEEEabrv,UWA,ownpub,PaperList}
		}{
			\printbibliography[heading=bibintoc]
		}
	\makeatother

	\appendices

	\section{Proof of \theoremname~\ref{thm:equivalence}}
		\label{sec:equivalence theorem proof}
		\begin{enumerate}
			\item	Suppose that $\bb{\vec{x}_{0}, \vec{y}_{0}} \in \mathcal{K}$ is a solution to \problemname~\eqref{prob:find_xy} for a given observation $\vec{z} = \vec{z}_{0} = \vec{S}\argd{\vec{x}_{0}, \vec{y}_{0}}$.
			Setting $\mat{W}_{0} = \vec{x}_{0} \tpose{\vec{y}_{0}}$ and using \eqref{eqn:set change} we have $\mat{W}_{0} \in \mathcal{K}'$.
			Using \eqref{eqn:lifted op} we get $\mathscr{S}\argd{\mat{W}_{0}} = \vec{S}\argd{\vec{x}_{0}, \vec{y}_{0}} = \vec{z}_{0}$ and $\rank{\mat{W}_{0}} = \rank{\vec{x}_{0} \tpose{\vec{y}_{0}}} \leq 1$.
			Thus, $\mat{W}_{0}$ is a feasible point of \problemname~\eqref{prob:rank} with rank at most one.
			As there exists a $\text{rank} \leq 1$ matrix $\mat{W}$ satisfying $\mathscr{S}\argd{\mat{W}} = \vec{z}$ and $\mat{W} \in \mathcal{K}'$, the solution of \problemname~\eqref{prob:rank} must be of rank one or less.
			
			\item	Any $\mat{W} \in \mathcal{K}'_{\opt} \subseteq \mathcal{K}'$ satisfies $\rank{\mat{W}} \leq 1$ (see proof of first part) and $\mathscr{S}\argd{\mat{W}} = \vec{z}$.
			Thus,
			\begin{subequations}
				\makeatletter
					\if@twocolumn
						\begin{align}
							\mathcal{K}'_{\opt}	& \subseteq \mathcal{K}' \bigcap \set{\mat{W} \in \setR^{m \times n}}{\rank{\mat{W}} \leq 1}	\notag \\
							& \hphantom{\subseteq \mathcal{K}'} \; \bigcap \set{\mat{W} \in \setR^{m \times n}}{\mathscr{S}\argd{\mat{W}} = \vec{z}}	\label{eqn:feasible set} \\
							& = \set{\vec{x} \tpose{\vec{y}}}{\bb{\vec{x}, \vec{y}} \in \mathcal{K}} \bigcap \set{\mat{W}}{\mathscr{S}\argd{\mat{W}} = \vec{z}}	\label{eqn:condense rank} \\
							& = \set{\vec{x} \tpose{\vec{y}}}{\bb{\vec{x}, \vec{y}} \in \mathcal{K}} \bigcap \set{\vec{x} \tpose{\vec{y}}}{\mathscr{S}\argd{\vec{x} \tpose{\vec{y}}} = \vec{z}}	\notag \\
							& = \set{\vec{x} \tpose{\vec{y}}}{\bb{\vec{x}, \vec{y}} \in \mathcal{K}} \bigcap \set{\vec{x} \tpose{\vec{y}}}{\vec{S}\argd{\vec{x}, \vec{y}} = \vec{z}}	\label{eqn:condense constraint} \\
							& = \set{\vec{x} \tpose{\vec{y}}}{\bb{\vec{x}, \vec{y}} \in \mathcal{K}_{\opt}},	\label{eqn:K opt}
						\end{align}
					\else
						\begin{align}
							\mathcal{K}'_{\opt}	& \subseteq \mathcal{K}' \bigcap \set{\mat{W} \in \setR^{m \times n}}{\rank{\mat{W}} \leq 1} \bigcap \set{\mat{W} \in \setR^{m \times n}}{\mathscr{S}\argd{\mat{W}} = \vec{z}}	\label{eqn:feasible set} \\
							& = \set{\vec{x} \tpose{\vec{y}}}{\bb{\vec{x}, \vec{y}} \in \mathcal{K}} \bigcap \set{\mat{W}}{\mathscr{S}\argd{\mat{W}} = \vec{z}}	\label{eqn:condense rank} \\
							& = \set{\vec{x} \tpose{\vec{y}}}{\bb{\vec{x}, \vec{y}} \in \mathcal{K}} \bigcap \set{\vec{x} \tpose{\vec{y}}}{\mathscr{S}\argd{\vec{x} \tpose{\vec{y}}} = \vec{z}}	\notag \\
							& = \set{\vec{x} \tpose{\vec{y}}}{\bb{\vec{x}, \vec{y}} \in \mathcal{K}} \bigcap \set{\vec{x} \tpose{\vec{y}}}{\vec{S}\argd{\vec{x}, \vec{y}} = \vec{z}}	\label{eqn:condense constraint} \\
							& = \set{\vec{x} \tpose{\vec{y}}}{\bb{\vec{x}, \vec{y}} \in \mathcal{K}_{\opt}},	\label{eqn:K opt}
						\end{align}
					\fi
				\makeatother
			\end{subequations}
			where \eqref{eqn:condense rank} is due to \eqref{eqn:set change}, \eqref{eqn:condense constraint} is due to \eqref{eqn:lifted op} and \eqref{eqn:K opt} is true because \problemname~\eqref{prob:find_xy} is a feasibility problem.
			
			\item	The feasible set for \problemname~\eqref{prob:rank} is $\mathcal{K}' \bigcap \set{\mat{W} \in \setR^{m \times n}}{\mathscr{S}\argd{\mat{W}} = \vec{z}}$.
			From the proof of second part, we know that
			\makeatletter
				\if@twocolumn
					\begin{multline}
						\mathcal{K}' \bigcap \set{\mat{W}}{\rank{\mat{W}} \leq 1} \bigcap \set{\mat{W}}{\mathscr{S}\argd{\mat{W}} = \vec{z}}	\\
						= \set{\vec{x} \tpose{\vec{y}}}{\bb{\vec{x}, \vec{y}} \in \mathcal{K}_{\opt}}.
					\end{multline}
				\else
					\begin{equation}
						\mathcal{K}' \bigcap \set{\mat{W}}{\rank{\mat{W}} \leq 1} \bigcap \set{\mat{W}}{\mathscr{S}\argd{\mat{W}} = \vec{z}}
						= \set{\vec{x} \tpose{\vec{y}}}{\bb{\vec{x}, \vec{y}} \in \mathcal{K}_{\opt}}.
					\end{equation}
				\fi
			\makeatother
			Thus, clearly
			\begin{equation}
				\set{\vec{x} \tpose{\vec{y}}}{\bb{\vec{x}, \vec{y}} \in \mathcal{K}_{\opt}} \subseteq \mathcal{K}' \bigcap \set{\mat{W}}{\mathscr{S}\argd{\mat{W}} = \vec{z}}.
				\label{eqn:K opt feasible set}
			\end{equation}
			
			We shall prove the contrapositive statement in each direction.
			First assume that $\cc{\mat{0}} \subsetneq \set{\vec{x} \tpose{\vec{y}}}{\bb{\vec{x}, \vec{y}} \in \mathcal{K}_{\opt}}$.
			By~\eqref{eqn:K opt feasible set}, $\mat{0}$ is a feasible point for \problemname~\eqref{prob:rank} and thus $\rank{\mat{0}} = 0$ is the optimal value for this problem.
			Since every $\mat{W} \in \setR^{m \times n} \setminus \cc{\mat{0}}$ has a rank strictly greater than zero we conclude that $\mathcal{K}'_{\opt} = \cc{\mat{0}} \neq \set{\vec{x} \tpose{\vec{y}}}{\bb{\vec{x}, \vec{y}} \in \mathcal{K}_{\opt}}$.
			Conversely, suppose that $\mathcal{K}'_{\opt} \neq \set{\vec{x} \tpose{\vec{y}}}{\bb{\vec{x}, \vec{y}} \in \mathcal{K}_{\opt}}$.
			Since $\mathcal{K}'_{\opt} \subseteq \set{\vec{x} \tpose{\vec{y}}}{\bb{\vec{x}, \vec{y}} \in \mathcal{K}_{\opt}}$ (see proof of second part), $\exists \bb{\vec{x}_{0}, \vec{y}_{0}} \in \mathcal{K}_{\opt}$ such that $\vec{x}_{0} \tpose{\vec{y}_{0}} \notin \mathcal{K}'_{\opt}$.
			By \eqref{eqn:K opt feasible set}, $\vec{x}_{0} \tpose{\vec{y}_{0}}$ is a feasible point for \problemname~\eqref{prob:rank} and hence the optimal value of this problem is strictly less than $\rank{\vec{x}_{0} \tpose{\vec{y}_{0}}} \leq 1$.
			The only way for this to be possible is to have $\rank{\vec{x}_{0} \tpose{\vec{y}_{0}}} = 1$ and the optimal value of \problemname~\eqref{prob:rank} as zero.
			Since the only matrix of rank zero is the all zero matrix, we conclude that $\cc{\mat{0}} = \mathcal{K}'_{\opt} \subsetneq \set{\vec{x} \tpose{\vec{y}}}{\bb{\vec{x}, \vec{y}} \in \mathcal{K}_{\opt}}$.
		\end{enumerate}

	\section{Proof of \corollaryname~\ref{cor:equivalence}}
		\label{sec:equivalence corollary proof}
		From \eqref{eqn:rank-k null space}, \eqref{eqn:feasible set} and \eqref{eqn:K opt} we have
		\begin{equation}
			\mathcal{K}' \bigcap \mathcal{N}\argd{\mathscr{S}, 1} = \set{\vec{x} \tpose{\vec{y}}}{\bb{\vec{x}, \vec{y}} \in \mathcal{K}_{\opt}\argd{\vec{0}}}.
			\label{eqn:rank-1 null space and feasible set}
		\end{equation}
		We shall prove the contrapositive statements.
		First assume that $\cc{\mat{0}} \subsetneq \mathcal{K}' \bigcap \mathcal{N}\argd{\mathscr{S}, 1}$.
		Using \eqref{eqn:rank-1 null space and feasible set}, we have $\cc{\mat{0}} \subsetneq \set{\vec{x} \tpose{\vec{y}}}{\bb{\vec{x}, \vec{y}} \in \mathcal{K}_{\opt}\argd{\vec{0}}}$ and the last part of \theoremname~\ref{thm:equivalence} implies that $\mathcal{K}'_{\opt}\argd{\vec{0}} \neq \set{\vec{x} \tpose{\vec{y}}}{\bb{\vec{x}, \vec{y}} \in \mathcal{K}_{\opt}\argd{\vec{0}}}$.
		Since $\mathcal{K}_{\opt}\argd{\vec{0}}$ is nonempty, $\vec{0} \in \set{\vec{S}\argd{\vec{x}, \vec{y}}}{\bb{\vec{x}, \vec{y}} \in \mathcal{K}}$.
		Thus, \problemsname~\eqref{prob:find_xy} and \eqref{prob:rank} are not equivalent (equivalence fails for $\vec{z} = \vec{0}$).
		Conversely, suppose that $\exists \, \vec{z} \in \set{\vec{S}\argd{\vec{x}, \vec{y}}}{\bb{\vec{x}, \vec{y}} \in \mathcal{K}}$ resulting in $\mathcal{K}'_{\opt}\argd{\vec{z}} \neq \set{\vec{x} \tpose{\vec{y}}}{\bb{\vec{x}, \vec{y}} \in \mathcal{K}_{\opt}\argd{\vec{z}}}$.
		Using last part of \theoremname~\ref{thm:equivalence}, we have $\cc{\mat{0}} \subsetneq \set{\vec{x} \tpose{\vec{y}}}{\bb{\vec{x}, \vec{y}} \in \mathcal{K}_{\opt}\argd{\vec{z}}}$, which is possible only if $\vec{z} = \mathscr{S}\argd{\mat{0}} = \vec{0}$.
		Now using \eqref{eqn:rank-1 null space and feasible set} we get $\cc{\mat{0}} \subsetneq \mathcal{K}' \bigcap \mathcal{N}\argd{\mathscr{S}, 1}$.

	\section{Proof of \propositionname~\ref{prop:ident_easy}}
		\label{sec:ident_easy proposition proof}
		\problemname~\eqref{prob:rank} fails if and only if it admits more than one optimal solution.
		
		Let $\mathcal{N}\argd{\mathscr{S}, 2} \bigcap \mathcal{M} = \cc{\mat{0}}$ and for the sake of contradiction suppose that $\mat{W}_1 \in \mathcal{K}'$ and $\mat{W}_2 \in \mathcal{K}'$ denote two solutions to \problemname~\eqref{prob:rank} for some observation $\vec{z}$, so that $\bb{\mat{W}_1 - \mat{W}_2} \in \mathcal{M}$.
		Then, $\mathscr{S}\argd{\mat{W}_1} = \mathscr{S}\argd{\mat{W}_2}$ so that $\bb{\mat{W}_1 - \mat{W}_2}$ is in the null space of $\mathscr{S}$.
		But, $\rank{\mat{W}_1 - \mat{W}_2} \leq \rank{\mat{W}_1} + \rank{\mat{W}_2} \leq 2$ so that we have $\mat{W}_1 - \mat{W}_2 = \mat{0}$ and \problemname~\eqref{prob:rank} has a unique solution.
		
		Conversely, let \problemname~\eqref{prob:rank} have a unique solution for every observation $\vec{z} = \vec{S}\argd{\vec{x}, \vec{y}}$.
		For the sake of contradiction, suppose that there is a matrix $\mat{Y}$ in $\mathcal{N}\argd{\mathscr{S}, 2} \bigcap \mathcal{M} \setminus \cc{\mat{0}}$.
		Since $\mat{Y} \in \mathcal{M}$, $\exists \mat{Y}_1, \mat{Y}_2 \in \mathcal{K}'$ such that $\mat{Y} = \mat{Y}_1 - \mat{Y}_2$.
		Further, $\mat{Y} \neq \mat{0}$ is in the null space of $\mathscr{S}$, so that $\vec{z} = \mathscr{S}\argd{\mat{Y}_1} = \mathscr{S}\argd{\mat{Y}_2}$ with $\mat{Y}_{1} \neq \mat{Y}_{2}$ implying that $\mat{Y}_1$ and $\mat{Y}_2$ are both valid solutions to \problemname~\eqref{prob:rank} for the observation $\vec{z}$.
		Since $\mathcal{K}' = \set{\vec{x} \tpose{\vec{y}}}{\bb{\vec{x}, \vec{y}} \in \mathcal{K}}$, $\exists \bb{\vec{x}_{1}, \vec{y}_{1}}, \bb{\vec{x}_{2}, \vec{y}_{2}} \in \mathcal{K}$ such that $\vec{S}\argd{\vec{x}_{1}, \vec{y}_{1}} = \mathscr{S}\argd{\mat{Y}_1}$ and $\vec{S}\argd{\vec{x}_{2}, \vec{y}_{2}} = \mathscr{S}\argd{\mat{Y}_2}$, so that $\vec{z} = \mathscr{S}\argd{\mat{Y}_1} = \mathscr{S}\argd{\mat{Y}_2}$ is a valid observation.
		This violates the unique solution assumption on \problemname~\eqref{prob:rank} for the valid observation $\vec{z}$.
		Hence $\mathcal{N}\argd{\mathscr{S}, 2} \bigcap \mathcal{M} = \cc{\mat{0}}$, completing the proof.

	\section{Proof of \theoremname~\ref{thm:suff_ident}}
		\label{sec:suff_ident theorem proof}
		Let $\mat{M}^{\ast} \in \mathcal{K}'$ be a solution to \problemname~\eqref{prob:rank} such that $\mat{M}^{\ast} \neq \mat{M}$.
		Since $\mat{M}$ is a valid solution to \problemname~\eqref{prob:rank}, we have $\rank{\mat{M}^{\ast}} = \rank{\mat{M}} = 1$ and $\mat{X} = \mat{M} - \mat{M}^{\ast} \in \mathcal{N}\argd{\mathscr{S}, 2} \bigcap \mathcal{M} \setminus \cc{\mat{0}}$.
		If $\mat{M}^{\ast} = \sigma_{\ast} \vec{u}_{\ast} \tpose{\vec{v}_{\ast}}$, then $\mathcal{R}\argd{\mat{X}} = \spn{\vec{v}, \vec{v}_{\ast}}$ and $\mathcal{C}\argd{\mat{X}} = \spn{\vec{u}, \vec{u}_{\ast}}$.
		This contradicts the assumption that at least one of $\vec{u} \notin \mathcal{C}\argd{\mat{X}}$ or $\vec{v} \notin \mathcal{R}\argd{\mat{X}}$ is true and completes the proof.

	\section{Proof of \corollaryname~\ref{cor:equal singular values}}
		\label{sec:equal singular values corollary proof}
		We start with the ``if'' part.
		For $\mat{M}$ to be identifiable, we need $\rank{\mat{M} - \mat{X}} > 1$ for every matrix $\mat{X} \neq \mat{M}$ in the null space of $\mathscr{S}\argd{\cdot}$ that also satisfies $\mat{M} - \mat{X} \in \mathcal{K}'$.
		Since $\rank{\mat{M} - \mat{X}} \geq \rank{\mat{X}} - \rank{\mat{M}} = \rank{\mat{X}} - 1$, it is sufficient to consider matrices $\mat{X}$ with $\rank{\mat{X}} \leq 2$.
		Thus, for identifiability of $\mat{M}$, we need $\rank{\mat{M} - \mat{X}} > 1$, $\forall \mat{X} \in \mathcal{N}\argd{\mathscr{S}, 2} \bigcap \bb{\mat{M} - \mathcal{K}'} \setminus \cc{\mat{0}}$.
		Using $\mathcal{N}\argd{\mathscr{S}, 1} \bigcap \mathcal{M} = \cc{\mat{0}}$ and $\mat{X} \in \mathcal{N}\argd{\mathscr{S}, 2} \bigcap \bb{\mat{M} - \mathcal{K}'} \setminus \cc{\mat{0}}$, we have $\vec{u} \in \mathcal{C}\argd{\mat{X}}$ and $\vec{v} \in \mathcal{R}\argd{\mat{X}}$ and by assumption, we have $\sigma_1\argd{\mat{X}} = \sigma_2\argd{\mat{X}}$.
		Let $\mat{X} = \sigma_{\ast} \vec{u}_1 \tpose{\vec{v}_1} + \sigma_{\ast} \vec{u}_2 \tpose{\vec{v}_2}$ and $\vec{u} = \alpha_1 \vec{u}_1 + \alpha_2 \vec{u}_2$, $\vec{v} = \alpha_3 \vec{v}_1 + \alpha_4 \vec{v}_2$ for some $\alpha_1, \alpha_2, \alpha_3, \alpha_4 \in \setR$ with $\alpha_1^2 + \alpha_2^2 = \alpha_3^2 + \alpha_4^2 = 1$.
		It is easy to check that $\mat{X}$ has the following equivalent singular value decompositions,
		\begin{equation}
			\makeatletter
				\if@twocolumn
					\begin{split}
						\mat{X}	& = \sigma_{\ast} \vec{u}_1 \tpose{\vec{v}_1} + \sigma_{\ast} \vec{u}_2 \tpose{\vec{v}_2} \\
						& = \sigma_{\ast}\bb{\alpha_1 \vec{u}_1 + \alpha_2 \vec{u}_2} \tpose{\bb{\alpha_1 \vec{v}_1 + \alpha_2 \vec{v}_2}} \\
						& \quad {} + \sigma_{\ast} \bb{\alpha_2 \vec{u}_1 - \alpha_1 \vec{u}_2} \tpose{\bb{\alpha_2 \vec{v}_1 - \alpha_1 \vec{v}_2}}.
					\end{split}
				\else
					\mat{X}	= \sigma_{\ast} \vec{u}_1 \tpose{\vec{v}_1} + \sigma_{\ast} \vec{u}_2 \tpose{\vec{v}_2}
					= \sigma_{\ast}\bb{\alpha_1 \vec{u}_1 + \alpha_2 \vec{u}_2} \tpose{\bb{\alpha_1 \vec{v}_1 + \alpha_2 \vec{v}_2}} + \sigma_{\ast} \bb{\alpha_2 \vec{u}_1 - \alpha_1 \vec{u}_2} \tpose{\bb{\alpha_2 \vec{v}_1 - \alpha_1 \vec{v}_2}}.
				\fi
			\makeatother
			\label{eqn:alt_svd}
		\end{equation}
		Using the representations for $\vec{u}$ and $\vec{v}$, we have,
		\begin{equation}
			\makeatletter
				\if@twocolumn
					\begin{split}
						\mat{M} - \mat{X}	& = - \sigma_{\ast} \bb{\alpha_2 \vec{u}_1 - \alpha_1 \vec{u}_2} \tpose{\bb{\alpha_2 \vec{v}_1 - \alpha_1 \vec{v}_2}} \\
						& \quad {} + \bb{\alpha_1 \vec{u}_1 + \alpha_2 \vec{u}_2} \Bd{\bb{\sigma \alpha_3 - \sigma_{\ast} \alpha_1} \vec{v}_1} \\
						& \quad \hphantom{\bb{\alpha_1 \vec{u}_1 + \alpha_2 \vec{u}_2}} \quad + \tpose{\dB{\bb{\sigma \alpha_4 - \sigma_{\ast} \alpha_2} \vec{v}_2}}.
					\end{split}
				\else
					\mat{M} - \mat{X}	= -\sigma_{\ast} \bb{\alpha_2 \vec{u}_1 - \alpha_1 \vec{u}_2} \tpose{\bb{\alpha_2 \vec{v}_1 - \alpha_1 \vec{v}_2}} + \bb{\alpha_1 \vec{u}_1 + \alpha_2 \vec{u}_2}
					\tpose{\BB{\bb{\sigma \alpha_3 - \sigma_{\ast} \alpha_1} \vec{v}_1 + \bb{\sigma \alpha_4 - \sigma_{\ast} \alpha_2} \vec{v}_2}}.
				\fi
			\makeatother
			\label{eqn:mstar}
		\end{equation}
		As the column vectors $\vec{u} = \alpha_1 \vec{u}_1 + \alpha_2 \vec{u}_2$ and $\vec{u}' = \alpha_2 \vec{u}_1 - \alpha_1 \vec{u}_2$ on the right hand side of \eqref{eqn:mstar} are linearly independent, $\rank{\mat{M} - \mat{X}} = 1$ is possible if and only if every column of $\mat{M} - \mat{X}$ combines $\vec{u}$ and $\vec{u}'$ in the same ratio.
		This means that the row vectors on the r.h.s.~of \eqref{eqn:mstar} are scalar multiples of each other.
		Thus, for $\rank{\mat{M} - \mat{X}} = 1$ it is necessary that
		\begin{subequations}
			\begin{align}
				\frac{\sigma \alpha_3 - \sigma_{\ast} \alpha_1}{\alpha_2}	& = \frac{\sigma \alpha_4 - \sigma_{\ast} \alpha_2}{-\alpha_1}	\label{eqn:ratio_1} \\
				\shortintertext{or equivalently,}
				\sigma \alpha_1 \alpha_3 + \sigma \alpha_2 \alpha_4	& = \sigma_{\ast} \alpha_1^2 + \sigma_{\ast} \alpha_2^2 = \sigma_{\ast}	\label{eqn:ratio_2} \\
				\shortintertext{which is not possible unless,}
				\alpha_1 \alpha_3 + \alpha_2 \alpha_4	& = \frac{\sigma_{\ast}}{\sigma} > 0.	\label{eqn:ratio_3}
			\end{align}
		\end{subequations}
		So, $\alpha_1 \alpha_3 + \alpha_2 \alpha_4 \leq 0$ implies that $\rank{\mat{M} - \mat{X}} > 1$.
		As $\mat{X} \in \mathcal{N}\argd{\mathscr{S}, 2} \bigcap \bb{\mat{M} - \mathcal{K}'} \setminus \cc{\mat{0}}$ is arbitrary, $\mat{M}$ is identifiable by \problemname~\eqref{prob:rank}.
		
		Next we prove the ``only if'' part.
		Let $\mat{M}$ be identifiable and $\mat{X} \in \mathcal{N}\argd{\mathscr{S}, 2} \bigcap \bb{\mat{M} - \mathcal{K}'} \setminus \cc{\mat{0}}$ so that $\mat{M} - \mat{X}$ is feasible for \problemname~\eqref{prob:rank}.
		As before, we have $\vec{u} \in \mathcal{C}\argd{\mat{X}}$, $\vec{v} \in \mathcal{R}\argd{\mat{X}}$ and $\sigma_1\argd{\mat{X}} = \sigma_2\argd{\mat{X}}$.
		If $\mat{X} = \sigma_{\ast} \vec{u}_1 \tpose{\vec{v}_1} + \sigma_{\ast} \vec{u}_2 \tpose{\vec{v}_2}$, then $\vec{u} = \alpha_1 \vec{u}_1 + \alpha_2 \vec{u}_2$, $\vec{v} = \alpha_3 \vec{v}_1 + \alpha_4 \vec{v}_2$ for some $\alpha_1, \alpha_2, \alpha_3, \alpha_4 \in \setR$ with $\alpha_1^2 + \alpha_2^2 = \alpha_3^2 + \alpha_4^2 = 1$.
		It is simple to check that \eqref{eqn:alt_svd} and \eqref{eqn:mstar} are valid.
		We shall now assume $\epsilon = \alpha_1 \alpha_3 + \alpha_2 \alpha_4 > 0$ and arrive at a contradiction.
		Since multiplying a matrix by a nonzero scalar does not change its row or column space and scales every nonzero singular value in the same ratio, we can take $\sigma_{\ast} = \sigma \epsilon$ without violating any assumptions on $\mat{X}$.
		Thus, $\alpha_1 \alpha_3 + \alpha_2 \alpha_4 = \sigma_{\ast}/\sigma$ and we have $\eqref{eqn:ratio_3} \implies \eqref{eqn:ratio_2} \implies \eqref{eqn:ratio_1} \implies \rank{\mat{M} - \mat{X}} = 1$ (the last implication is due to \eqref{eqn:mstar}) thus contradicting the identifiability of $\mat{M}$.

	\section{Proof of \lemmaname~\ref{lem:Markov estimate}}
		\label{sec:Markov estimate lemma proof}
		Using assumption~\aref{itm:cov id}, we have
		\begin{equation}
			\makeatletter
				\if@twocolumn
					\begin{split}
						\expect{\twonorm{\vec{x}}^{2}}
						& = \expect{\tpose{\vec{x}} \vec{x}} = \expect{\trace{\vec{x} \tpose{\vec{x}}}} \\
						& = \trace{\expect{\vec{x} \tpose{\vec{x}}}} = \trace{\eye} = m.
					\end{split}
				\else
					\expect{\twonorm{\vec{x}}^{2}} = \expect{\tpose{\vec{x}} \vec{x}}
					= \expect{\trace{\vec{x} \tpose{\vec{x}}}} = \trace{\expect{\vec{x} \tpose{\vec{x}}}}
					= \trace{\eye} = m.
				\fi
			\makeatother
			\label{eqn:expect norm squared}
		\end{equation}
		Hence,
		\begin{subequations}
			\begin{align}
				\SwapAboveDisplaySkip
				\expect{\twonorm{\mat{P}_{\mathcal{C}\bb{\mat{X}}} \vec{u}}^{2}}
				& = \frac{1}{m} \expect{\twonorm{\vec{x}}^{2}} \expect{\twonorm{\mat{P}_{\mathcal{C}\bb{\mat{X}}} \vec{u}}^{2}}	\label{eqn:using expect norm squared} \\
				& = \frac{1}{m} \expect{\twonorm{\vec{x}}^{2} \twonorm{\mat{P}_{\mathcal{C}\bb{\mat{X}}} \vec{u}}^{2}}	\label{eqn:using independence} \\
				& = \frac{1}{m} \expect{\twonorm{\mat{P}_{\mathcal{C}\bb{\mat{X}}} \vec{x}}^{2}}	\label{eqn:relating u and x} \\
				& = \frac{1}{m} \expect{\tpose{\vec{x}} \mat{P}_{\mathcal{C}\bb{\mat{X}}} \vec{x}}	\label{eqn:square of projection matrix} \\
				& = \frac{1}{m} \expect{\trace{\mat{P}_{\mathcal{C}\bb{\mat{X}}} \vec{x} \tpose{\vec{x}}}}	\notag \\
				& = \frac{1}{m} \trace{\mat{P}_{\mathcal{C}\bb{\mat{X}}} \expect{\vec{x} \tpose{\vec{x}}}}	\label{eqn:E commutes with Tr and P} \\
				& = \frac{1}{m} \trace{\mat{P}_{\mathcal{C}\bb{\mat{X}}} \eye}	\label{eqn:using cov id} \\
				& \leq \frac{2}{m}	\label{eqn:using rank-2 X}
			\end{align}
			\label{eqn:square norm projected estimate}%
		\end{subequations}
		where \eqref{eqn:using expect norm squared} follows from \eqref{eqn:expect norm squared}, \eqref{eqn:using independence} and \eqref{eqn:relating u and x} are true because $\vec{u} = \vec{x}/\twonorm{\vec{x}}$ and assumption~\aref{itm:factor marginal} implies independence of $\twonorm{\vec{x}}$ and $\vec{u}$, \eqref{eqn:square of projection matrix} is true since $\mat{P}^{2} = \mat{P}$ for any projection matrix $\mat{P}$, \eqref{eqn:E commutes with Tr and P} is true since expectation operator commutes with trace and projection operators, \eqref{eqn:using cov id} follows from assumption~\aref{itm:cov id} and, \eqref{eqn:using rank-2 X} is true since $\mat{X} \in \mathcal{N}\argd{\mathscr{S}, 2} \bigcap \mathcal{M} \setminus \cc{\mat{0}}$ is a matrix of rank at most two.

		Finally, applying Markov inequality to the non-negative random variable $\twonorm{\mat{P}_{\mathcal{C}\bb{\mat{X}}} \vec{u}}^{2}$ and using the computed estimate of $\expect{\twonorm{\mat{P}_{\mathcal{C}\bb{\mat{X}}} \vec{u}}^{2}}$ from \eqref{eqn:square norm projected estimate} gives
		\begin{equation}
			\makeatletter
				\if@twocolumn
					\begin{split}
						\Pr\bb{\twonorm{\mat{P}_{\mathcal{C}\bb{\mat{X}}} \vec{u}}^{2} \geq 1 - \delta}
						& \leq \frac{\expect{\twonorm{\mat{P}_{\mathcal{C}\bb{\mat{X}}} \vec{u}}^{2}}}{1 - \delta} \\
						& = \frac{2}{m \bb{1 - \delta}}.
					\end{split}
				\else
					\Pr\bb{\twonorm{\mat{P}_{\mathcal{C}\bb{\mat{X}}} \vec{u}}^{2} \geq 1 - \delta}
					\leq \frac{\expect{\twonorm{\mat{P}_{\mathcal{C}\bb{\mat{X}}} \vec{u}}^{2}}}{1 - \delta}
					= \frac{2}{m \bb{1 - \delta}}.
				\fi
			\makeatother
		\end{equation}
		We have thus established \eqref{eqn:Markov 1}.
		Using the exact same sequence of steps for the random vector $\vec{v}$ gives the bound in~\eqref{eqn:Markov 2}.

	\section{Proof of \lemmaname~\ref{lem:Markov estimate special}}
		\label{sec:Markov estimate special lemma proof}
		Notice that \eqref{eqn:square of projection matrix}-\eqref{eqn:using rank-2 X} in the proof of \lemmaname~\ref{lem:Markov estimate} in \appendixname~\ref{sec:Markov estimate lemma proof} does not use assumption~\aref{itm:factor marginal}.
		Hence, reusing the same arguments we get
		\begin{equation}
			\expect{\twonorm{\mat{P}_{\mathcal{C}\bb{\mat{X}}} \vec{x}}^{2}} = \expect{\tpose{\vec{x}} \mat{P}_{\mathcal{C}\bb{\mat{X}}} \vec{x}} \leq 2.
			\label{eqn:expect projected norm square}
		\end{equation}
		Thus, we have
		\begin{subequations}
			\makeatletter
				\if@twocolumn
					\begin{align}
						\Pr & \bb{\twonorm{\mat{P}_{\mathcal{C}\bb{\mat{X}}} \vec{u}}^{2} \geq 1 - \delta}	\notag \\
						& = \Pr\bb{\twonorm{\mat{P}_{\mathcal{C}\bb{\mat{X}}} \vec{x}}^{2} \geq \bb{1 - \delta} \twonorm{\vec{x}}^{2}}	\label{eqn:relate u and x} \\
						& \leq \Pr\bb{\twonorm{\mat{P}_{\mathcal{C}\bb{\mat{X}}} \vec{x}}^{2} \geq \bb{1 - \delta} r_{\vec{x}}^{2}}	\label{eqn:bound norm x} \\
						& \leq \frac{\expect{\twonorm{\mat{P}_{\mathcal{C}\bb{\mat{X}}} \vec{x}}^{2}}}{r_{\vec{x}}^{2} \bb{1 - \delta}}	\label{eqn:Markov inequality} \\
						& \leq \frac{2}{r_{\vec{x}}^{2} \bb{1 - \delta}}	\label{eqn:bound norm projected x}
					\end{align}
				\else
					\begin{align}
						\SwapAboveDisplaySkip
						\Pr\bb{\twonorm{\mat{P}_{\mathcal{C}\bb{\mat{X}}} \vec{u}}^{2} \geq 1 - \delta}
						& = \Pr\bb{\twonorm{\mat{P}_{\mathcal{C}\bb{\mat{X}}} \vec{x}}^{2} \geq \bb{1 - \delta} \twonorm{\vec{x}}^{2}}	\label{eqn:relate u and x} \\
						& \leq \Pr\bb{\twonorm{\mat{P}_{\mathcal{C}\bb{\mat{X}}} \vec{x}}^{2} \geq \bb{1 - \delta} r_{\vec{x}}^{2}}	\label{eqn:bound norm x} \\
						& \leq \frac{\expect{\twonorm{\mat{P}_{\mathcal{C}\bb{\mat{X}}} \vec{x}}^{2}}}{r_{\vec{x}}^{2} \bb{1 - \delta}}	\label{eqn:Markov inequality} \\
						& \leq \frac{2}{r_{\vec{x}}^{2} \bb{1 - \delta}}	\label{eqn:bound norm projected x}
					\end{align}
				\fi
			\makeatother
			\label{eqn:derivation}%
		\end{subequations}
		where \eqref{eqn:relate u and x} is true since $\vec{u} = \vec{x}/\twonorm{\vec{x}}$, \eqref{eqn:bound norm x} holds because of assumption~\aref{itm:abs lower bound}, \eqref{eqn:Markov inequality} follows from applying Markov inequality to the non-negative random variable $\twonorm{\mat{P}_{\mathcal{C}\bb{\mat{X}}} \vec{x}}^{2}$ and, \eqref{eqn:bound norm projected x} follows from \eqref{eqn:expect projected norm square}.
		Thus, the derivation \eqref{eqn:derivation} establishes \eqref{eqn:Markov 1 special}.
		Using the same sequence of steps for the random vector $\vec{v}$ gives the bound in \eqref{eqn:Markov 2 special}.

	\section{Proof of \theoremname~\ref{thm:whp_suff_ident}}
		\label{sec:whp_suff_ident theorem proof}
		For any constant $\delta \in \bb{0, 1}$, let $\mathcal{A}\argd{\delta}$ denote the event that $\exists \mat{X} \in \mathcal{N}\argd{\mathscr{S}, 2} \bigcap \mathcal{M} \setminus \cc{\mat{0}}$ satisfying both $\twonorm{\vec{u} - \mat{P}_{\mathcal{C}\bb{\mat{X}}} \vec{u}}^{2} \leq \delta$ and $\twonorm{\vec{v} - \mat{P}_{\mathcal{R}\bb{\mat{X}}} \vec{v}}^{2} \leq \delta$.
		We note that $\mathcal{A}\argd{\delta}$ constitutes a non-decreasing sequence of sets as $\delta$ increases.
		Hence, using continuity of the probability measure from above we have,
		\begin{subequations}
			\begin{align}
				\Pr\argd{\mathcal{A}\argd{0}}	& \leq \Pr\argd{\mathcal{A}\argd{\delta}}	\label{eqn:inequality} \\
				\shortintertext{for any $\delta \in \bb{0,1}$, and}
				\Pr\argd{\mathcal{A}\argd{0}}	& = \lim_{\delta \to 0} \Pr\argd{\mathcal{A}\argd{\delta}}.	\label{eqn:limit equality}
			\end{align}
		\end{subequations}
		Note that $\mathcal{A}\argd{0}$ denotes the event that $\exists \mat{X} \in \mathcal{N}\argd{\mathscr{S}, 2} \bigcap \mathcal{M} \setminus \cc{\mat{0}}$ satisfying both $\vec{u} \in \mathcal{C}\bb{\mat{X}}$ and $\vec{v} \in \mathcal{R}\bb{\mat{X}}$ which is a ``hard'' event.
		The event $\mathcal{A}\argd{0}^{\comp}$ corresponds precisely to the sufficient conditions of \theoremname~\ref{thm:suff_ident}.
		Hence, it is sufficient to obtain an appropriate lower bound for $\Pr\argd{\mathcal{A}\argd{0}^{\comp}}$ to make our desired statement.
		Drawing inspiration from \eqref{eqn:inequality} and \eqref{eqn:limit equality}, we shall upper bound $\Pr\argd{\mathcal{A}\argd{0}}$ by $\Pr\argd{\mathcal{A}\argd{\delta}}$.
		
		For any given $\mat{X} \in \mathcal{N}\argd{\mathscr{S}, 2} \bigcap \mathcal{M} \setminus \cc{\mat{0}}$ we have,
		\begin{subequations}
			\begin{align}
				\MoveEqLeft[1] \Pr\argd{\twonorm{\vec{u} - \mat{P}_{\mathcal{C}\bb{\mat{X}}} \vec{u}}^{2} \leq \delta, \twonorm{\vec{v} - \mat{P}_{\mathcal{R}\bb{\mat{X}}} \vec{v}}^{2} \leq \delta}	\notag \\
				& = \Pr\argd{\twonorm{\vec{u}}^{2} - \twonorm{\mat{P}_{\mathcal{C}\bb{\mat{X}}} \vec{u}}^{2} \leq \delta, \twonorm{\vec{v}}^{2} - \twonorm{\mat{P}_{\mathcal{R}\bb{\mat{X}}} \vec{v}}^{2} \leq \delta} \label{eqn:orthogonal decomposition} \\
				& = \Pr\argd{\twonorm{\mat{P}_{\mathcal{C}\bb{\mat{X}}} \vec{u}}^{2} \geq 1 - \delta, \twonorm{\mat{P}_{\mathcal{R}\bb{\mat{X}}} \vec{v}}^{2} \geq 1 - \delta}	\label{eqn:unit norm vectors} \\
				& = \Pr\argd{\twonorm{\mat{P}_{\mathcal{C}\bb{\mat{X}}} \vec{u}}^{2} \geq 1 - \delta} \Pr\argd{\twonorm{\mat{P}_{\mathcal{R}\bb{\mat{X}}} \vec{v}}^{2} \geq 1 - \delta}	\label{eqn:by independence} \\
				& \leq \frac{4}{mn \bb{1 - \delta}^{2}}	\label{eqn:from lemma}
			\end{align}
			\label{eqn:multiplicative estimate}%
		\end{subequations}
		where \eqref{eqn:orthogonal decomposition} is true because $\eye - \mat{P}_{\mathcal{C}\bb{\mat{X}}}$ (respectively $\eye - \mat{P}_{\mathcal{R}\bb{\mat{X}}}$) is the orthogonal projection matrix onto the orthogonal complement space of $\mathcal{C}\argd{\mat{X}}$ (respectively $\mathcal{R}\argd{\mat{X}}$), \eqref{eqn:unit norm vectors} is true because we have $\twonorm{\vec{u}} = \twonorm{\vec{v}} = 1$, \eqref{eqn:by independence} is true by independence of $\vec{u}$ and $\vec{v}$, and \eqref{eqn:from lemma} comes from applying \lemmaname~\ref{lem:Markov estimate}.
		
		Next we employ union bounding over all $\mat{X} \in \mathcal{N}\argd{\mathscr{S}, 2} \bigcap \mathcal{M} \setminus \cc{\mat{0}}$ representing distinct pairs of column and row subspaces $\bb{\mathcal{C}\argd{\mat{X}}, \mathcal{R}\argd{\mat{X}}}$ to upper bound $\Pr\argd{\mathcal{A}\argd{\delta}}$.
		We denote the number of these distinct pairs of $\bb{\mathcal{C}\argd{\mat{X}}, \mathcal{R}\argd{\mat{X}}}$ over $\mat{X} \in \mathcal{N}\argd{\mathscr{S}, 2} \bigcap \mathcal{M} \setminus \cc{\mat{0}}$ by $f_{\mathscr{S}, \mathcal{M}}\argd{m,n}$.
		
		Finally, using \eqref{eqn:multiplicative estimate} we have
		\begin{subequations}
			\makeatletter
				\if@twocolumn
					\begin{align}
						\MoveEqLeft[1] \Pr\argd{\mathcal{A}\argd{\delta}}	\notag \\
						& \leq \sum_{\bb{\mathcal{C}\argd{\mat{X}}, \mathcal{R}\argd{\mat{X}}}} \Pr\argd{\twonorm{\mat{P}_{\mathcal{C}\bb{\mat{X}}^{\perp}} \vec{u}}^{2} \leq \delta, \twonorm{\mat{P}_{\mathcal{R}\bb{\mat{X}}^{\perp}} \vec{v}}^{2} \leq \delta}	\label{eqn:union bound finite} \\
						& = f_{\mathscr{S}, \mathcal{M}}\argd{m,n} \Pr\argd{\twonorm{\mat{P}_{\mathcal{C}\bb{\mat{X}}^{\perp}} \vec{u}}^{2} \leq \delta, \twonorm{\mat{P}_{\mathcal{R}\bb{\mat{X}}^{\perp}} \vec{v}}^{2} \leq \delta}	\notag \\
						& \leq \frac{4 f_{\mathscr{S}, \mathcal{M}}\argd{m,n}}{mn \bb{1 - \delta}^{2}}	\label{eqn:lemma 1 result}
					\end{align}
				\else
					\begin{align}
						\Pr\argd{\mathcal{A}\argd{\delta}}
						& \leq \sum_{\bb{\mathcal{C}\bb{\mat{X}}, \mathcal{R}\bb{\mat{X}}}} \Pr\argd{\twonorm{\mat{P}_{\mathcal{C}\bb{\mat{X}}^{\perp}} \vec{u}}^{2} \leq \delta, \twonorm{\mat{P}_{\mathcal{R}\bb{\mat{X}}^{\perp}} \vec{v}}^{2} \leq \delta}	\label{eqn:union bound finite} \\
						& = f_{\mathscr{S}, \mathcal{M}}\argd{m,n} \Pr\argd{\twonorm{\mat{P}_{\mathcal{C}\bb{\mat{X}}^{\perp}} \vec{u}}^{2} \leq \delta, \twonorm{\mat{P}_{\mathcal{R}\bb{\mat{X}}^{\perp}} \vec{v}}^{2} \leq \delta}	\notag \\
						& \leq \frac{4 f_{\mathscr{S}, \mathcal{M}}\argd{m,n}}{mn \bb{1 - \delta}^{2}}	\label{eqn:lemma 1 result}
					\end{align}
				\fi
			\makeatother
			\label{eqn:union bound}%
		\end{subequations}
		where \eqref{eqn:union bound finite} is an union bounding step.
		Hence,
		\begin{subequations}
			\begin{align}
				\Pr\argd{\mathcal{A}\argd{0}^{\comp}}
				& = 1 - \Pr\argd{\mathcal{A}\argd{0}}	\notag \\
				& \geq 1 - \Pr\argd{\mathcal{A}\argd{\delta}}	\label{eqn:relaxation} \\
				& \geq 1 - \frac{4 f_{\mathscr{S}, \mathcal{M}}\argd{m,n}}{mn \bb{1 - \delta}^{2}}	\label{eqn:relaxation value} \\
				& \geq 1 - \frac{4 f_{\mathscr{S}, \mathcal{M}}\argd{m,n}}{mn \bb{1 - \delta'}}
			\end{align}
		\end{subequations}
		where \eqref{eqn:relaxation} is from \eqref{eqn:inequality}, \eqref{eqn:relaxation value} is from \eqref{eqn:union bound} and $\delta' = 1 - \bb{1 - \delta}^{2} \in \bb{0, 1}$.

	\section{Proof of \corollaryname~\ref{cor:whp_suff_ident}}
		\label{sec:whp_suff_ident corollary proof}
		The proof is essentially to that of \theoremname~\ref{thm:whp_suff_ident} in \appendixname~\ref{sec:whp_suff_ident theorem proof} with one important difference: we use \lemmaname~\ref{lem:Markov estimate special} instead of \lemmaname~\ref{lem:Markov estimate} when bounding the right hand side of \eqref{eqn:by independence}.
		This gives us the bound
		\makeatletter
			\if@twocolumn
				\begin{multline}
					\Pr\argd{\twonorm{\vec{u} - \mat{P}_{\mathcal{C}\bb{\mat{X}}} \vec{u}}^{2} \leq \delta, \twonorm{\vec{v} - \mat{P}_{\mathcal{R}\bb{\mat{X}}} \vec{v}}^{2} \leq \delta}	\\
					\leq \frac{4}{r_{\vec{x}}^{2}\argd{m} r_{\vec{y}}^{2}\argd{n} \bb{1 - \delta}^{2}}
				\end{multline}
			\else
				\begin{equation}
					\Pr\argd{\twonorm{\vec{u} - \mat{P}_{\mathcal{C}\bb{\mat{X}}} \vec{u}}^{2} \leq \delta, \twonorm{\vec{v} - \mat{P}_{\mathcal{R}\bb{\mat{X}}} \vec{v}}^{2} \leq \delta}
					\leq \frac{4}{r_{\vec{x}}^{2}\argd{m} r_{\vec{y}}^{2}\argd{n} \bb{1 - \delta}^{2}}
				\end{equation}
			\fi
		\makeatother
		which leads to the bound
		\begin{equation}
			\Pr\argd{\mathcal{A}\argd{\delta}} \leq \frac{4 f_{\mathscr{S}, \mathcal{M}}\argd{m,n}}{r_{\vec{x}}^{2}\argd{m} r_{\vec{y}}^{2}\argd{n} \bb{1 - \delta}^{2}}
			\label{eqn:union bound special}
		\end{equation}
		in place of \eqref{eqn:lemma 1 result}.
		Finally,
		\begin{subequations}
			\begin{align}
				\SwapAboveDisplaySkip
				\Pr\argd{\mathcal{A}\argd{0}^{\comp}}
				& = 1 - \Pr\argd{\mathcal{A}\argd{0}}	\notag \\
				& \geq 1 - \Pr\argd{\mathcal{A}\argd{\delta}}	\label{eqn:relaxation special} \\
				& \geq 1 - \frac{4 f_{\mathscr{S}, \mathcal{M}}\argd{m,n}}{r_{\vec{x}}^{2}\argd{m} r_{\vec{y}}^{2}\argd{n} \bb{1 - \delta}^{2}}	\label{eqn:relaxation value special} \\
				& \geq 1 - \frac{4 f_{\mathscr{S}, \mathcal{M}}\argd{m,n}}{r_{\vec{x}}^{2}\argd{m} r_{\vec{y}}^{2}\argd{n} \bb{1 - \delta'}}
			\end{align}
		\end{subequations}
		where \eqref{eqn:relaxation special} is from \eqref{eqn:inequality}, \eqref{eqn:relaxation value special} is from \eqref{eqn:union bound special} and $\delta' = 1 - \bb{1 - \delta}^{2} \in \bb{0, 1}$.

	\section{Proof of \lemmaname~\ref{lem:Gaussian Chernoff estimate}}
		\label{sec:Gaussian Chernoff estimate proof}
		This is a Chernoff-type bound.
		We set
		\begin{equation}
			\makeatletter
				\if@twocolumn
					\begin{split}
						Y	& = \twonorm{\mat{P}_{\mathcal{C}\bb{\mat{X}}} \vec{x}}^{2} - \bb{1 - \delta} \twonorm{\vec{x}}^{2}	\\
							& = \delta \twonorm{\mat{P}_{\mathcal{C}\bb{\mat{X}}} \vec{x}}^{2} - \bb{1 - \delta} \twonorm{\mat{P}_{\mathcal{C}\bb{\mat{X}}^{\perp}} \vec{x}}^{2}
					\end{split}
				\else
					Y = \twonorm{\mat{P}_{\mathcal{C}\bb{\mat{X}}} \vec{x}}^{2} - \bb{1 - \delta} \twonorm{\vec{x}}^{2}
					= \delta \twonorm{\mat{P}_{\mathcal{C}\bb{\mat{X}}} \vec{x}}^{2} - \bb{1 - \delta} \twonorm{\mat{P}_{\mathcal{C}\bb{\mat{X}}^{\perp}} \vec{x}}^{2}
				\fi
			\makeatother
			\label{eqn:ortho split}
		\end{equation}
		and compute the bound
		\begin{equation}
			\Pr\bb{Y \geq 0} \leq \expect{\exp\argd{tY}}
			\label{eqn:chernoff}
		\end{equation}
		that holds for all values of the parameter $t$ for which the right hand side of \eqref{eqn:chernoff} exists.
		Using properties of Gaussian random vectors under linear transforms, we have $\mat{P}_{\mathcal{C}\bb{\mat{X}}} \vec{x}$ and $\mat{P}_{\mathcal{C}\bb{\mat{X}}^{\perp}} \vec{x}$ as statistically independent Gaussian random vectors implying
		\makeatletter
			\if@twocolumn
				\begin{multline}
					\expect{\exp\argd{t\delta \twonorm{\mat{P}_{\mathcal{C}\bb{\mat{X}}} \vec{x}}^{2} - t\bb{1 - \delta} \twonorm{\mat{P}_{\mathcal{C}\bb{\mat{X}}^{\perp}} \vec{x}}^{2}}}	\\
					= \expect{\exp\argd{t\delta Z_{1}}} \cdot \expect{\exp\argd{- t\bb{1 - \delta} Z_{2}}}.
					\label{eqn:independent split}
				\end{multline}
			\else
				\begin{equation}
					\expect{\exp\argd{t\delta \twonorm{\mat{P}_{\mathcal{C}\bb{\mat{X}}} \vec{x}}^{2} - t\bb{1 - \delta} \twonorm{\mat{P}_{\mathcal{C}\bb{\mat{X}}^{\perp}} \vec{x}}^{2}}}
					= \expect{\exp\argd{t\delta Z_{1}}} \cdot \expect{\exp\argd{- t\bb{1 - \delta} Z_{2}}}.
					\label{eqn:independent split}
				\end{equation}
			\fi
		\makeatother
		with,
		\begin{equation}
			Z_{1} = \twonorm{\mat{P}_{\mathcal{C}\bb{\mat{X}}} \vec{x}}^{2} \quad \text{and} \quad
			Z_{2} = \twonorm{\mat{P}_{\mathcal{C}\bb{\mat{X}}^{\perp}} \vec{x}}^{2}.
		\end{equation}
		Since $\mathcal{N}\argd{\mathscr{S}, 1} \bigcap \mathcal{M} = \cc{\mat{0}}$, both $\mathcal{C}\argd{\mat{X}}$ and $\mathcal{R}\argd{\mat{X}}$ are two dimensional spaces.
		On rotating coordinates to the basis given by $\cc{\mathcal{C}\argd{\mat{X}}, \mathcal{C}\argd{\mat{X}}^{\perp}}$, it can be seen that $Z_{1}$ is the sum of squares of two \iid~standard Gaussian random variables and hence has a $\chi^{2}$ distribution with two DoF.
		By the same argument, $Z_{2}$ is a $\chi^{2}$ distributed random variable with $(m - 2)$ DoF.
		Recall that the moment generating function of a $\chi^{2}$ distributed random variable $Z$ with $k$ DoF is given by
		\begin{equation}
			\expect{\exp\bb{tZ}} = \bb{1-2t}^{-k/2}, \quad \forall t < 1/2.
			\label{eqn:chi square mgf}
		\end{equation}
		Using \eqref{eqn:ortho split}, \eqref{eqn:chernoff}, \eqref{eqn:independent split} and \eqref{eqn:chi square mgf} we have the bound
		\makeatletter
			\if@twocolumn
				\begin{multline}
					\Pr\bb{Y \geq 0} \leq \bb{1-2t\delta}^{-1} \bb{1-2t\bb{1-\delta}}^{-\bb{m-2}/2} \\
					= \exp\BB{-\bb{\frac{m - 2}{2}} \log \bb{1-2t\bb{1-\delta}} - \log \bb{1-2t\delta}}
					\label{eqn:parametrized exponent}
				\end{multline}
			\else
				\begin{equation}
					\begin{split}
						\Pr\bb{Y \geq 0}	& \leq \bb{1-2t\delta}^{-1} \bb{1-2t\bb{1-\delta}}^{-\bb{m-2}/2}	\\
						& = \exp\BB{-\bb{\frac{m - 2}{2}} \log \bb{1-2t\bb{1-\delta}} - \log \bb{1-2t\delta}}
					\end{split}
					\label{eqn:parametrized exponent}
				\end{equation}
			\fi
		\makeatother
		which can be optimized over $t$.
		It can be verified by differentiation that the best bound is obtained for
		\begin{equation}
			t^{\ast} = \frac{m-2}{2 \delta m} - \frac{1}{m \bb{1-\delta}}.
		\end{equation}
		Plugging this value of $t$ into \eqref{eqn:parametrized exponent} and using \eqref{eqn:ortho split} we get the desired result.

	\section{Proof of \lemmaname~\ref{lem:Bernoulli Chernoff estimate}}
		\label{sec:Bernoulli Chernoff estimate proof}
		This is also a Chernoff-type bound.
		Although the final results of \lemmasname~\ref{lem:Gaussian Chernoff estimate} and~\ref{lem:Bernoulli Chernoff estimate} look quite similar, we cannot reuse the manipulations in \appendixname~\ref{sec:Gaussian Chernoff estimate proof} for this proof and proceed by a slightly different route (also applicable to other subgaussian distributions) since the symmetric Bernoulli distribution does not share the rotational invariance property of the multivariate standard normal distribution.
		Let $\cc{\vec{c}_{1}, \vec{c}_{2}}$ denote an orthonormal basis for $\mathcal{C}\argd{\mat{X}}$ and set
		\begin{equation}
			Y = \twonorm{\mat{P}_{\mathcal{C}\bb{\mat{X}}} \vec{x}}^{2} - \bb{1 - \delta} \twonorm{\vec{x}}^{2}.
		\end{equation}
		Notice that $\twonorm{\vec{x}} = \sqrt{m}$, so we have
		\begin{subequations}
			\begin{align}
				\Pr\bb{Y \geq 0}	& = \Pr\bb{\twonorm{\mat{P}_{\mathcal{C}\bb{\mat{X}}} \vec{x}}^{2} \geq m \bb{1 - \delta}}	\notag \\
				& = \Pr\bb{\abs{\ip{\vec{c}_{1}}{\vec{x}}}^{2} + \abs{\ip{\vec{c}_{2}}{\vec{x}}}^{2} \geq m \bb{1 - \delta}}	\notag \\
				& \leq \Pr\bb{\bigcup_{j = 1,2} \cc{\abs{\ip{\vec{c}_{j}}{\vec{x}}}^{2} \geq \frac{m}{2} \bb{1 - \delta}}}	\label{eqn:union bound 1} \\
				& \leq 2 \Pr\bb{\abs{\ip{\vec{c}}{\vec{x}}}^{2} \geq \frac{m}{2} \bb{1 - \delta}}	\label{eqn:union bound 2} \\
				& = 2 \Pr\bb{\abs{\ip{\vec{c}}{\vec{x}}} \geq \sqrt{\frac{m}{2} \bb{1 - \delta}}}	\notag \\
				& = 4 \Pr\bb{\ip{\vec{c}}{\vec{x}} \geq \sqrt{\frac{m}{2} \bb{1 - \delta}}}	\label{eqn:symmetric distribution} \\
				& \leq 4 \exp \BB{\frac{t^{2}}{2} - t \sqrt{\frac{m}{2} \bb{1 - \delta}}} \label{eqn:chernoff bound}
			\end{align}
		\end{subequations}
		where \eqref{eqn:union bound 1} and \eqref{eqn:union bound 2} utilize elementary union bounds, $\vec{c}$ is a generic unit vector, \eqref{eqn:symmetric distribution} uses the symmetry of the distribution of $\vec{x}$ about the origin, and \eqref{eqn:chernoff bound} is the Chernoff bounding step that utilizes the following computation:
		\begin{subequations}
			\begin{align}
				\expect{\exp\bb{t \ip{\vec{c}}{\vec{x}}}} & = \expect{\exp\bb{\sum_{j = 1}^{m} t x_{j} c_{j}}}	\notag \\
				& = \expect{\prod_{j = 1}^{m} \exp\bb{t x_{j} c_{j}}}	\notag \\
				& = \prod_{j = 1}^{m} \expect{\exp\bb{t x_{j} c_{j}}}	\label{eqn:independence} \\
				& = \prod_{j = 1}^{m} \frac{e^{t c_{j}} + e^{-t c_{j}}}{2}	\label{eqn:symmetric Bernoulli} \\
				& = \prod_{j = 1}^{m} \sum_{k = 0}^{\infty} \frac{\bb{t c_{j}}^{2k}}{\bb{2k}!}	\label{eqn:series expansion} \\
				& < \prod_{j = 1}^{m} \sum_{k = 0}^{\infty} \frac{\bb{t c_{j}}^{2k}}{2^{k} k!}	\label{eqn:factorial bound} \\
				& = \prod_{j = 1}^{m} \exp\bb{t^{2} c_{j}^{2}/2}	\notag \\
				& = \exp \bb{\frac{t^{2}}{2} \sum_{j = 1}^{m} c_{j}^{2}}	\notag \\
				& = \exp \bb{\frac{t^{2}}{2}}	\label{eqn:unit norm}
			\end{align}
		\end{subequations}
		where \eqref{eqn:independence} uses independence of elements of $\vec{x}$, \eqref{eqn:symmetric Bernoulli} is true because each element of $\vec{x}$ has a symmetric Bernoulli distribution, \eqref{eqn:series expansion} uses the series expansion of the exponential function, \eqref{eqn:unit norm} follows from $\twonorm{\vec{c}} = 1$ and \eqref{eqn:factorial bound} is due to
		\begin{equation}
			\bb{2k}! = 2^{k} \prod_{r = 0}^{k-1} \bb{2r+1} > 2^{k} \prod_{r = 0}^{k-1} \bb{r+1} = 2^{k} k!.
		\end{equation}
		The bound in \eqref{eqn:chernoff bound} can be optimized over $t$ with the optimum being achieved at
		\begin{equation}
			t^{\ast} = \sqrt{\frac{m}{2} \bb{1 - \delta}}.
		\end{equation}
		Plugging this value of $t$ into \eqref{eqn:chernoff bound} gives the desired result.

	\section{Proof of \lemmaname~\ref{lem:metric entropy}}
		\label{sec:epsilon net proof}
		Consider the norm $\twoinfnorm{\cdot}$ on $\setR^{m \times 2}$ defined as
		\begin{equation}
			\twoinfnorm{\mat{Y}} = \max\cc{\twonorm{\vec{y}_{1}}, \twonorm{\vec{y}_{2}}}
		\end{equation}
		for all $\mat{Y} = \BB{\vec{y}_{1}, \vec{y}_{2}} \in \setR^{m \times 2}$.
		It is clear that $\mathcal{D}\argd{m}$ is the unit ball $\set{\mat{Y} \in \setR^{m \times 2}}{\twoinfnorm{\mat{Y}} \leq 1}$ of this norm, which is a convex body symmetric about the origin.
		Hence, using \eqref{eqn:covering number} we have the metric entropy of $\mathcal{D}\argd{m}$ \wrt~$\epsilon \mathcal{D}\argd{m}$ as $2m \log \Theta\argd{1/\epsilon}$.
		It is clear that $\mathcal{G}\argd{m} = \set{\mat{Y} \in \setR^{m \times 2}}{\tpose{\mat{Y}} \mat{Y} = \eye} \subsetneq \mathcal{D}\argd{m}$, implying that metric entropy of $\mathcal{G}\argd{m}$ \wrt~$\epsilon \mathcal{D}\argd{m}$ is $\leq 2m \log \Theta\argd{1/\epsilon}$.

		Let $\mat{Y} = \BB{\vec{y}_{1}, \vec{y}_{2}}$ and $\mat{Z} = \BB{\vec{z}_{1}, \vec{z}_{2}}$ be two elements from $\mathcal{G}\argd{m}$ such that $\mat{Y} - \mat{Z} \in \epsilon \mathcal{D}\argd{m}$, and let $\vec{x} \in \setR^{m}$ be arbitrary.
		Then,
		\begin{subequations}
			\begin{align}
				\SwapAboveDisplaySkip
				\twonorm{\mat{P}_{\mathcal{C}\bb{\mat{Y}}} \vec{x}}
				&	= \sqrt{\abs{\ip{\vec{y}_{1}}{\vec{x}}}^{2} + \abs{\ip{\vec{y}_{2}}{\vec{x}}}^{2}}	\notag \\
				&	= \sqrt{\sum_{j = 1,2} \abs{\ip{\vec{z}_{j}}{\vec{x}} + \ip{\vec{y}_{j} - \vec{z}_{j}}{\vec{x}}}^{2}}	\notag \\
				&	\leq \sqrt{\sum_{j = 1,2} \bb{\abs{\ip{\vec{z}_{j}}{\vec{x}}} + \abs{\ip{\vec{y}_{j} - \vec{z}_{j}}{\vec{x}}}}^{2}}	\label{eqn:abs val inside} \\
				&	\leq \sqrt{\sum_{j = 1,2} \bb{\abs{\ip{\vec{z}_{j}}{\vec{x}}} + \epsilon \twonorm{\vec{x}}}^{2}}	\label{eqn:epsilon ball} \\
				&	= \twonorm{\epsilon \twonorm{\vec{x}}	\begin{bmatrix}
																1 \\ 1
															\end{bmatrix}
						+	\begin{bmatrix}
								\abs{\ip{\vec{z}_{1}}{\vec{x}}} \\
								\abs{\ip{\vec{z}_{2}}{\vec{x}}}
							\end{bmatrix}}	\notag \\
				&	\leq \sqrt{2} \epsilon \twonorm{\vec{x}} + \twonorm{\mat{P}_{\mathcal{C}\bb{\mat{Z}}} \vec{x}}	\label{eqn:triangle inequality}
			\end{align}
			\label{eqn:control proj norm diff}%
		\end{subequations}
		where \eqref{eqn:abs val inside} is due to $\bb{x + y}^{2} \leq \bb{\abs{x} + \abs{y}}^{2}, \forall x,y \in \setR$, \eqref{eqn:epsilon ball} is due to the Cauchy-Schwartz inequality and the bound $\twonorm{\vec{y}_{j} - \vec{z}_{j}} \leq \epsilon, j = 1,2$ as $\mat{Y} - \mat{Z} \in \epsilon \mathcal{D}\argd{m}$, and \eqref{eqn:triangle inequality} is due to the triangle inequality.
		Since $\mat{Y}$ and $\mat{Z}$ are interchangeable in the derivation of \eqref{eqn:triangle inequality} and $\vec{x}$ is arbitrary, we immediately arrive at \eqref{eqn:projection bound}.

	\section{Proof of \theoremname~\ref{thm:whp_infinite}}
		\label{sec:whp_infinite theorem proof}
		We follow a proof strategy similar to that of \theoremname~\ref{thm:whp_suff_ident}.
		For any constant $\delta \in \bb{0, 1}$, let $\mathcal{B}_{c}\argd{\delta}$ (respectively $\mathcal{B}_{r}\argd{\delta}$) denote the event that $\exists \mat{X} \in \mathcal{N}\argd{\mathscr{S}, 2} \bigcap \mathcal{M} \setminus \cc{\mat{0}}$ satisfying, $\twonorm{\mat{P}_{\mathcal{C}\bb{\mat{X}}} \vec{x}}^{2} \geq \bb{1 - \delta} \twonorm{\vec{x}}^{2}$ (respectively $\twonorm{\mat{P}_{\mathcal{R}\bb{\mat{X}}} \vec{y}}^{2} \geq \bb{1 - \delta} \twonorm{\vec{y}}^{2}$), and let $\mathcal{A}\argd{\delta}$ denote the event that $\exists \mat{X} \in \mathcal{N}\argd{\mathscr{S}, 2} \bigcap \mathcal{M} \setminus \cc{\mat{0}}$ satisfying both $\twonorm{\mat{P}_{\mathcal{C}\bb{\mat{X}}} \vec{x}}^{2} \geq \bb{1 - \delta} \twonorm{\vec{x}}^{2}$ and $\twonorm{\mat{P}_{\mathcal{R}\bb{\mat{X}}} \vec{y}}^{2} \geq \bb{1 - \delta} \twonorm{\vec{y}}^{2}$.
		We note that $\mathcal{A}\argd{\delta}$ constitutes a non-decreasing sequence of sets as $\delta$ increases.
		Hence, using continuity of the probability measure from above we have,
		\begin{subequations}
			\begin{align}
				\Pr\argd{\mathcal{A}\argd{0}}	& \leq \Pr\argd{\mathcal{A}\argd{\delta}}	\label{eqn:inequality copy} \\
				\shortintertext{for any $\delta \in \bb{0,1}$, and}
				\Pr\argd{\mathcal{A}\argd{0}}	& = \lim_{\delta \to 0} \Pr\argd{\mathcal{A}\argd{\delta}}.	\label{eqn:limit equality copy}
			\end{align}
		\end{subequations}
		Note that $\mathcal{A}\argd{0}$ denotes the event that $\exists \mat{X} \in \mathcal{N}\argd{\mathscr{S}, 2} \bigcap \mathcal{M} \setminus \cc{\mat{0}}$ satisfying both $\vec{x} \in \mathcal{C}\bb{\mat{X}}$ and $\vec{y} \in \mathcal{R}\bb{\mat{X}}$ which is a ``hard'' event.
		The event $\mathcal{A}\argd{0}^{\comp}$ corresponds precisely to the sufficient conditions of \theoremname~\ref{thm:suff_ident}.
		Hence, it is sufficient to obtain an appropriate lower bound for $\Pr\argd{\mathcal{A}\argd{0}^{\comp}}$, or alternatively, upper bound $\Pr\argd{\mathcal{A}\argd{0}}$ using \eqref{eqn:inequality copy}.
		It is straightforward to see that
		\begin{subequations}
			\begin{align}
				\SwapAboveDisplaySkip
				\Pr\argd{\mathcal{A}\argd{\delta}}
				& \leq \Pr\argd{\mathcal{B}_{r}\argd{\delta} \bigcap \mathcal{B}_{c}\argd{\delta}}	\label{eqn:decompose probabilities a}	\\
				& = \Pr\argd{\mathcal{B}_{r}\argd{\delta}} \Pr\argd{\mathcal{B}_{c}\argd{\delta}},	\label{eqn:decompose probabilities}
			\end{align}
		\end{subequations}
		where \eqref{eqn:decompose probabilities a} is because $\mathcal{A}\argd{\delta}$ happens only when $\mathcal{B}_{r}\argd{\delta}$ and $\mathcal{B}_{c}\argd{\delta}$ are caused by the same matrix $\mat{X} \in \mathcal{N}\argd{\mathscr{S}, 2} \bigcap \mathcal{M} \setminus \cc{\mat{0}}$, and \eqref{eqn:decompose probabilities} is due to mutual independence between $\vec{x}$ and $\vec{y}$.
		
		We have $\vec{x}$ and $\vec{y}$ drawn component-wise \iid~from a symmetric Bernoulli distribution.
		For any given $\mat{Y} \in \mathcal{N}\argd{\mathscr{S}, 2} \bigcap \mathcal{M} \setminus \cc{\mat{0}}$ we have a bound on $\Pr\argd{\twonorm{\mat{P}_{\mathcal{C}\bb{\mat{Y}}} \vec{x}} \geq \sqrt{1 - \delta}\twonorm{\vec{x}}}$ from \lemmaname~\ref{lem:Bernoulli Chernoff estimate}.
		We focus on the union bounding step to compute $\Pr\argd{\mathcal{B}_{c}\argd{\delta}}$.
		The proof of \lemmaname~\ref{lem:metric entropy} assures us that as long as $\mat{Y}, \mat{Z} \in \mathcal{G}\argd{m} \bigcap \set{\mathcal{C} \bb{\mat{X}}}{\mat{X} \in \mathcal{N} \argd{\mathscr{S}, 2}  \bigcap \mathcal{M} \setminus \cc{\mat{0}}}$ are close enough, \ie~within the same $\epsilon \mathcal{D}\argd{m}$ ball for some $1 > \epsilon \geq \epsilon_{0} > 0$, we are guaranteed tight control over $\abs{\twonorm{\mat{P}_{\mathcal{C}\bb{\mat{Y}}} \vec{x}} - \twonorm{\mat{P}_{\mathcal{C}\bb{\mat{Z}}} \vec{x}}}$ for any arbitrary $\vec{x}$.
		In fact, using \eqref{eqn:control proj norm diff} we have the bound
		\makeatletter
			\if@twocolumn
				\begin{multline}
					\Pr\argd{\exists\mat{Y} \in \mat{Z} + \epsilon \mathcal{D}\argd{m}, \twonorm{\mat{P}_{\mathcal{C}\bb{\mat{Y}}} \vec{x}} \geq \sqrt{1 - \delta}\twonorm{\vec{x}}}	\\
					\leq \Pr\argd{\twonorm{\mat{P}_{\mathcal{C}\bb{\mat{Z}}} \vec{x}} \geq \bb{\sqrt{1 - \delta} - \sqrt{2}\epsilon}\twonorm{\vec{x}}}.
					\label{eqn:bound from lemma 4}
				\end{multline}
			\else
				\begin{equation}
					\Pr\argd{\exists\mat{Y} \in \mat{Z} + \epsilon \mathcal{D}\argd{m}, \twonorm{\mat{P}_{\mathcal{C}\bb{\mat{Y}}} \vec{x}} \geq \sqrt{1 - \delta}\twonorm{\vec{x}}}
					\leq \Pr\argd{\twonorm{\mat{P}_{\mathcal{C}\bb{\mat{Z}}} \vec{x}} \geq \bb{\sqrt{1 - \delta} - \sqrt{2}\epsilon}\twonorm{\vec{x}}}.
					\label{eqn:bound from lemma 4}
				\end{equation}
			\fi
		\makeatother
		Letting $\mat{Z}_{k} \in \setR^{m \times 2}$ denote the center of the $k^\thp$ $\epsilon \mathcal{D}\argd{m}$ ball we have $k$ ranging from 1 to $\exp\BB{p_{c} \log \Theta\argd{1/\epsilon}}$.
		We thus have $\Pr \argd{\mathcal{B}_{c}\argd{\delta}}$ upper bounded by
		\begin{subequations}
			\begin{align}
				\MoveEqLeft[1] \sum_{k} \Pr\argd{\exists\mat{Y} \in \mat{Z}_{k} + \epsilon \mathcal{D}\argd{m}, \twonorm{\mat{P}_{\mathcal{C}\bb{\mat{Y}}} \vec{x}} \geq \sqrt{1 - \delta}\twonorm{\vec{x}}}	\label{eqn:union bounding} \\
				& \leq \sum_{k} \Pr\argd{\twonorm{\mat{P}_{\mathcal{C}\bb{\mat{Z}_{k}}} \vec{x}} \geq \bb{\sqrt{1 - \delta} - \sqrt{2}\epsilon}\twonorm{\vec{x}}}	\label{eqn:pointwise bound}\\
				& \leq \exp\BB{p_{c} \log \Theta\argd{\frac{1}{\epsilon}}} \Pr\argd{\twonorm{\mat{P}_{\mathcal{C}\bb{\mat{Z}}} \vec{x}} \geq \sqrt{1 - \delta'}\twonorm{\vec{x}}}	\label{eqn:intermediate bound} \\
				& \leq 4\exp\BB{p_{c} \log \Theta\argd{\frac{1}{\epsilon}} - \frac{m\bb{1 - \delta'}}{4}}	\label{eqn:final bound}
			\end{align}
		\end{subequations}
		where \eqref{eqn:union bounding} is from an elementary union bound, \eqref{eqn:pointwise bound} is from \eqref{eqn:bound from lemma 4}, \eqref{eqn:intermediate bound} uses
		\begin{equation}
			\delta' = 1 - \bb{\sqrt{1 - \delta} - \sqrt{2}\epsilon}^{2},
			\label{eqn:delta prime}
		\end{equation}
		with $\mat{Z}$ being generic, and \eqref{eqn:final bound} is true due to \lemmaname~\ref{lem:Bernoulli Chernoff estimate}.

		Replicating a similar sequence of steps to bound $\Pr\argd{\mathcal{B}_{r}\argd{\delta}}$, one readily obtains the bound
		\begin{equation}
			\Pr\argd{\mathcal{B}_{r}\argd{\delta}} \leq 4\exp\BB{p_{r} \log \Theta\argd{\frac{1}{\epsilon}} - \frac{n\bb{1 - \delta'}}{4}}	\label{eqn:final bound rows}
		\end{equation}
		with $\delta'$ given by \eqref{eqn:delta prime}.
		Hence, combining \eqref{eqn:inequality copy}, \eqref{eqn:decompose probabilities}, \eqref{eqn:final bound} and \eqref{eqn:final bound rows} we get
		\begin{equation}
			\makeatletter
				\if@twocolumn
					\begin{split}
						\Pr\argd{\mathcal{A}\argd{0}}	& \leq \Pr\argd{\mathcal{B}_{r}\argd{\delta}} \Pr\argd{\mathcal{B}_{c}\argd{\delta}}	\\
							& \leq 16\exp\BB{\bb{p_{c} + p_{r}} \log \Theta\argd{\frac{1}{\epsilon}} - \bb{m + n}\frac{1 - \delta'}{4}}
					\end{split}
				\else
					\Pr\argd{\mathcal{A}\argd{0}}
					\leq \Pr\argd{\mathcal{B}_{r}\argd{\delta}} \Pr\argd{\mathcal{B}_{c}\argd{\delta}}
					\leq 16\exp\BB{\bb{p_{c} + p_{r}} \log \Theta\argd{\frac{1}{\epsilon}} - \bb{m + n}\frac{1 - \delta'}{4}}
				\fi
			\makeatother
			\label{eqn:advertised bound 1}
		\end{equation}
		which yields the desired bound for $\Pr\argd{\mathcal{A}\argd{0}^\comp}$ when $p = p_{c} + p_{r}$.

	\section{Proof of \theoremname~\ref{thm:whp_infinite_Gaussian}}
		\label{sec:whp_infinite_Gaussian theorem proof}
		We have $\vec{x}$ and $\vec{y}$ drawn component-wise \iid~from a standard Gaussian distribution.
		The proof is essentially similar to that of \theoremname~\ref{thm:whp_infinite} with one important difference (beside replacing all occurrences of $\mathcal{N}\bb{\mathscr{S}, 2} \bigcap \mathcal{M}$ by $\mathcal{N}\bb{\mathscr{S}, 2}$ and $\epsilon$ assuming values in $\bb{0,1}$): we use the bound given by \lemmaname~\ref{lem:Gaussian Chernoff estimate} instead of \lemmaname~\ref{lem:Bernoulli Chernoff estimate} when evaluating $\Pr\argd{\twonorm{\mat{P}_{\mathcal{C}\bb{\mat{Z}}} \vec{x}} \geq \sqrt{1 - \delta'}\twonorm{\vec{x}}}$ in \eqref{eqn:intermediate bound}.
		This gives us the bounds
		\begin{subequations}
			\begin{align}
				\Pr\argd{\mathcal{B}_{c}\argd{\delta'}}	& \leq C'\argd{m,\delta'}\exp\BB{p_{c} \log \Theta\argd{\frac{1}{\epsilon}} -m \log \frac{1}{\sqrt{\delta'}}}	\\
				\shortintertext{and (analogously),}
				\Pr\argd{\mathcal{B}_{r}\argd{\delta'}}	& \leq C'\argd{n,\delta'} \exp\BB{p_{r} \log \Theta\argd{\frac{1}{\epsilon}} - n \log \frac{1}{\sqrt{\delta'}}}
			\end{align}
		\end{subequations}
		where
		\makeatletter
			\if@twocolumn
				\begin{equation}
					\begin{split}
						C'\argd{m,\delta'}	& = \exp\BB{2 \log m - \frac{2}{m} + 2 - \log \frac{2 \delta'}{1 - \delta'}}	\\
						& = 0.5 \exp\bb{2} \bb{\frac{1-\delta'}{\delta'}} \exp\BB{2 \log m - \frac{2}{m}}	\\
						& = \bb{\frac{1-\delta'}{\delta'}} \Theta\argd{m^{2}},
					\end{split}
				\end{equation}
			\else
				\begin{equation}
					\begin{split}
						C'\argd{m,\delta'}	& = \exp\BB{2 \log m - \frac{2}{m} + 2 - \log \frac{2 \delta'}{1 - \delta'}}	\\
						& = 0.5 \exp\bb{2} \bb{\frac{1-\delta'}{\delta'}} \exp\BB{2 \log m - \frac{2}{m}}
						= \bb{\frac{1-\delta'}{\delta'}} \Theta\argd{m^{2}},
					\end{split}
				\end{equation}
			\fi
		\makeatother
		As in \eqref{eqn:advertised bound 1}, we have
		\makeatletter
			\if@twocolumn
				\begin{equation}
					\begin{split}
						\Pr\argd{\mathcal{A}\argd{0}}	& \leq \Pr\argd{\mathcal{B}_{r}\argd{\delta'}} \Pr\argd{\mathcal{B}_{c}\argd{\delta'}}	\\
						& \leq C'\argd{m,\delta'} C'\argd{n,\delta'} \exp\BB{\bb{p_{r} + p_{c}} \log \Theta\argd{\frac{1}{\epsilon}}}	\\
						& \hphantom{\leq C'\argd{m,\delta'}} \times \exp\BB{-\bb{m + n} \log \frac{1}{\sqrt{\delta'}}}
					\end{split}
				\end{equation}
			\else
				\begin{equation}
					\begin{split}
						\Pr\argd{\mathcal{A}\argd{0}}
						& \leq \Pr\argd{\mathcal{B}_{r}\argd{\delta'}} \Pr\argd{\mathcal{B}_{c}\argd{\delta'}}	\\
						& \leq C'\argd{m,\delta'} C'\argd{n,\delta'} \exp\BB{\bb{p_{r} + p_{c}} \log \Theta\argd{\frac{1}{\epsilon}}} \exp\BB{-\bb{m + n} \log \frac{1}{\sqrt{\delta'}}}
					\end{split}
				\end{equation}
			\fi
		\makeatother
		which gives the desired bound, since $p = p_{c} + p_{r}$ and
		\begin{equation}
			C'\argd{m,\delta'} C'\argd{n,\delta'} = \bb{\frac{1-\delta'}{\delta'}}^{2} \Theta\argd{m^{2}} \Theta\argd{n^{2}} = C\argd{m,n,\delta'}.
		\end{equation}

	\section{Proof of \propositionname~\ref{prop:rank-2 nullspace}}
		\label{sec:rank-2 null space proposition proof}
		Let $\mat{X}$ admit a factorization as in \eqref{eqn:rank-2 nullspace}.
		Then,
		\begin{equation}
			\mat{X} =	\underbrace{\begin{bmatrix}
										\vec{0}	&	\vec{u}\tpose{\vec{v}}	\\
										0	&	\tpose{\vec{0}}
									\end{bmatrix}}_{\mat{X}_{1}}
					+	\underbrace{\begin{bmatrix}
										\tpose{\vec{0}}		&	0	\\
										- \vec{u} \tpose{\vec{v}}	&	\vec{0}
									\end{bmatrix}}_{\mat{X}_{2}}
			\label{eqn:anti-diagonal representation}
		\end{equation}
		and we see that the matrix $\mat{X}_{2}$ is obtained by shifting down the elements of the matrix $\mat{X}_{1}$ by one unit along the anti diagonals, and then flipping the sign of each element.
		Since the convolution operator $\mathscr{S}\argd{\cdot}$ sums elements along the anti diagonals (see \figurename~\ref{fig:lifting example} for illustration), the representation of $\mat{X}$ as in \eqref{eqn:anti-diagonal representation} immediately implies that $\mathscr{S}\argd{\mat{X}} = \vec{0}$.
		Since \eqref{eqn:rank-2 nullspace} implies that $\rank{\mat{X}} \leq 2$ so we have $\mat{X} \in \mathcal{N}\argd{\mathscr{S}, 2}$.
\end{document}